\documentclass[useAMS,usenatbib]{mn2e}
\usepackage{graphicx}
\usepackage[figuresright]{rotating}
\usepackage{subfigure}
\usepackage{amssymb}

\newcommand{\ii}{\'\i}
\newcommand{\ion}[2]{#1{\sc #2}} 

\usepackage{color}

\title[The interacting systems AM\,2058--381 and AM\,1228--260]
{Photometry and dynamics of the minor mergers AM\,1228-260 and AM\,2058-381}

\author[Hernandez-Jimenez et al.]
{J.~A.~Hernandez-Jimenez$^1$\thanks{E-mail:hernandez.jimenez@ufrgs.br}, 
M.~G.~Pastoriza$^1$,  C.~Bonatto$^1$,  I.~Rodrigues$^2$, \and A.~C.~Krabbe$^2$, Cl\'audia ~Winge  \\
$^1$ Instituto de F\ii sica, Universidade Federal do Rio Grande do Sul, Av.~Bento Gon\c{c}alves,9500, Cep 91501-970, Porto Alegre, RS, Brazil\\
$^2$ Universidade do Vale do Para\'iba, Av. Shishima Hifumi, 2911, Cep 12244-000, S\~ao Jos\'e dos Campos, SP, Brazil\\
}
 
\begin{document}

\date{Accepted -. Received -.}

\pagerange{\pageref{firstpage}--\pageref{lastpage}} \pubyear{2006}

\maketitle

\begin{abstract}

We investigate interaction effects on the dynamics and morphology of the galaxy 
pairs AM\,2058-381 and AM\,1228-260. This work is based on  $r'$ images and  
long-slit spectra obtained with the Gemini Multi-Object Spectrograph at the 
Gemini South Telescope. The luminosity ratio between the main (AM\,2058A) 
and secondary (AM\,2058B) components of the first pair  is  a factor of   
$\sim$ 5,  while for the other pair, the  main (AM\,1228A) component is 
20 times more luminous than the secondary (AM\,1228B). The four galaxies have 
pseudo-bulges, with  a S\'ersic index $n<2$. Their observed radial velocities 
profiles (RVPs) present several irregularities. The receding side of the RVP of 
AM\,2058A is displaced with respect to the velocity field model, while there is a 
strong evidence that AM\,2058B is a tumbling body, rotating along its major 
axis. The RVPs for AM\,1228A indicate a misalignment between the kinematic and 
photometric major axes. The RVP for AM\,1228B is quite perturbed, very likely due 
to the interaction with AM\,1228A. NFW halo parameters for AM\,2058A are similar to 
those of the Milky Way and M\,31. The halo mass of AM\,1228A is roughly 10\% that of 
AM\,2058A. The mass-to-light (M/L) of AM\,2058 agrees with the mean value derived for late-type 
spirals, while the low M/L for AM\,1228A may be due to the intense star 
formation ongoing in this galaxy.

\end{abstract}


\section{Introduction}

Within the $\lambda$CDM cosmology framework, mergers or interactions play a fundamental
role in the formation, growth and subsequent galactic evolution 
\citep[e.g.,][ and references therein]{somerville01,hopkins10}. Indeed, as shown 
in merger trees of hierarchical models of galaxy formation, the galactic growth is driven by accretion 
of other galaxies, most often minor companions  \citep[e.g.,][]{cole00,wechsler02,bedorf12}. Despite  
their importance, these minor mergers have been less studied than major merger interactions  
\citep{schwarzkopf00}. From the observational point of view, the statistical samples 
show a bias favouring  major mergers, due to  the large magnitude 
differences between galaxies and the magnitude limit set by redshift 
\citep{woods07}. On the other hand, numerical simulations also show a trend 
to study major interactions, since the computational cost is 
larger for minor mergers, due to the higher resolution
required to model the small companions \citep{hernquist95,barnes09}.

Nevertheless, there have been significant advances in understanding minor mergers.
For instance, numerical simulations indicate that they can trigger star formation and  
transform the morphologies of galaxies \citep[e.g.,][]{mihos94,hernquist95,naab03,cox08,qu11}. 
These results have been confirmed by observational studies 
\citep[e.g.,][]{larson78,kennicutt87,donzelli97,barton00,lambas03,woods07,lambas12}.

On the other hand, minor mergers are also recognized as potential agents to drive 
the morphological evolution of galaxies. For example, as a result of a satellite  accretion, 
the galactic discs can become warped and heated \citep[e.g.,][]{quinn93,walker96} 
or  inner structures can be created, such as discs, 
rings and spiral arms \citep[e.g.,][]{eliche11}.
Furthermore, the interaction with a small companion can generate all 
kinds of phenomenons seen in majors cases, such as tidal tails,
bridges, rings, as well as form or destruct bars or spiral arms 
(e.g., \citealt{salo93,mihos97,irapa1999,irapa2000,thies01,krabbe08,krabbe11}).
In addition, the velocity fields of the large galaxy 
often  shows asymmetries and irregularities due to the interaction with 
the smaller companion \citep[e.g.,][]{rubin91,rubin99,dale01,mendes03,fuentes04,krabbe08,hernandez13}. 
Such distortions are seen in the rotation curves as  
significantly rising or falling profiles on the side 
pointing towards the companion galaxy, or pronounced velocity bumps, which are stronger 
at perigalacticum passages and decline 0.5 Gyr after that \citep{kronberger06}. 

The kinematic and photometric effects caused by minor mergers strongly depend on structural parameters,
such as morphological type (bulge, disc, bar, etc.),  baryonic-to-dark mass ratios, and
orbital parameters, such as retrograde, prograde,  inclination and 
coplanar orbits \citep{hernquist95,berentzen03,cox08,eliche11}. Thus, obtaining photometric and kinematic informations 
on minor merger systems is useful for understanding the effects that interaction may have on each component. 
 The decomposition of  the surface brightness profile can be used to infer the stellar mass distribution. 
Rotation curves are used to constrain models of dark matter distribution 
\citep{vanalbada85,carignan85,kent87,blais01}.

\begin{figure*}
\centering
\includegraphics*[width=0.8\textwidth]{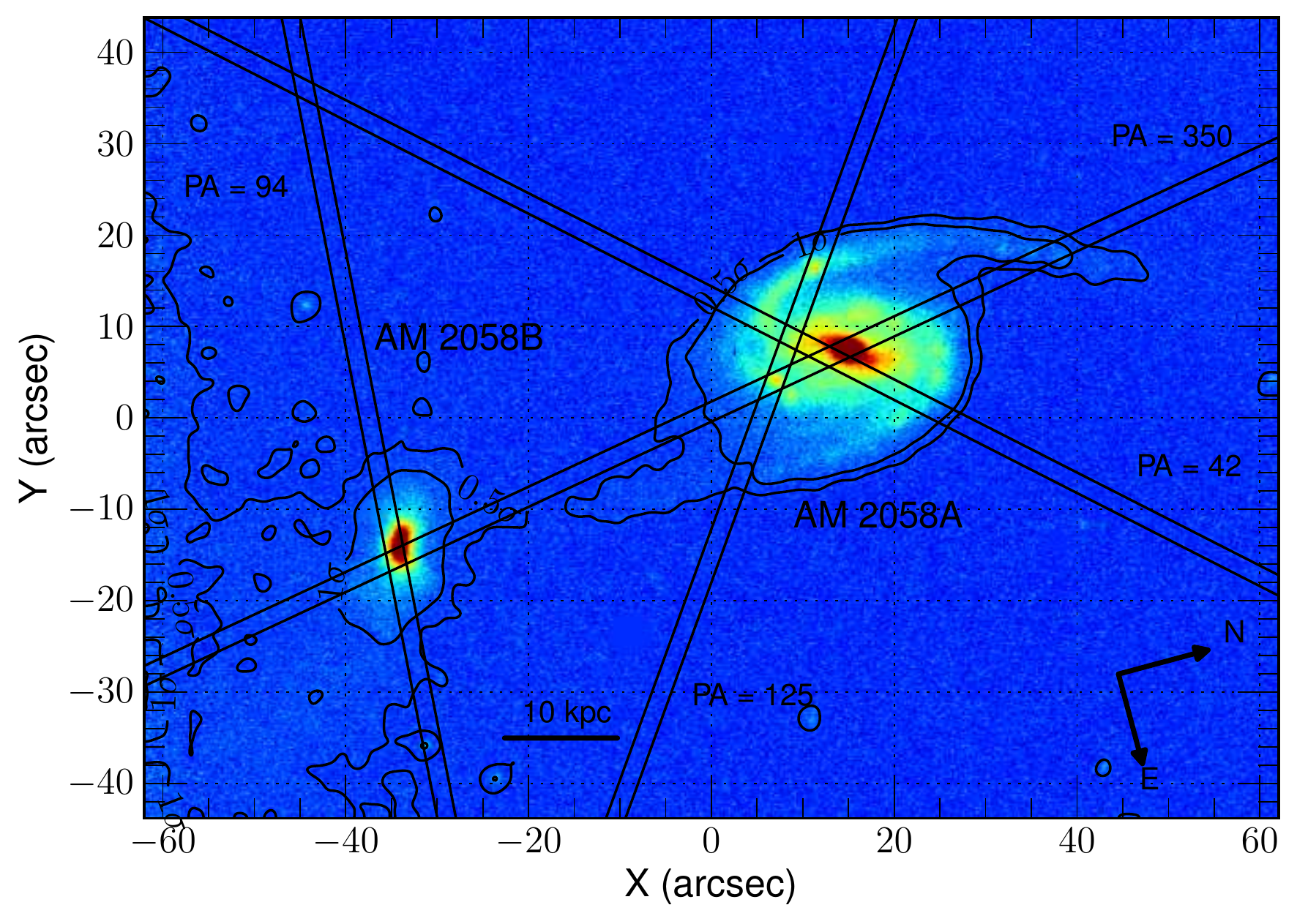}  			
\includegraphics*[width=0.7\textwidth]{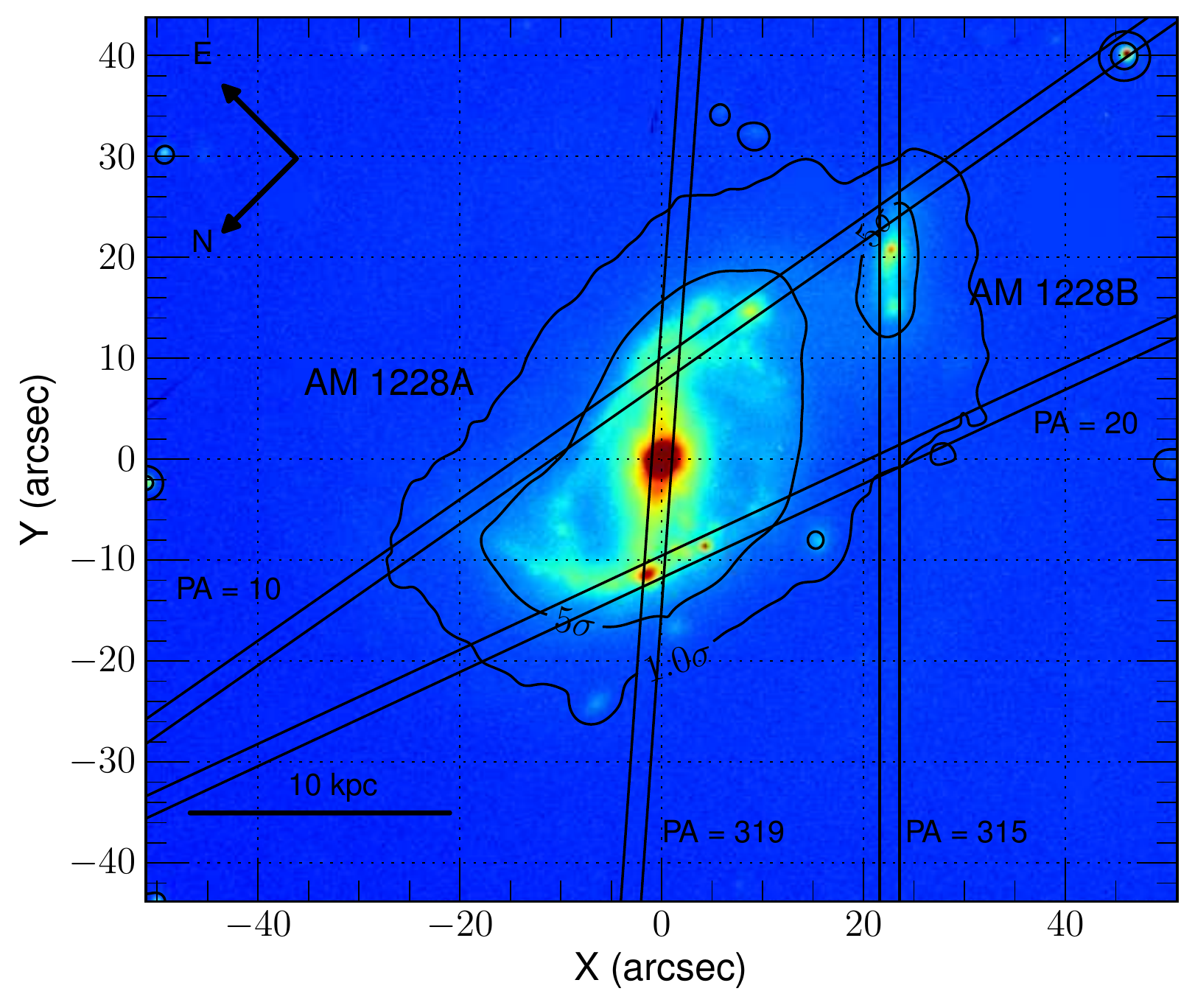}  		
\caption{$r'$ images with the observed slit positions of AM\,2058-381 (top)  and AM\,1228-260 (bottom).
Isophotes with  values above the sky are traced to show the
tidal structures in AM\,2058-381 and AM\,1228-260.}
\label{contours}
\end{figure*}

In order to investigate the interaction effects on kinematic and photometric 
properties of minor merger components, we have selected several systems 
from  \citet{donzelli97} and \citet{winge15} 
samples of interacting galaxies taken from the Arp-Madore catalogue 
\citep{arp87}. These pairs consist of a main 
galaxy (component A) and a companion (component B) that has about half or less 
the diameter of component A. 
The pairs lack basic information, such as morphological types, 
magnitudes and redshifts. Optical spectroscopic properties (e.g, star 
formation rates, diagnostic diagrams, stellar population) of these samples have 
been already studied by \citet{donzelli97}, \citet{pastoriza99} and \citet{winge15}. 
From their samples, we have selected systems in which the main component has a  
well-defined spiral structure, so that the effect of the interaction in the arms is 
easily seen, and the galactic disc has an inclination ($i$) with respect to the plane 
of the sky of 30{\degr}$\leq i \leq$70{\degr}. In addition, these systems have different
separations between the components, morphological distortions and likely
interaction stages.  Long-slit spectroscopy 
and images of these systems were obtained  with the Gemini Multi-Object 
Spectrograph (GMOS) at Gemini South Telescope. 
Previous results from this project have been presented 
for the systems AM\,2306-721 \citep{krabbe08}, AM\,2322-821 \citep{krabbe11} 
and AM\,1219-430 \citep{hernandez13}. Along these works, we have 
developed a robust methodology to obtain the kinematic and photometric 
properties of the galaxies in minor mergers. Such properties are valuable 
constraints for numerical simulations in case studies in order to understand the 
specific mechanisms that drive the collision in an interaction of unequal mass 
galaxies. In this paper, we present the results for two 
other pairs, AM\,2058-381 from \citet{donzelli97}, and AM\,1228-260 from 
\citet{winge15}.  Fig. \ref{contours} shows the $r'$ images of both pairs. 
These systems show different projected  separations between the pair members. 
For AM\,2058-381,  there is a projected distance between galaxy centres of 
$\sim$ 43.3\,kpc ($\sim$ 4.4 diameters of the main galaxy), 
while for AM\,1228-260, the projected distance is $\sim$ 11.9\,kpc 
($\sim$ 2 diameters of the main galaxy).

AM\,2058-381 is composed by a large spiral galaxy (hereafter AM\,2058A) with two
arms, and a small peanut shape companion (hereafter AM\,2058B) (Fig. \ref{contours}).
\citet{ferreiro04}  found that AM\,2058A 
presents bright \ion{H}{ii} regions distributed along the spiral arms.
The ages of these regions are in the range of $5.2\times10^6 < t <
6.7\times10^6$ yr \citep{ferreiro08}. The integrated colours 
of AM\,2058A and AM\,2058B are
rather blue with (B$-$V) = 0.6 and (B$-$V) = 0.4, respectively, indicating an 
enhancement of star formation in both galaxies. \citet{krabbe14} studied
the electron density for this system, and found a wide variation of the electron 
density across AM\,2058A with $33<N_e<911$\,cm$^{-3}$. On the other hand, for AM\,2058B the 
electron densities are relatively low, with a mean value of
$N_e= 86\pm33$\,cm$^{-3}$, which is compatible with that found for  
giant extragalactic  \ion{H}{ii} regions. The 
metallicity gradient in AM\,2058A has a shallow slope
when compared with those of typical isolated spiral galaxies  (Rosa et al. 2014). 
Such flat metallicity gradient has been found in several
interacting galaxies \citep[e.g.,][]{krabbe08,kewley10,krabbe11,rosa14}, and 
may result from  the  interaction that induces  gas inflow from
the external disc towards the central region of the galaxies \citep{dalcanton07,perez11}.

AM\,1228-260 is composed by  a large barred spiral (hereafter AM\,1228A)
and a dwarf galaxy (hereafter AM\,1228B) (see Fig. \ref{contours}).
The main galaxy is classified as an extreme IRAS galaxy \citep{vandenbroek91},  with far-infrared luminosity 
$L_{FIR} = 4\times10^{10}\,L_{\odot}$, and a high luminosity  ratio, $L_{FIR}/L_B\sim8$, 
indicating intense star formation activity. In addition, $H_{\alpha}$ images 
of this system show the main galaxy with luminous \ion{H}{ii} regions along to the spiral arms, 
while the secondary galaxy looks like an irregular galaxy with two dominant  \ion{H}{ii} regions. 
Both galaxies are also rather blue with (B$-$V) = 0.52 and (B$-$V) = 0.66 for
AM\,1228A and AM\,1228B, respectively. 
 
This paper is organized as follows: in Sect. \ref{datared} we provide 
details on the observations and data reduction, photometric calibrations, 
and image restoration. Sect. \ref{photana} gives the integrated magnitudes 
of the galaxies, and describes the morphological analysis and the photometric 
decomposition of the surface brightness profiles. Sect. \ref{vel} describes 
the gas kinematics. In Sect. \ref{massmodel}, we present the bulge, disc 
and halo components used to model the velocity field. In Sect. \ref{fitmodel}, 
we discuss the fit to the velocity field and its results, such as mass distribution 
in the galaxies, and  the determination of the mass-to-light (M/L) ratio of each 
component and halo parameters. Finally, the conclusions are summarized in 
Sect. \ref{final}. Throughout this paper, we adopt the Hubble constant as 
$H_0$=73\,km\,s$^{-1}$\,Mpc$^{-1}$ \citep{spergel07}.

\section{Observations and data reduction}
\label{datared}

\begin{table*}
\caption{Journal of image observations }
\label{observ_ima}
\begin{tabular}{clccc}
\noalign{\smallskip}
\hline
\noalign{\smallskip}
Galaxy  & Date (UT) & Exp. time (s) & Filter & $\Delta \lambda$ (\AA) \\
\noalign{\smallskip}
\hline
\noalign{\smallskip}
AM\,2058-381   &  2007--05--11  &  3$\times$40    & $r^{\prime}$ (G0326)       & 4562--6980  \\
\hline
AM\,1228-260   &  2011--03--20  &  2$\times$30    & $r^{\prime}$ (G0326)      & 4562--6980   \\
               &  2011--03--29  &  1$\times$30    & $r^{\prime}$ (G0326)      & 4562--6980   \\
               &  2011--04--14  &  2$\times$30    & $r^{\prime}$ (G0326)      & 4562--6980   \\
               &  2011--04--15  &  1$\times$30    & $r^{\prime}$ (G0326)      & 4562--6980   \\

\noalign{\smallskip}
\hline
\noalign{\smallskip}
\end{tabular}
\end{table*}

\begin{table*}
\caption{Journal of long-slit observations}
\label{observ_spec}
\begin{tabular}{	clccc}
\noalign{\smallskip}
\hline
\noalign{\smallskip}
Galaxy  &  Date (UT) & Exp. time (s) & PA (\degr)& $\Delta \lambda$ (\AA) \\
\noalign{\smallskip}
\hline
\noalign{\smallskip}

AM\,2058-381   &      2007--05--20  &  4$\times$600    & 42       & 4280--7130  \\
               &      2007--05--24  &  4$\times$600    & 125      & 4280--7130   \\
               &      2007--05--26  &  4$\times$600    & 94       & 4280--7130   \\
               &      2007--05--30  &  4$\times$600    & 350      & 4280--7130   \\
\hline
AM\,1228-260   &      2011--03--20  &  2$\times$900    & 319      & 4449--7312  \\
               &      2011--03--20  &  2$\times$900    & 315      & 4449--7312   \\
               &      2011--03--29  &  2$\times$900    & 20       & 4449--7312   \\
               &      2011--04--14  &  2$\times$900    & 10       & 4449--7312   \\
\noalign{\smallskip}
\hline
\noalign{\smallskip}
\end{tabular}
\end{table*}

This paper is based on $r'$ images and  long-slit spectra  obtained with the GMOS at Gemini South 
Telescope, as part of the poor weather programmes GS-2007A-Q-76 and GS-2011A-Q-90.

Imaging and spectroscopic data reductions were carried out  using 
the {\sc \small{gemini.gmos}} package as well as  generic {\sc \small{iraf}}\footnote{{\sc \small{iraf}} is distributed by 
the National Optical Astronomy Observatories, which is operated by the Association of
Universities for Research in Astronomy, Inc. (AURA) under cooperative agreement with 
the National Science Foundation.} tasks.

As part of the standard target acquisition procedure, we 
obtained sets of short exposure time $r'$ images. 
The journal of observations is presented in Table \ref{observ_ima}. 
The  images were  binned by 2 pixels, 
resulting in a spatial scale of 0.146 arcsec pixel$^{-1}$. They were  processed using
standard procedures (bias subtraction and flat-fielding) and combined to obtain the final $r'$ images. 
The seeing was calculated using {\sc \small{gemseeing}} task of {\sc \small{gemini.gmos}} package. 
This task  derives the median  value  of the full width high maximum for the fields star in the  observed images 
by fitting a Moffat profile. Delivered image quality of $\sim0.82$ and $\sim0.75$ arcsec were 
estimated for  $r'$ combined final images of AM\,2058-381 and AM\,1228-260, respectively.

Spectra were obtained  with the B600 grating
plus the 1 arcsec slit, which gives a spectral resolution of 5.5\,\AA. 
The frames were binned on-chip by 4 and 2 pixels in the spatial and wavelength directions, respectively, 
resulting in a spatial scale of 0.288 arcsec pixel$^{-1}$, and dispersion of 0.9\,\AA\,pixel$^{-1}$. 

Spectra at four different position angles (PAs) were taken for each system. 
Fig. \ref{contours} shows the slit positions
over-plotted on $r'$ images for AM\,2058-381 (top panel) and AM\,1228-260 (bottom).
Dates,  exposure times, PAs and spectral ranges  of spectroscopic observations are listed in  
Table \ref{observ_spec}. Exposures times were limited to minimize the effects of cosmic rays,
and several frames were obtained for each slit position to achieve high signal-to-noise ratio.

We followed the standard procedure for spectroscopy reduction by applying 
bias correction, flat-fielding, cosmic ray cleaning, sky subtraction,
wavelength and relative flux calibrations. In order to increase the
signal-to-noise ratio, the spectra were extracted by summing over
four rows. Thus, each spectrum represents an aperture of 1 $\times$
1.17 arcsec$^2$.  

The distance to each galaxy pair was taken as the radial velocity
measured at the nucleus of the main component (see  Sect.  \ref{vel}).
We obtained  distances of $\sim$ 167 and $\sim$ 80\,Mpc for    
AM\,2058-381 and AM\,1228-260, respectively; thus, 
the apertures samples regions of 809\,$\times$\,946\,pc$^2$
and 388\,$\times$\,454\,pc$^2$ for each pair, respectively.

\subsection{Photometric calibration}
\label{calima}

Since the data were taken in non-photometric conditions, foreground stars 
from United States Naval Observatory-B1.0 Catalogue \citep[USNO-B;][]{monet03}
present in the field-of-view of the images, were used to calibrate the data. 
 Point spread function (PSF) photometry of these stars was performed using the {\sc \small{psf}} task within 
{\sc \small{iraf/daophot}}. We applied the bandpass transformation given by \citet{monet03} 
to convert the J and F photographic magnitudes 
to $r'$ magnitude in the Sloan Digital Sky Survey (SDSS) photometry system. 
Then, the zero-points for the image
were found to be $m_0=27.28\pm 0.08$ and $m_0=27.83\pm 0.09$  
for  AM\,2058-381 and AM\,1228-260, respectively.

\subsection{Sky background}
\label{skyback}

The sky background levels of the $r'$ images were adopted as the mean value of 
several boxes of $60\times60$ pixels, located far from stars and galaxies in the field-of-view.
The statistical standard deviation ($\sigma$) of the sky background around the mean value was also computed for
these regions, to be used as an estimate of the sky noise, and we adopt 
the value of 1\,$\sigma$ to define the limiting detection level for each system. 
Table \ref{sigma} shows the detection limits, in magnitudes per
square arc-second, of the $r'$ images measured at 1, 2, and 3\,$\sigma$ for pairs AM\,2058-381 and AM\,1228-260.

\subsection{Image restoration}
\label{restima}

One way to enhance star-forming features and morphological structures in images 
is by means of image restoration. In this work, we use the   
Lucy--Richardson (L-R) algorithm \citep{richardson72,lucy74} to
deconvolve the $r'$ images. 
\citet{hernandez13} applied this algorithm with success on images 
of the pair AM\,1219-430 to resolve candidates  star-formation knots in 
several \ion{H}{ii} regions.  With respect to the procedure, we obtained a PSF
 model for the images, and used the {\sc \small{lucy}} task within {\sc \small{iraf/stsdas}}.
The restored data were properly normalized, and  the  integrated flux  in  the  image  was conserved.  
Like any restoration technique, the L-R algorithm 
can introduce spurious information. One of those well know 
artefacts is the appearance of a negative moat around very high contrast 
point sources \citep{pogge02}. This effect is a problem for
images with strongly saturated nuclei, which is here the case of the nucleus of AM\,1228A.
Therefore, the image for this galaxy was not restored. 
The deconvolved  images for AM\,2058A, AM\,2058B and AM\, 1228B are shown in the left-panels of Fig. \ref{elmeimages}.
As described above, the star-forming regions and substructures were enhanced in the images of all
galaxies, particularly, the bright bar shows up in the restored image of AM\,2058A.

\begin{table}
\caption{Sky background levels }
\label{sigma}
\begin{tabular}{@{}clcccc}
\noalign{\smallskip}
\hline
\noalign{\smallskip}
Galaxy  & 1$\sigma$  & 2$\sigma$ & 3$\sigma$  \\
\noalign{\smallskip}
\hline
\noalign{\smallskip}
AM\,1228-260 &   23.32 &  22.57  & 22.13 \\
AM\,2058-381 &   22.91 & 22.16  & 21.72 \\

\noalign{\smallskip}
\hline
\noalign{\smallskip}
\end{tabular}
\end{table}

\section{Photometric Analysis}
\label{photana}
 
 Tidal structures found in pairs
are important clues to trace galactic encounter, as well as of the internal structure
of the galaxy. They also serve for these systems as constraint to a numerical simulation.    
In order to detect tidal structures, we plot isophotes with different 
$\sigma$ levels over the images (see Fig. \ref{contours}). We found for 
AM\,1228-260, at 1\,$\sigma$ brighter than  the sky background, 
a common isophote enclosing   the   members. This tidal structure is broken up 
at 5$\sigma$  in individual isophotes for each galaxy. 
On the other hand, the pair AM\,2058-381 does not show any connecting 
structure between the members above the  1\,$\sigma$ level. However, by relaxing the 
above criteria of 1\,$\sigma$ as detection limit, we found that the main 
galaxy shows two symmetric long tidal tails at  the 0.5\,$\sigma$ level, as 
shown in Fig. \ref{contours} (top panel). 

Table \ref{magpar} lists the integrated apparent ($m_{\scriptsize{\mbox{T}}}$) 
$r'$ magnitudes for the individual galaxies.
For the AM\,1228-260 system, the magnitudes of the components A and B were obtained 
by integrating the flux inside the isophote at a 5\,$\sigma$ level above the sky 
background, thus excluding the common envelope contribution. For the AM\,2058-381, 
the magnitudes of the components were estimating integrating all flux 
above the 1\,$\sigma$ level of the sky background. The surface brightness of those 
limiting isophotes (5\,$\sigma$ and 1\,$\sigma$, respectively) is 
also given in Table \ref{magpar} as $\mu_{lim}$. 
The absolute magnitudes ($M_{\scriptsize{\mbox{T}}}$)  were corrected for the Galactic extinction 
using the infrared-based dust map from \cite{schlafly11}, and
the luminosities ($L_r$)  were estimated by adopting the solar absolute 
$r'$  magnitude of 4.76 \citep{blanton03}.  The  total $r'$ luminosities of 
these systems, obtained integrating all light above the sky background, 
correspond to $7.3\times10^{10}$ and $4.1\times10^{10}\,L_{\odot}$ 
for AM\,2058-381 and AM\,1228-260, respectively.

\begin{table}
\caption{Total magnitudes and luminosities}
\label{magpar}
\begin{tabular}{lcccccc}
\noalign{\smallskip}
\hline
\noalign{\smallskip}
Galaxy &  $m_{\scriptsize{\mbox{T}}}$ & $M_{\scriptsize{\mbox{T}}}$ &  $L_r/\mbox{L}_{\odot}$ & $\mu_{lim}$ ($mag\,arcsec^{-2}$) \\
\noalign{\smallskip}
\hline
\noalign{\smallskip}
AM\,2058A    & 14.09 & $-22.14$  & $5.73\times10^{10}$ & 22.91 \\
AM\,2058B    & 15.88 & $-20.35$  & $1.10\times10^{10}$ & 22.91 \\
Tidal tails  & 16.74 & $-19.19$  & $3.80\times10^{9}$  & 23.63 \\
\hline
AM\,1228A  & 13.24 & $-21.46$  & $3.08\times10^{10}$ & 21.58 \\
AM\,1228B  & 16.58 & $-18.12$  & $1.42\times10^{9}$  & 21.58 \\
Envelope   & 14.27 & $-20.06$  & $8.48\times10^{9}$  & 23.32 \\
\hline
MW$^{(a)}$  & - & $-21.17$  & $2.34\times10^{10}$ & - \\
LMC$^{(a)}$        & - & $-18.60$  & $2.21\times10^{9}$ & - \\
SMC$^{(a)}$        & - & $-17.20$  & $6.08\times10^{8}$ & - \\                      
\noalign{\smallskip}
\hline
\noalign{\smallskip}
\end{tabular}
{\bf Note:} $^{(a)}$ values taken from \citet{robotham12}.
\end{table}

We compared the photometric luminosities of our systems
with those of a well known minor merger, the Milky Way (MW) and Large and  Small 
Magellanic Clouds (LMC and SMC). Their  $r'$ absolute magnitudes and luminosities 
are also listed in  Table \ref{magpar}. AM\,2058A 
is twice more luminous than the MW, while AM\,2058B  
is about five times more luminous than the LMC. Thus, this pair is a 
very luminous  minor merger when compared to the MW system.
In contrast, the main and secondary galaxies in the AM\,1228-260 system present 
luminosities similar to the MW and LMC, respectively.

Comparing the luminosities  of the components in both systems, 
we found that the secondary galaxy in AM\,1228-260 has 5\% of the 
luminosity of the main galaxy in this pair, making it similar in 
terms of luminosity and projected distance ($\sim11.9$\,kpc, or about two 
diameters of the main galaxy), to the barred spiral 
NGC\,1097 and its small companion \citep{garcia03}.
For AM2058-381, the secondary is much brighter, reaching 20\% 
the luminosity of the main component. 

The magnitudes of the tidal structures in AM\,1228-260 and AM\,2058-260 have
been obtained by integrating the flux between the 
1\,$\sigma$--5\,$\sigma$ and 0.5\,$\sigma$--1\,$\sigma$ isophotes, 
respectively (Table \ref{magpar}). The contribution of the tidal 
structures to the total luminosity of the systems are 20 and 5\% 
for AM\,1228-260 and AM\,2058-260, respectively. The contribution to the total
luminosity of the tidal structure of the first pair is comparable with the tidal tails
of the Antennas pair (NGC\,4038/4039) \citep{hibbard01}.

\subsection{Symmetrization method}
\label{elme}

\begin{figure*}
\centering
\includegraphics*[width=0.8\textwidth]{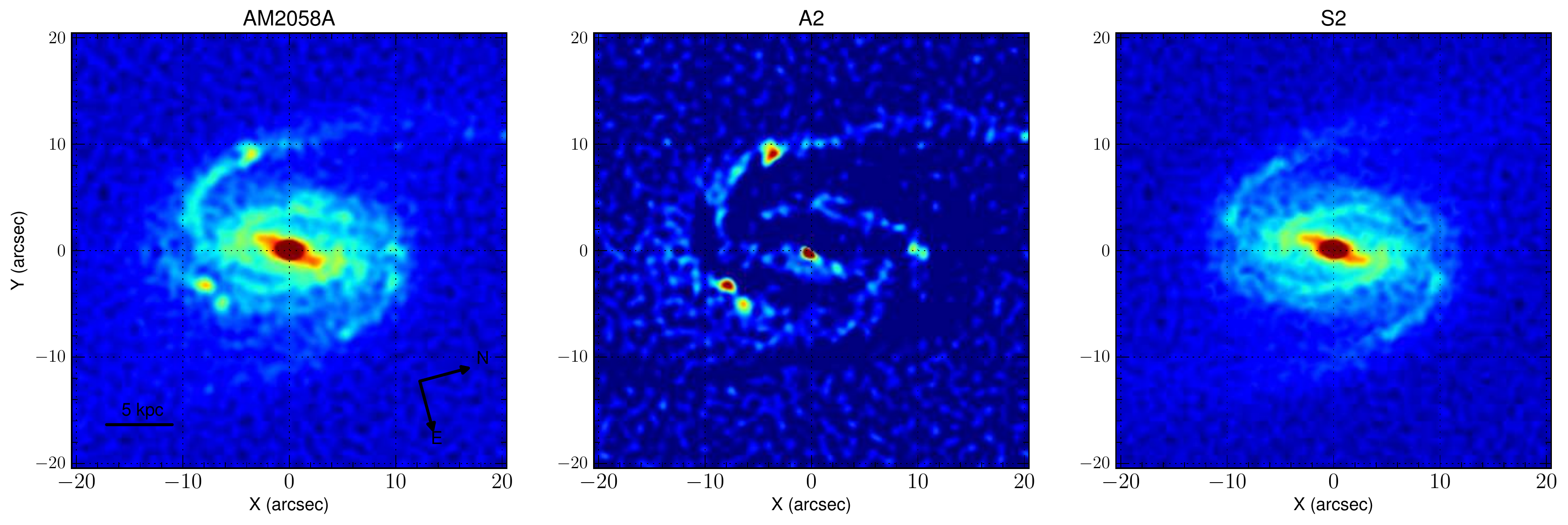} 
\includegraphics*[width=0.8\textwidth]{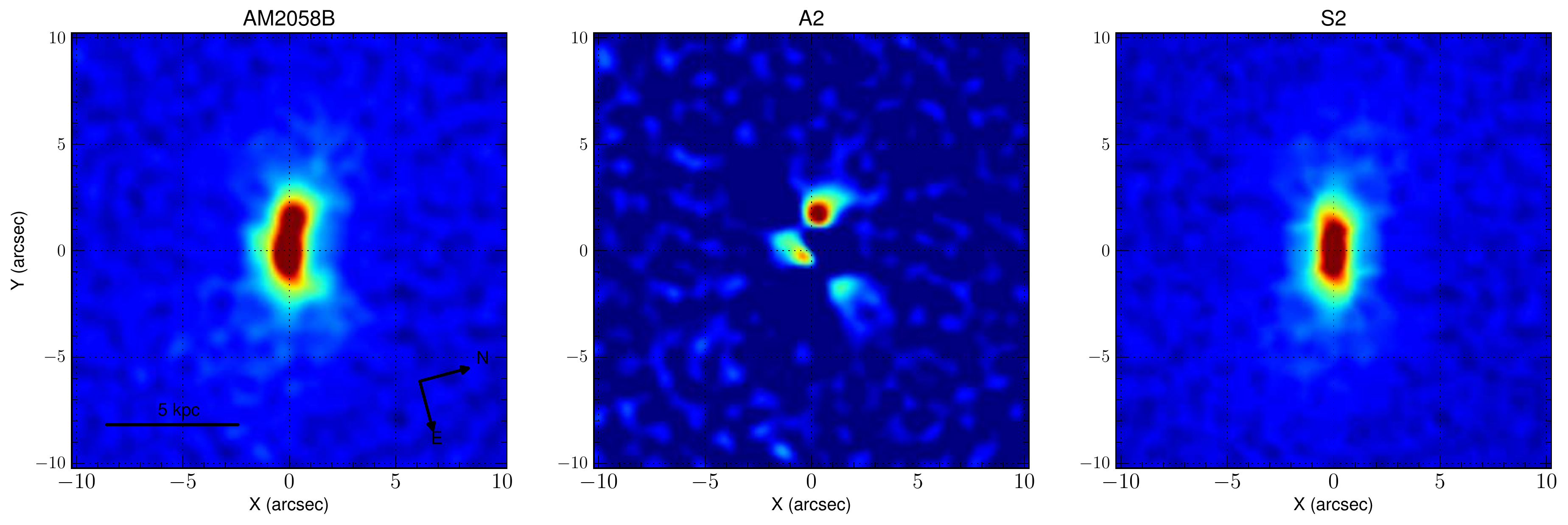}  
\includegraphics*[width=0.8\textwidth]{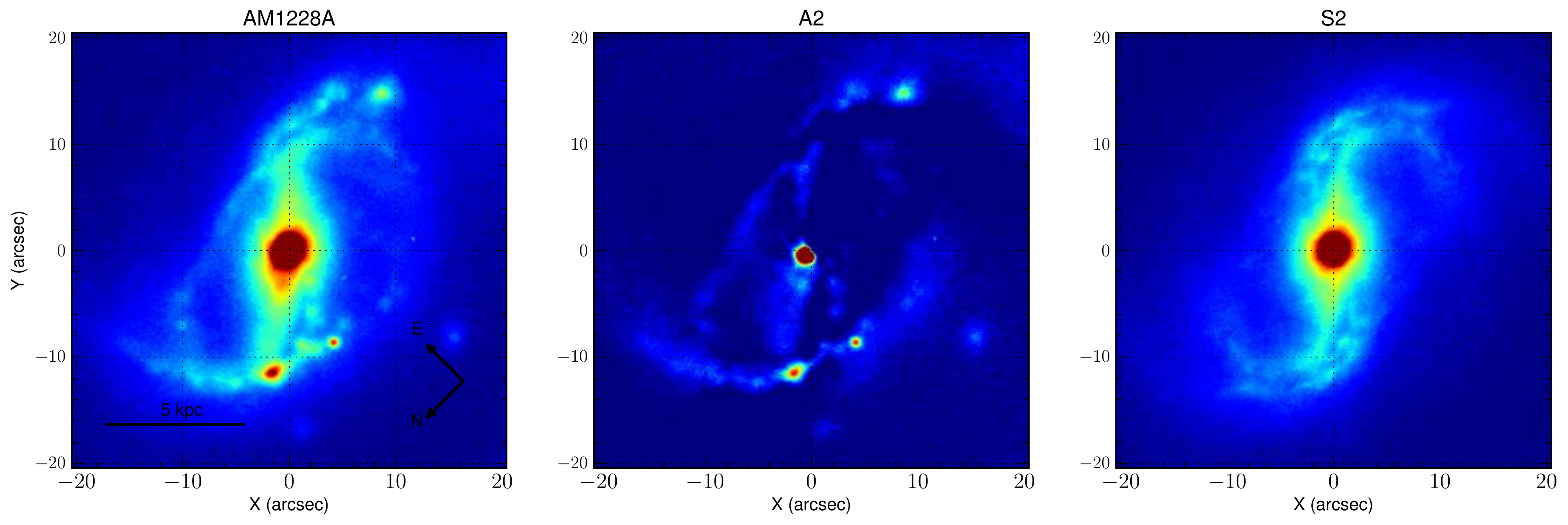}  
\includegraphics*[width=0.8\textwidth]{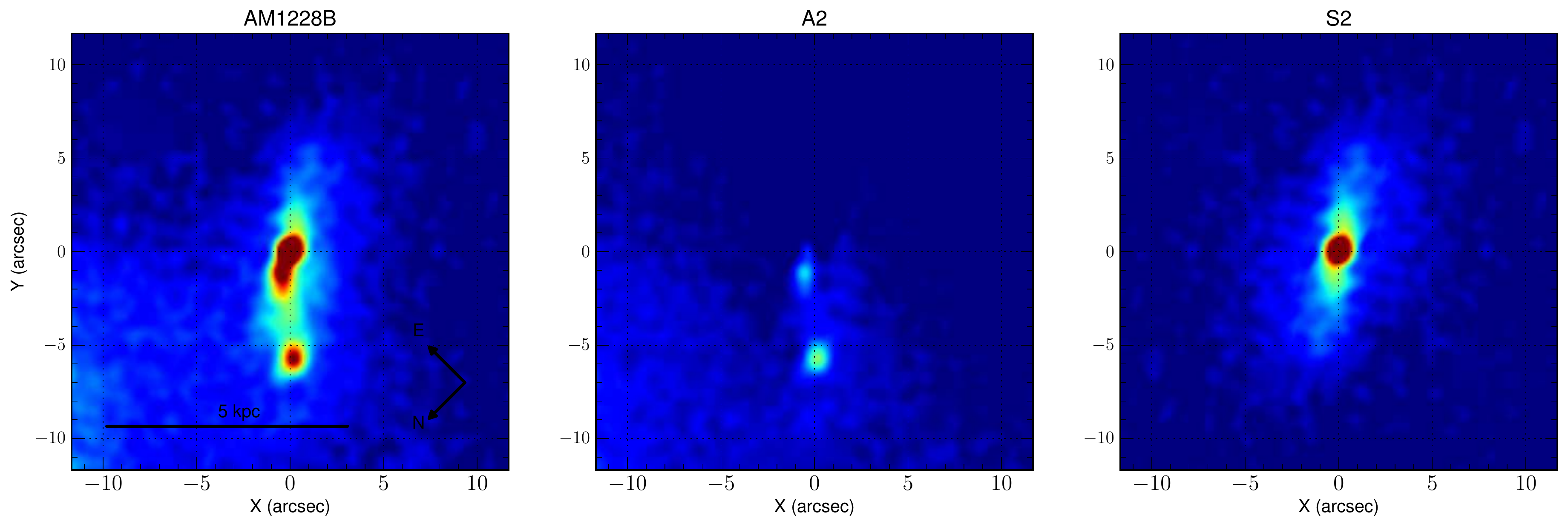}   				     
\caption{Image restoration and symmetrization for the main and secondary galaxies of the two systems.  
Left panels: L-R deconvolved images (except for AM\,1228A, which shows observed 
image, see text); middle and right panels: $A_{2}$ and $S_2$ images obtained from 
the symmetrization analysis.}
\label{elmeimages}
\end{figure*}

In order to subtract the morphological perturbations induced by the interaction, we used 
the symmetrization method of \citet{eem92} and the procedure outlined by \cite{hernandez13}. 
The method retrieves the two-fold symmetric and asymmetric aspects of the spiral galaxy pattern
by making successive image rotations and subtractions. 
The asymmetric image (hereafter $A_{2}$)  is obtained by subtracting from 
the observed image the same image rotated by $\pi$. On the other hand, the symmetric image 
(hereafter $S_{2}$) is obtained by subtracting the asymmetric image  from the observed one. The  
$S_{2}$ image would reveal the non-perturbed spiral pattern and disc. Figure \ref{elmeimages} 
shows the deconvolved  $r'$ images of the galaxies, the $A_{2}$ and $S_{2}$ images.

The  $A_{2}$  image of AM\,2058A shows a tidal arm to the west and a 
pseudo-ring in the disc, as well as
three large \ion{H}{ii} region complexes. The brightest one is on the 
tidal arm, while the others are in the South-East part of the ring. On the 
other hand, the $S_{2}$  image presents two symmetric arms, starting in 
the outer part of the disc. The  $S_{2}$  image reveals 
a faint ring around the bar. The analysis of the surface brightness 
profile confirms the existence of that structure (Sect. \ref{profphot}). 

The  $A_{2}$  image of AM\,2058B reveals
three high surface brightness knots. The one located at 1.42\,kpc W of the galaxy nucleus 
is very luminous when compared to the other two.
The $S_{2}$ image  ``digs up'' the disc structure and a boxy pseudo  bulge.

The $A_{2}$ image of AM\,1228A  shows a distorted ring around a bar, 
as well as an over-density in the North-West part of the 
bar. The over-density at North of the bulge
might be a giant  \ion{H}{ii} region. The $S_{2}$ image allows us to correctly 
classify the morphological type as ovally distorted barred spiral SABc. 
On the other hand, the $A_{2}$  image of AM\,1228B shows a very conspicuous 
North-West \ion{H}{ii} region  at 2.7\, kpc from the nucleus. We also see 
at North in this image, part of the weak common structure of the members. 
The  $S_{2}$ image reveals the underlying disc and bulge for this galaxy. 

The correct determination of the inclination and orientation of a galactic 
disc is not a straightforward task \citep[e.g.,][]{grosbol85,barbera04}, and
even more difficult for interacting systems due to  the morphological 
perturbations. One  advantage 
of the symmetrization method is that the $S_{2}$ images help to reveal the underlying 
galaxy disc. From those, we adopted as the position angle (PA) and inclination  
$i$ of the discs, the mean of the respective values of the 
most external isophotes. The calculated  values are listed in 
Table \ref{ipapar}. Another advantage of the $S_{2}$ images is that
they allow for a more clear classification of the  morphological type of 
the galaxies from the non-perturbed structures. The main components, 
AM\,2058A and AM\,1228A can both be classified as Sc galaxy types
(AM\,1228A is further identified as a SABc, as discussed above), while the secondary 
components, AM\,2058B and  AM\,1228B, are  S0 and Sd types, respectively.

\begin{table}
\caption{Inclination and position angle }
\label{ipapar}
\begin{tabular}{lcc}
\noalign{\smallskip}
\hline
\noalign{\smallskip}
Galaxy & $i$ (\degr)  & PA  (\degr)  \\
\noalign{\smallskip}
\hline
\noalign{\smallskip}
   AM\,2058A    & $58.1\degr\pm0.2\degr$    & $18.9\degr\pm0.5\degr$ \\
   AM\,2058B    & $70.2\degr\pm0.2\degr$    & $79\degr\pm0.1\degr$  \\
\hline
   AM\,1228A    & $63.6\degr\pm0.7\degr$    & $162.1\degr\pm0.5\degr$ \\
   AM\,1228B    & $69.4\degr\pm0.2\degr$    & $151.3\degr\pm0.1\degr$  \\
\noalign{\smallskip}
\hline
\noalign{\smallskip}
\end{tabular}
\end{table}

\subsection{Light profiles}
\label{profphot}

\begin{figure*}
\includegraphics*[width=\columnwidth]{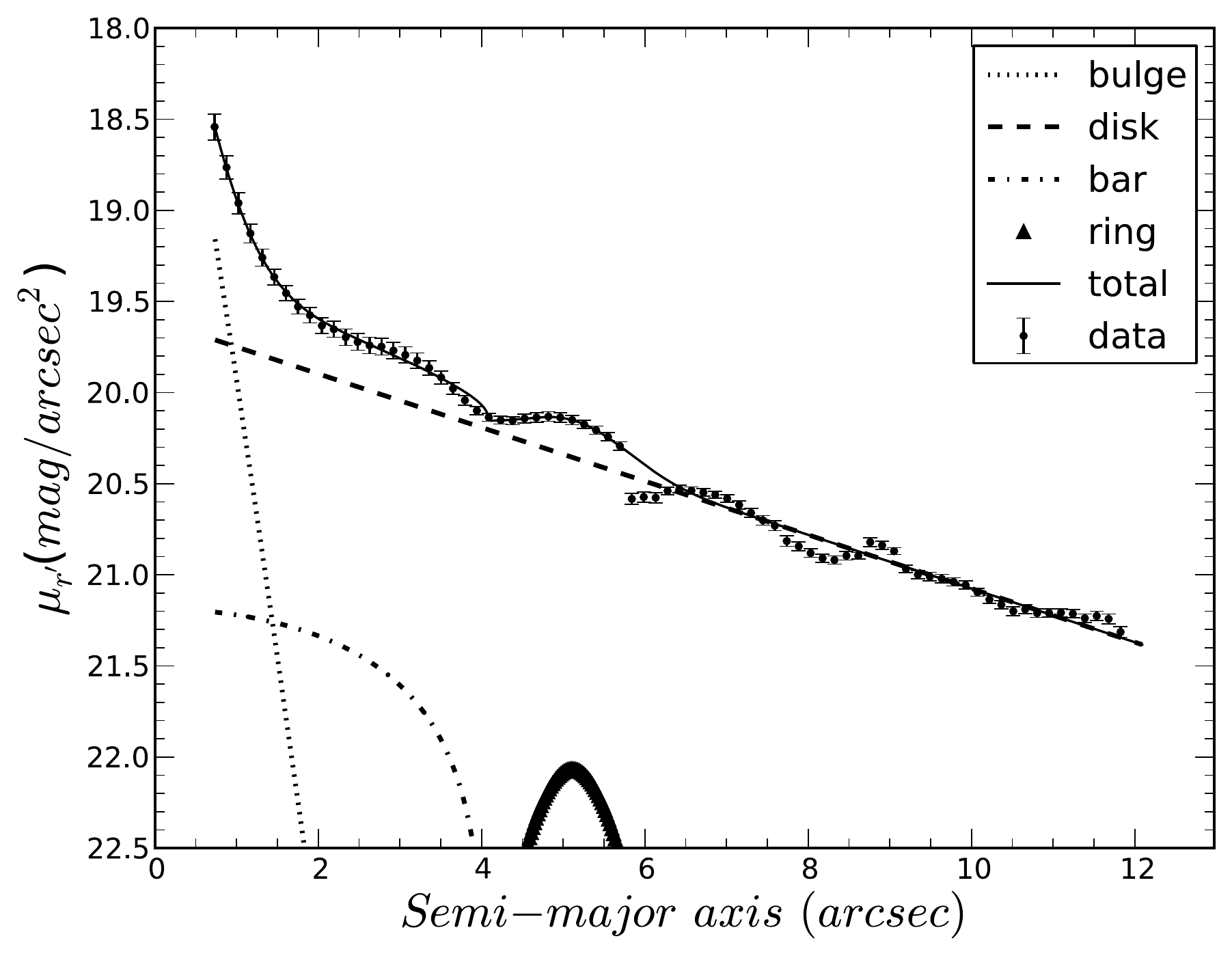}	
\includegraphics*[width=\columnwidth]{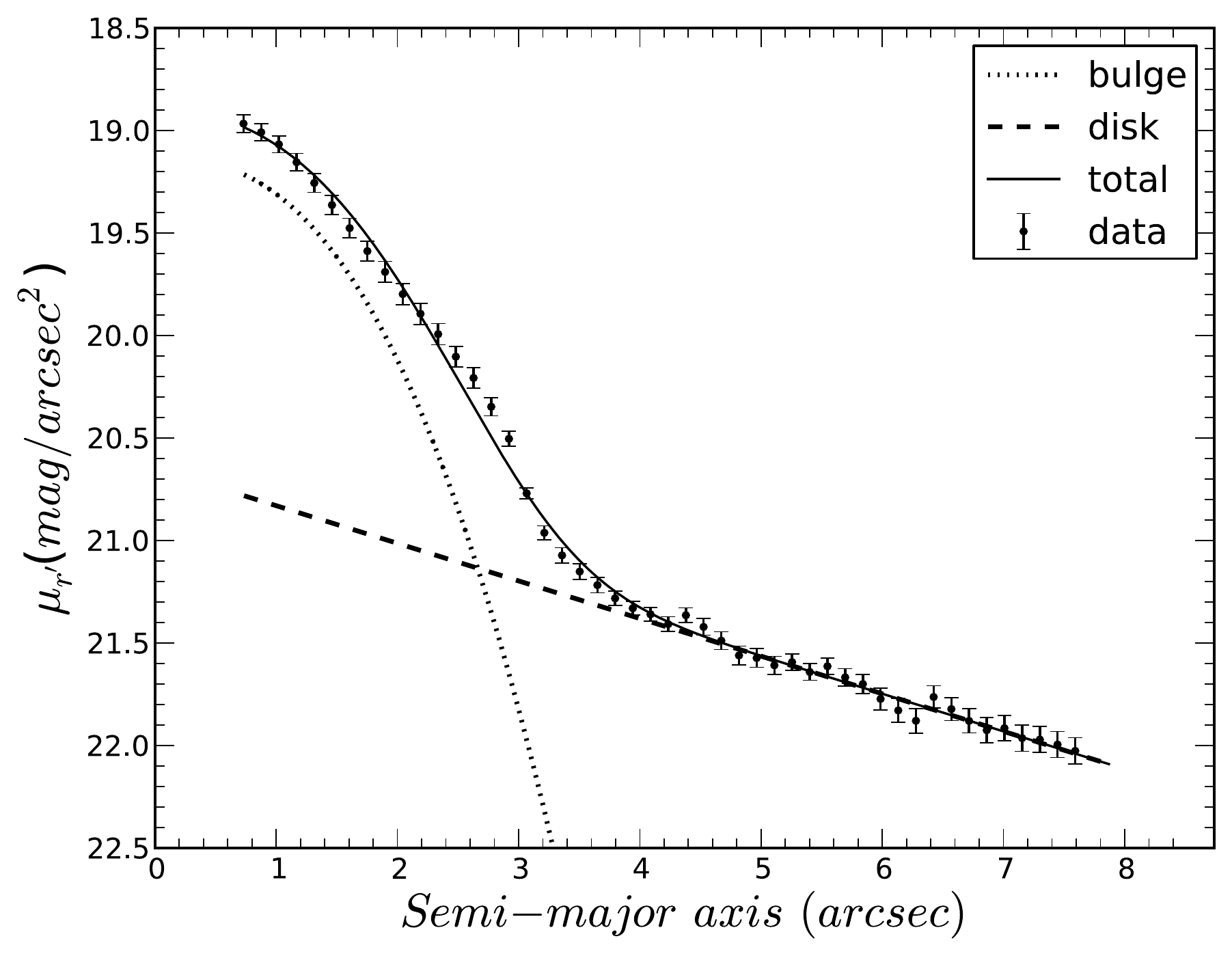}		
\includegraphics*[width=\columnwidth]{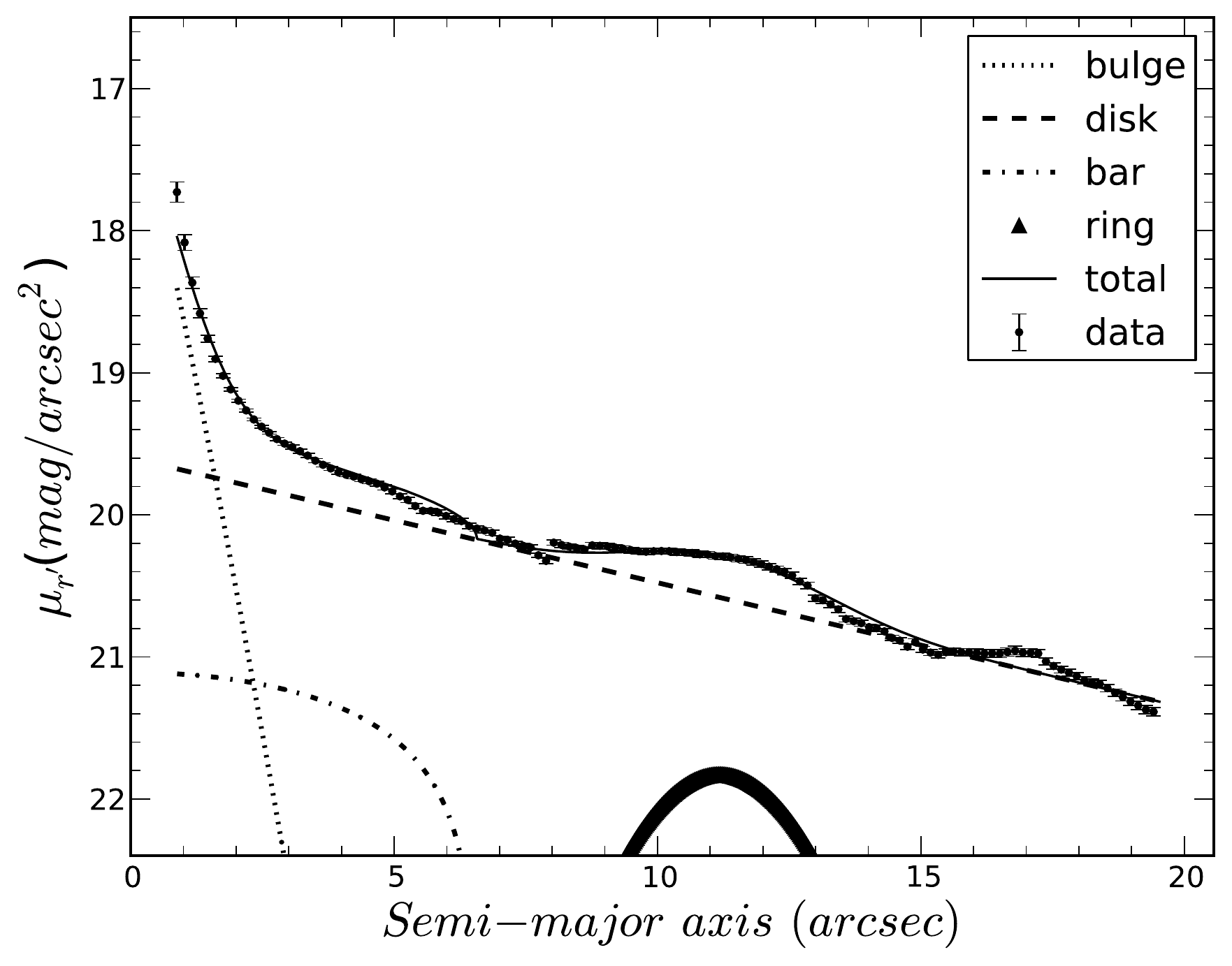}		
\includegraphics*[width=\columnwidth]{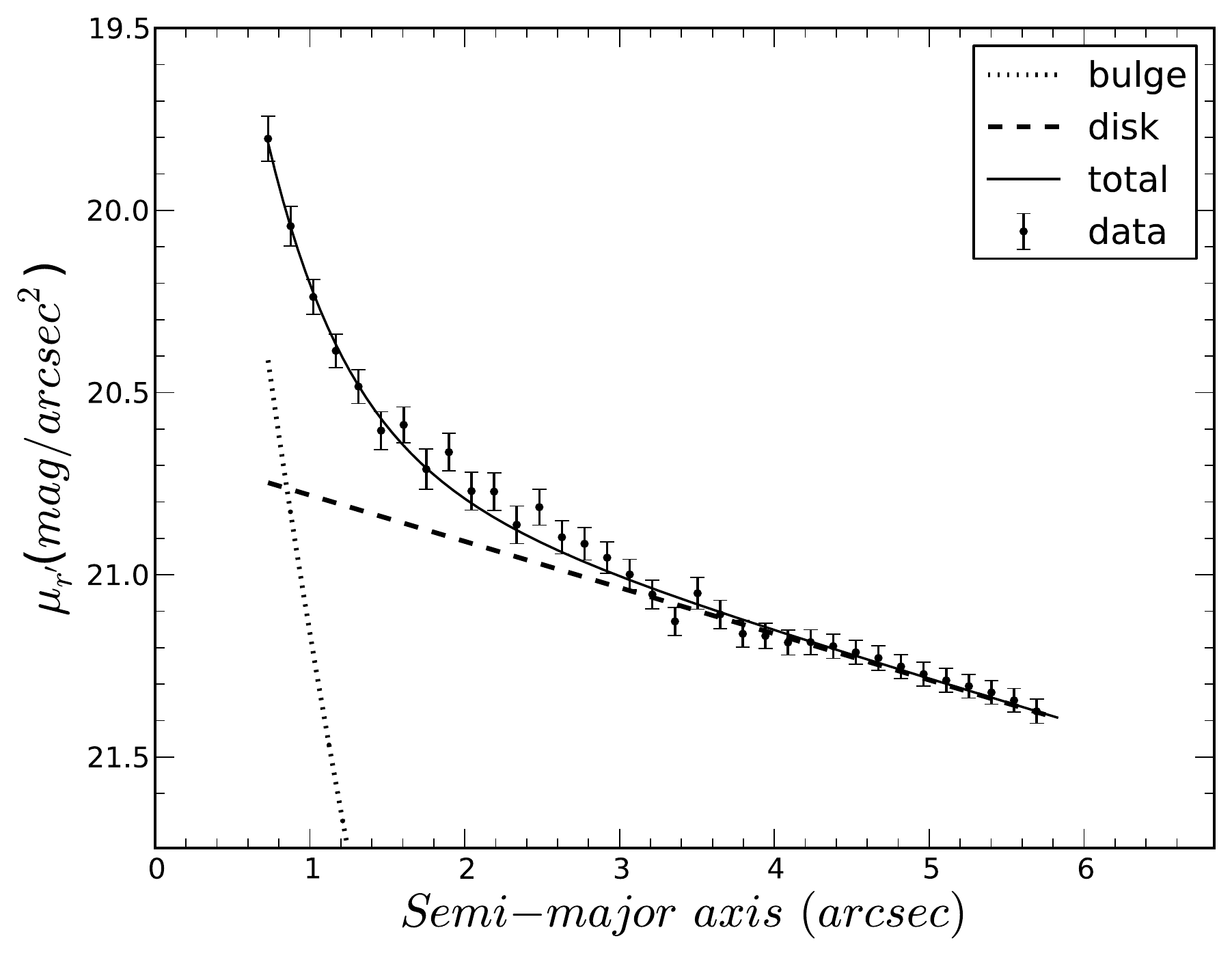}						     
\caption{Structural decomposition of the surface brightness profiles of  AM\,2058A (top-left panel),  AM\,2058B (top-right),
AM\,1228A (bottom-left) and  AM\,1228B (bottom-right).}
\label{prof}
\end{figure*}

In order to derive the  $r'$ surface brightness profiles of the $S_{2}$ 
images, we used the {\sc \small{ellipse}} task of {\sc \small{iraf/stsdas}} \citep{jedrzejewski87} and
followed the same procedure as \citet{hernandez13}, which is based on the methodology of \citet{cabrera04}. 
{\sc \small{ellipse}} fits the isophotal contours with a mean ellipse,
parametrized with values of PA, ellipticity and coordinates of the centre. The best fits 
were achieved by fixing the centre 
positions. During the fitting process, we adopted a clipping factor of 
20\% for the brightest pixels in each annulus
to avoid pixels of star formation regions. We also visually inspected the 
ellipse fits to each galaxy to insure that the position angle at a given 
semi-major radius was not artificially twisted by any star 
formation region, and we noted that 20\% clipping was good 
enough to isophote fit.

To represent the surface brightness profiles, we assume that the surface
luminosity of a galaxy is the sum of the luminosities of each individual 
component. We have used different profiles for the different components: 
an exponential law for the disc \citep{freeman70}, the \citeauthor{sersic68} profile for the bulge 
component \citep{sersic68}, an elliptical profile for the bars \citep{freeman66}, and the \citet{buta96} 
profile to represent a ring. 
The bulge and disc profile can be formally expressed as 
\begin{equation}
I(r)=I_{b}\exp \left[k_{n} \left( \frac{r}{r_{e}} \right)^{\frac{1}{n}}\right],\\
k_{n}=2n-0.324,
\label{sersiclaw}
\end{equation}
and
\begin{equation}
I(r)=I_{d} \exp \left[-\left( \frac{r}{r_{d}} \right)\right].
\label{exponentialdisk}
\end{equation}
where $I_{b}$ and $r_{e}$ are the bulge central intensity and effective radius,
and $I_{d}$ and  $r_{d}$ are the disc central intensity and  the scale length. 
The bar and ring components profiles are given by 

\begin{equation}
I(r)=I_{bar}\left[1-\left(\frac{r}{r_{bar}}\right)^{2}\right]^{1/2},
\label{bar}
\end{equation}
and
\begin{equation}
I(r)=I_{ring} \exp\left[-\frac{1}{2}\left(\frac{r-r_{ring}}{\sigma_{ring}}\right)^{2}\right].
\label{ring}
\end{equation}

The procedure to decompose the surface brightness profiles is described below.
First, the disc component was fitted and subtracted from the original profile. Then, 
the bulge component is fitted to the
residuals, and subtracted from the observed profile. The process 
(fitting then subtracting disc and bulge components) is repeated, and after some iterations, a 
stable set of parameters for the two main components is obtained. Those two are 
then subtracted from the observed profile, and the secondary components (bar and ring) are obtained.
Then, these components  are subtracted
from the observed profile, and the bulge and disc are fitted
again. The process continues until convergence of the parameters is achieved 
\citep[for more details, see][]{hernandez13}. 

Figure \ref{prof} presents the decomposition of the surface brightness profiles of
the pair members of AM\,2058-381  and AM\,1228-260. 
The bulge and disc structural  parameters are listed in Table \ref{photpar},
while the  structural  parameters  for secondary components (bars and rings)
are given in Table \ref{secpar}. 

\begin{table*}
\caption{Structural parameters of the bulges and discs}
\label{photpar}
\begin{tabular}{lcccccccc}
\noalign{\smallskip}
\hline
\noalign{\smallskip}
       &        \multicolumn{4}{c}{Bulge} & & \multicolumn{3}{c}{Disc} \\ \cline{2-5}  \cline{7-9} 
\noalign{\smallskip}
Galaxy &   $\mu_{b}$ ($mag/$arcsec$^{2}$) & $r_{e}$ (arcsec)&  $r_{e}$ (kpc) & $n$ &  &$\mu_{d}$ ($mag/$arcsec$^{2}$) & $r_{d}$ (arcsec) &
$r_{d}$ (kpc)  \\
\noalign{\smallskip}
\hline
\noalign{\smallskip}
AM\,2058A  &  $17.27\pm0.58$ &  $0.63\pm0.025$ & $0.51$  &$0.90\pm0.08$& & $19.60\pm0.11$ & $7.37\pm0.26$ & $5.96$ \\
     
AM\,2058B  &  $19.13\pm0.07$ &  $1.56\pm0.01$  & $1.27$  &$0.41\pm0.02$ & & $20.66\pm0.08$ & $6.00\pm0.19$  & $4.86$  \\
        
\hline

AM\,1228A  &  $17.07\pm1.08$ &  $0.99\pm0.06$  & $0.38$  &$0.86\pm0.16$& & $19.60\pm0.28$ & $12.36\pm1.05$ & $4.80$ \\
          
AM\,1228B  &  $15.83\pm5.9$  &  $0.60\pm0.13$  & $0.23$  &$2.08\pm0.95$& & $20.66\pm0.07$ & $8.58\pm0.47$ & $3.33$ \\
\noalign{\smallskip}
\hline
\noalign{\smallskip}
\end{tabular}
\end{table*}

\begin{table*}
\caption{Structural parameters of the secondary components}
\label{secpar}
\begin{tabular}{lcccccc}
\noalign{\smallskip}
\hline
\noalign{\smallskip}
       &      \multicolumn{2}{c}{Bar} & & \multicolumn{3}{c}{Ring} \\ \cline{2-3}  \cline{5-7} 
\noalign{\smallskip}
Galaxy & $\mu_{bar}$ ($mag/$arcsec$^{2}$) & $r_{bar}$ (arcsec) &&  $\mu_{ring}$ ($mag/$arcsec$^{2}$) & $r_{ring}$ (arcsec) &    $\sigma_{ring}$ \\
\noalign{\smallskip}
\hline
\noalign{\smallskip}
AM\,2058A  &  $21.19\pm0.11$ & $4.09\pm0.18$ && $22.07\pm0.01$ & $5.11\pm0.01$  & $0.63\pm0.01$  \\
\hline        
AM\,1228A  & $21.11\pm0.44$ & $6.57\pm1.11$ && $21.83\pm0.04$ & $11.18\pm0.06$ & $1.73\pm0.08$  \\
       
\noalign{\smallskip}
\hline
\noalign{\smallskip}
\end{tabular}
\end{table*}

\begin{table*}
\caption{Luminosities of main and secondary components}
\label{contpar}
\begin{tabular}{lcccccccccccccc}
\noalign{\smallskip}
\hline
\noalign{\smallskip}
       &      \multicolumn{2}{c}{Bulge} && \multicolumn{2}{c}{Disc} && \multicolumn{2}{c}{Bar} && \multicolumn{2}{c}{Ring} && B/T &  B/D  \\
 \cline{2-3}  \cline{5-6}  \cline{8-9} \cline{11-12}  \cline{14-14} \cline{15-15} 
\noalign{\smallskip}
Galaxy  & $L_r/\mbox{L}_{\odot}$ & \% && $L_r/\mbox{L}_{\odot}$ & \% && $L_r/\mbox{L}_{\odot}$ & \% && $L_r/\mbox{L}_{\odot}$ & \% &&  &     \\
\noalign{\smallskip}
\hline
\noalign{\smallskip}

AM\,2058A & $1.75\times10^{9}$ & 2.8  && $5.78\times10^{10}$ & 90.8 && $2.77\times10^{9}$ & 4.3 && $1.36\times10^{9}$ & 2.1 && 0.03 & 0.03 \\

AM\,2058B & $5.73\times10^{9}$ & 34.6 && $1.10\times10^{10}$ & 65.4 &&   -                &  -   &&          -         &  -   && 0.34 & 0.52 \\
\hline
AM\,1228A & $1.38\times10^{9}$ & 3.0  && $3.88\times10^{10}$ & 85.2 &&  $1.93\times10^{9}$ & 4.2 && $2.84\times10^{9}$ & 6.2 && 0.03 & 0.04 \\

AM\,1228B & $1.53\times10^{8}$ & 6.6  && $2.15\times10^{9}$  & 92.4 &&    -               &  -   &&          -         &  -   && 0.06 & 0.07  \\
       
\noalign{\smallskip}
\hline
\noalign{\smallskip}
\end{tabular}
\end{table*}

The observed surface brightness profiles of AM\,2058A and AM\,1228A 
cannot be properly represented by a simple decomposition in bulge and disc components. 
Visual inspection of the $S_2$ images (see Fig. \ref{elmeimages}), as well as the variation of the geometrical 
parameters and  the surface profiles, indicate that these galaxies 
host bar and  ring  structures. The sum of the four adopted components 
fits well the observed profiles over almost all radii (Fig. \ref{prof}), 
although the reduced $\chi^2$  of 4.73 for AM\,1228A and  5.63 for AM\,2058A. 
These high values are due to the irregularities of the observed surface brightness  profiles. 
On the other hand, the surface brightness profiles of the secondary galaxies, AM\,2058B and AM\,1228B,   
are well fitted by two components, bulge and disc, with a reduced $\chi^2$ 
of $1.62$ and $0.72$, respectively.

The disc scale lengths and central magnitudes obtained for all galaxies 
(Table \ref{photpar}) agree well with the average values 
($r_{d}=3.8\pm2.1$\,kpc and $\mu_{d}=20.2\pm0.7\,mag/arcsec^{2}$) derived by  \citet{fathi10a} 
and \citet{fathi10b} for  a large sample of   
galaxies with no evidence of ongoing interaction or disturbed morphology.  
This indicates that the  symmetrization method is adequate 
to recover the unperturbed disc of the interacting galaxies. 
Regarding the bulge component, the resulting profiles have S\'ersic indexes typical of pseudo bulge ($n<2$)
\citep{kormendy04}. Pseudo-bulges, when compared to classical ones, tend 
to show younger stellar populations, kinematics supported by rotation, 
and less concentrated surface brightness profiles, similar to those of discs \citep{gadotti09}. 
Pseudo-bulges can be  formed on longer time-scales, via disc instabilities 
and secular evolution processes   caused by  non-asymmetric
structures (see \citeauthor{kormendy04} \citeyear{kormendy04}, for review), or  
tidal  interaction between galaxies. Both  perturbations cause gas  
to flow towards the galaxy centre  and  subsequent star formation,
resulting in a compact stellar component with high 
v/$\sigma$, which  leads to features typical of a pseudo-bulge \citep{weinzirl09}.
Therefore, we infer that the pseudo-bulges may be caused by the on-going 
interaction. In order to test these scenarios, it would be necessary to
perform a numerical simulation for these pairs, which will be done in a forthcoming paper.    

The derived photometric parameters are used to calculate the 
integrated luminosity for each component:
 
\begin{equation}
L=\int^{r_{max}}_{r_{min}} I(r)2\pi rdr,
\label{lens}
\end{equation} 
where $I(r)$ can be any of the profiles above defined. 
The integral limits, $r_{min}$ and $r_{max}$, are the 
minimum and maximum radii of the surface brightness profile. The
luminosities ($L_r$) found for each component in the fit, their contribution (in \%) to the total luminosity, 
the  bulge-to-total (B/T) and bulge-to-disc (B/D) luminosity ratios
are listed in Table \ref{contpar}.
The B/T ratios obtained for AM\,2058A, AM\,1228A and 
AM\,1228B  are very small, with values $<0.1$, but consistent 
with their  morphological classification as late-type spirals \citep[e.g.,][]{fisher08,weinzirl09}. 
For AM\,2058B, the  B/T ratio is 0.34, which is similar to those found for early-type galaxies.
The B/D ratios found for the main galaxies, AM\,2058A and AM\,1228A, 
are also in good agreement with the  reported average value of $\log\,$(B/D)$=-1.07_{-0.30}^{0.45}$ for Sc
galaxies \citep{graham08}. Similarly, the B/D ratios determined for the 
secondary galaxies, AM\,2058B and AM\,1228B, are within the ranges of values reported 
for  their respective morphological types, $\log\,$(B/D)$=-0.34_{-0.07}^{0.10}$  for S0 galaxies
and $\log\,$(B/D)$=-1.38_{-0.50}^{0.47}$ for Sd \citep{graham08}. 

The bar lengths in AM\,2058A and AM\,1228A are 3.3 and 2.5\,kpc, respectively.
These values are typically seen in late-type spirals \citep{elmegreen85,gadotti08}.
Even so, their contribution to the total luminosity is quite low: $\sim $4\% for  
both galaxies. The ring structure in 
AM\,1228A contributes with $\sim$6\% to the total luminosity, while  in AM\,2058A, it contributes 
with only $\sim$ 2\%.

\section{Ionized gas kinematics}
\label{vel}

Individual spectra were extracted along the slit positions 
in apertures of 1 $\times$ 1.17 arcsec$^2$. The radial velocity 
at each position was derived by averaging the resulting 
centroid of Gaussian  curves fitted to the profiles of 
the strongest emission lines ([\ion{N}{ii}] $\lambda 6548.04$, 
$\rm H\alpha$ $\lambda 6563$, [\ion{N}{ii}] $\lambda 6584$ and [\ion{S}{ii}] 
$\lambda 6717$). We adopted the radial velocity of the central aperture 
of each galaxy as systemic velocity. These   values are listed in 
Table  \ref{velsys}. The systemic velocities for the members of 
AM\,2058-381 are in agreement with the previous 
values found by \citet{donzelli97}.   

\begin{figure*}
\subfigure{\includegraphics*[width=\columnwidth]{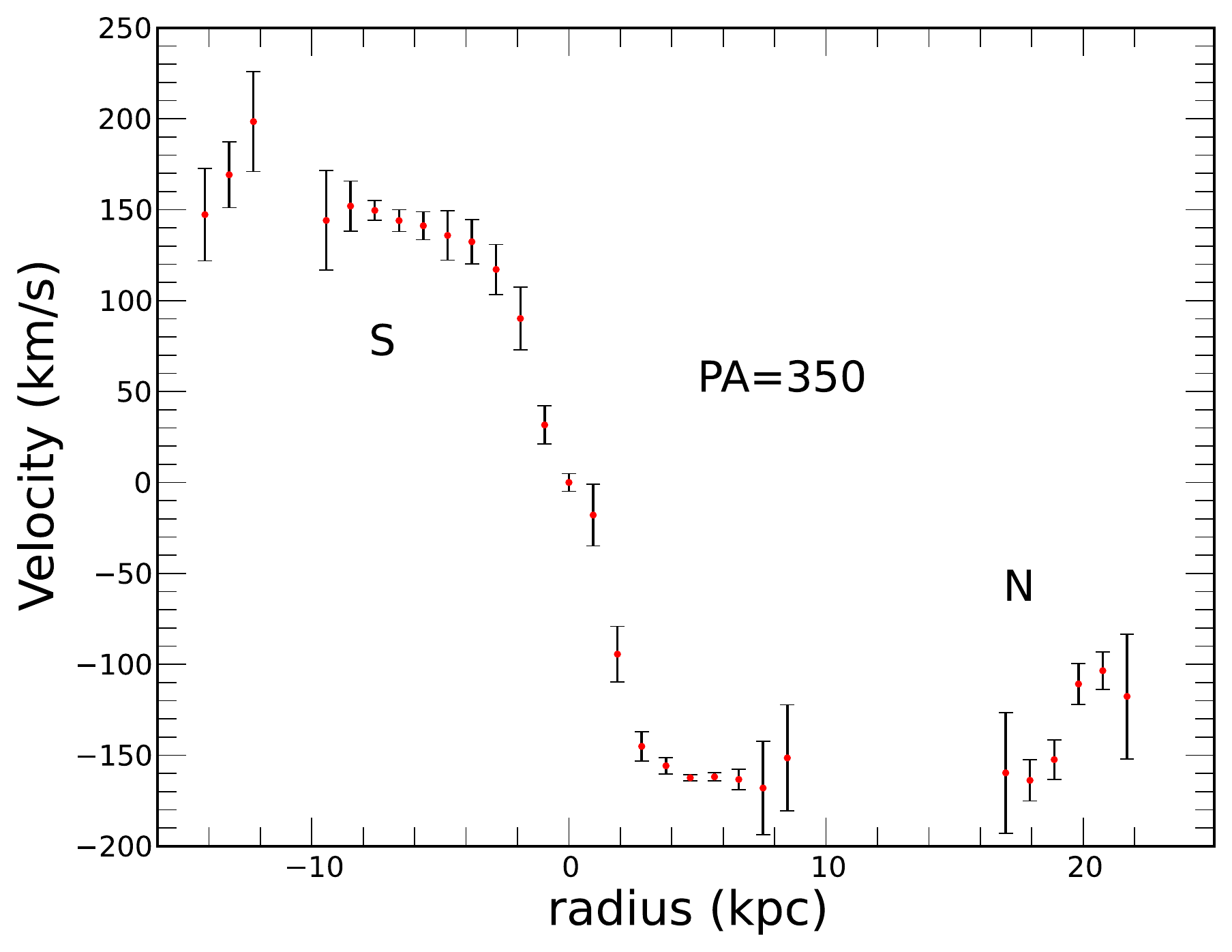}}	
\subfigure{\includegraphics*[width=\columnwidth]{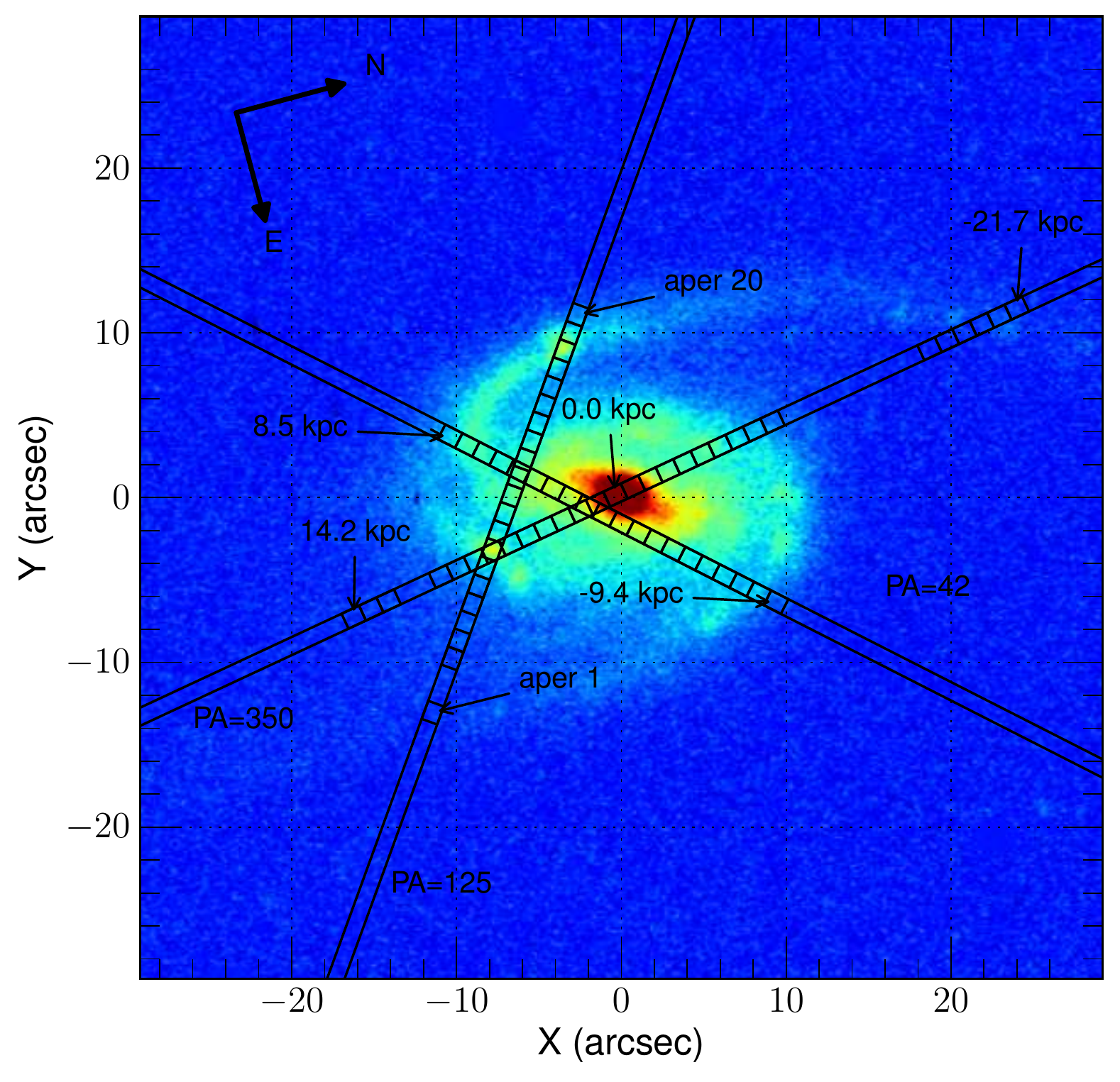}}
\subfigure{\includegraphics*[width=\columnwidth]{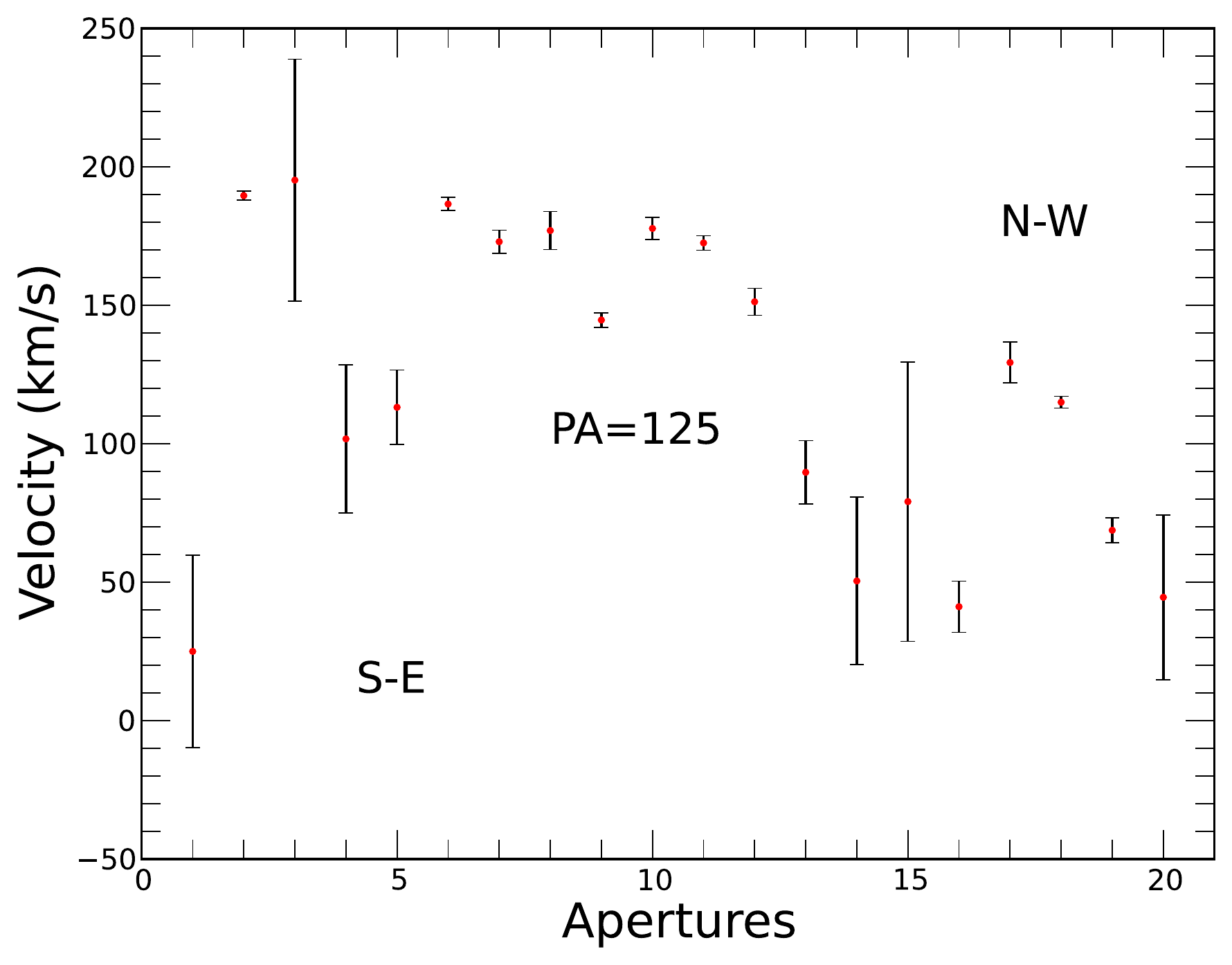}}	
\subfigure{\includegraphics*[width=\columnwidth]{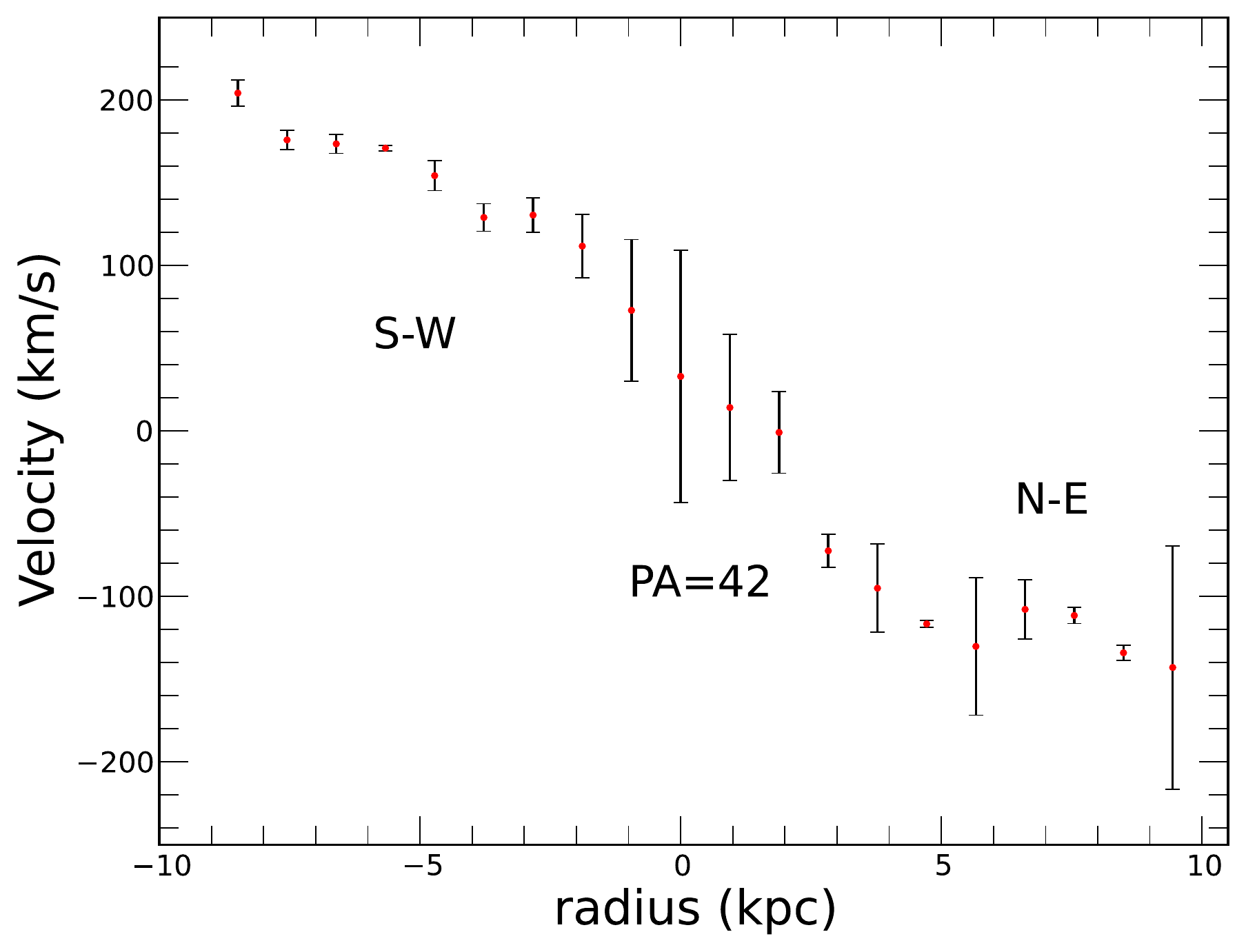}}
\caption{Kinematics along PA=350\degr (top-left panel), PA=125\degr (bottom-left) and 
PA=42\degr (bottom-right) in AM\,2058A. The velocity scale corresponds to the observed 
values after subtraction of the systemic velocity, without correction for inclination 
on the plane of the sky. The top-right panel shows the AM\,2058A image with the 
location of the slits and extracted apertures overlaid.}
\label{curveam2058A}
\end{figure*}

Figure \ref{curveam2058A} shows the AM\,2058A image with the three slit positions overlaid, 
and the radial velocity profiles (RVP) measured along the corresponding slits.
The RVP observed at PA=350\degr\, passed through the centre of the galaxy. The 
Northern and Southern sides of the curve (approaching and receding sides, respectively) are rather symmetric, 
with a steep rise in the inner  radii and a flattening trend in the outer regions, and a   
maximum velocity of $\pm$150\,km\,s$^{-1}$ at $\sim$ $\pm$10\,kpc. The RVP along the direction 
North-East to South-West (PA=42\degr)
is quite smooth, but asymmetric in velocity, reaching -120 and 200\,km\,s$^{-1}$
respectively. The velocity field  obtained along the slit with PA=125\degr\, shows wavelike form with 
different minimum and maximum. This slit position is located across the Western part of the 
disc and the North-Western spiral arm.  Similar effects were observed on the 
velocity field in the vicinity of the spiral arms in the interacting spiral galaxy NGC\,5427 \citep{alfaro01}.

\begin{figure*}
\subfigure{\includegraphics*[width=0.7\columnwidth]{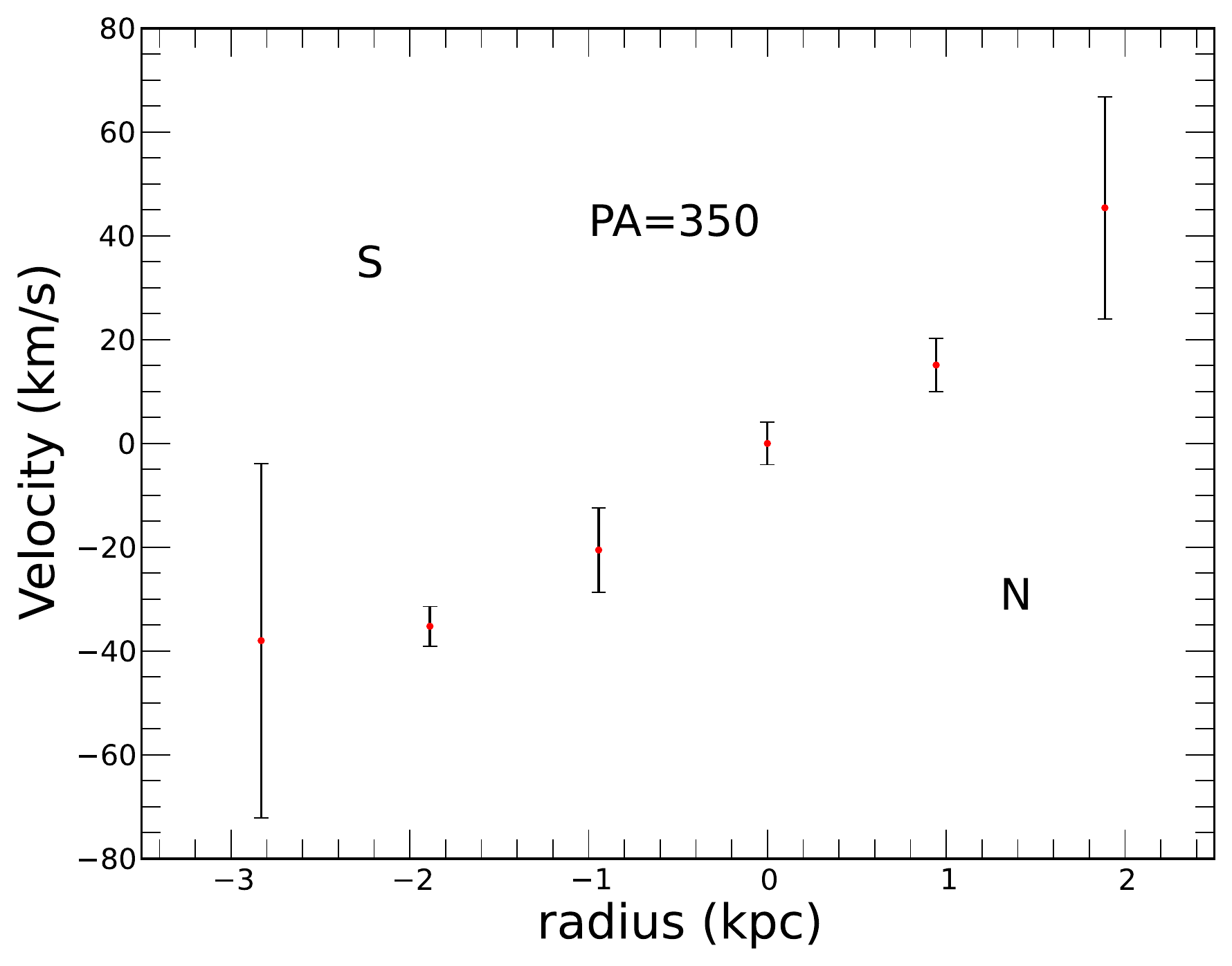}}
\subfigure{\includegraphics*[width=0.6\columnwidth]{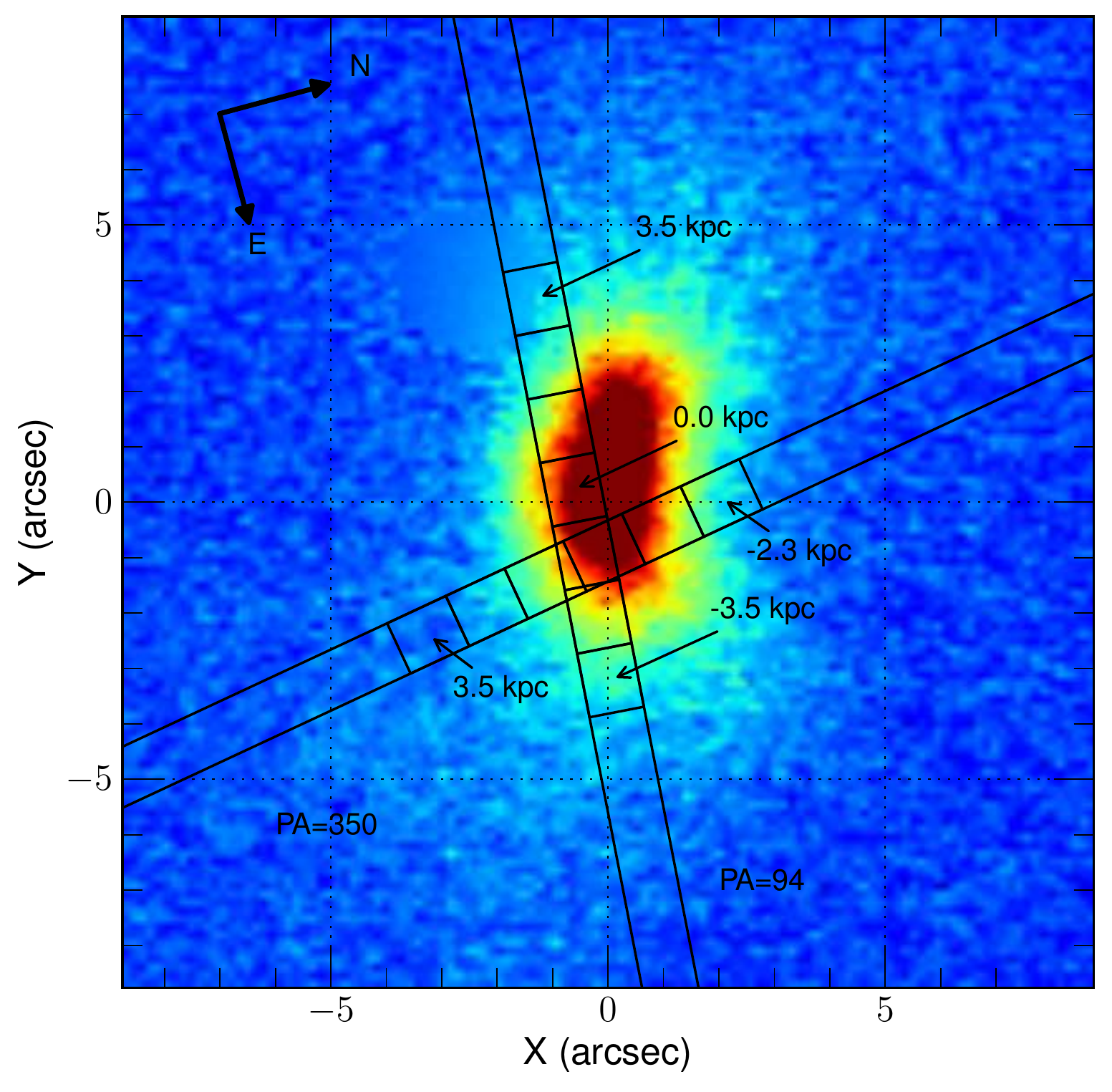}}
\subfigure{\includegraphics*[width=0.7\columnwidth]{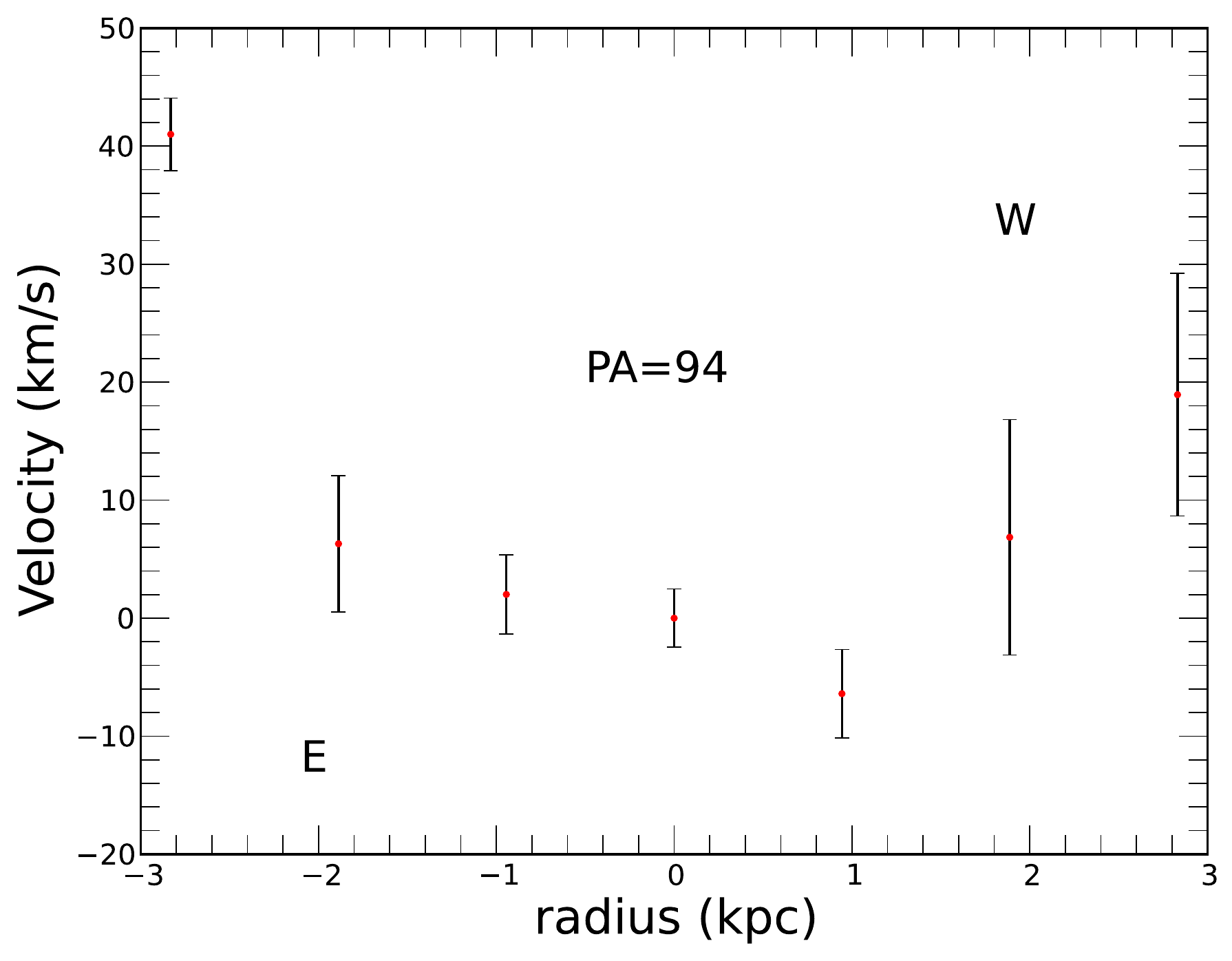}}								     
\caption{Same as Fig. \ref{curveam2058A}  for AM\,2058B and slits with 
PA=350\degr (right) and PA=94\degr (left).}
\label{curveam2058B}
\end{figure*}

Two slit positions (PA=350\degr\, and PA=94\degr) were observed  in AM\,2058B  
and their RVPs are shown in Fig. \ref{curveam2058B}. These RVPs have few points 
because of the small angular size of this galaxy, and  none of them through the 
galactic centre. The RVP along PA=350\degr\, is quite symmetric and has a linear 
behaviour with small slope. Both sides, approaching (South part) and receding (North), 
reach a  maximum velocity of $\pm$40\,km\,s$^{-1}$. In contrast, the RVP along PA=94\degr\, 
appear to be located along the zero-velocity line of this galaxy. This result is surprising,  because 
the velocity line-of-nodes should be aligned with the photometric major axis (PA=79\degr) and 
not with the photometric minor axis, which is the case for this galaxy.
 Could  AM\,2058B be a tumbling body, 
rotating along its major axis? To answer this question, a more detailed analysis of the 
velocity field would be  required (e.g., using integral field spectroscopy). 
However, if AM\,2058B is rotating like a solid body, with constant angular momentum,  
it would  explain the RVP linear behaviour  along PA=350\degr. Another question, could 
the misalignment of angular momenta of AM\,2058B be caused by the main companion?  
In a recent work, \citet{cen14} studied the evolution of angular  momenta 
in galaxies in cosmological simulations, and found that the spin changes 
direction frequently due to tidal interaction with nearby companions.

\begin{figure*}
\subfigure{\includegraphics*[width=\columnwidth]{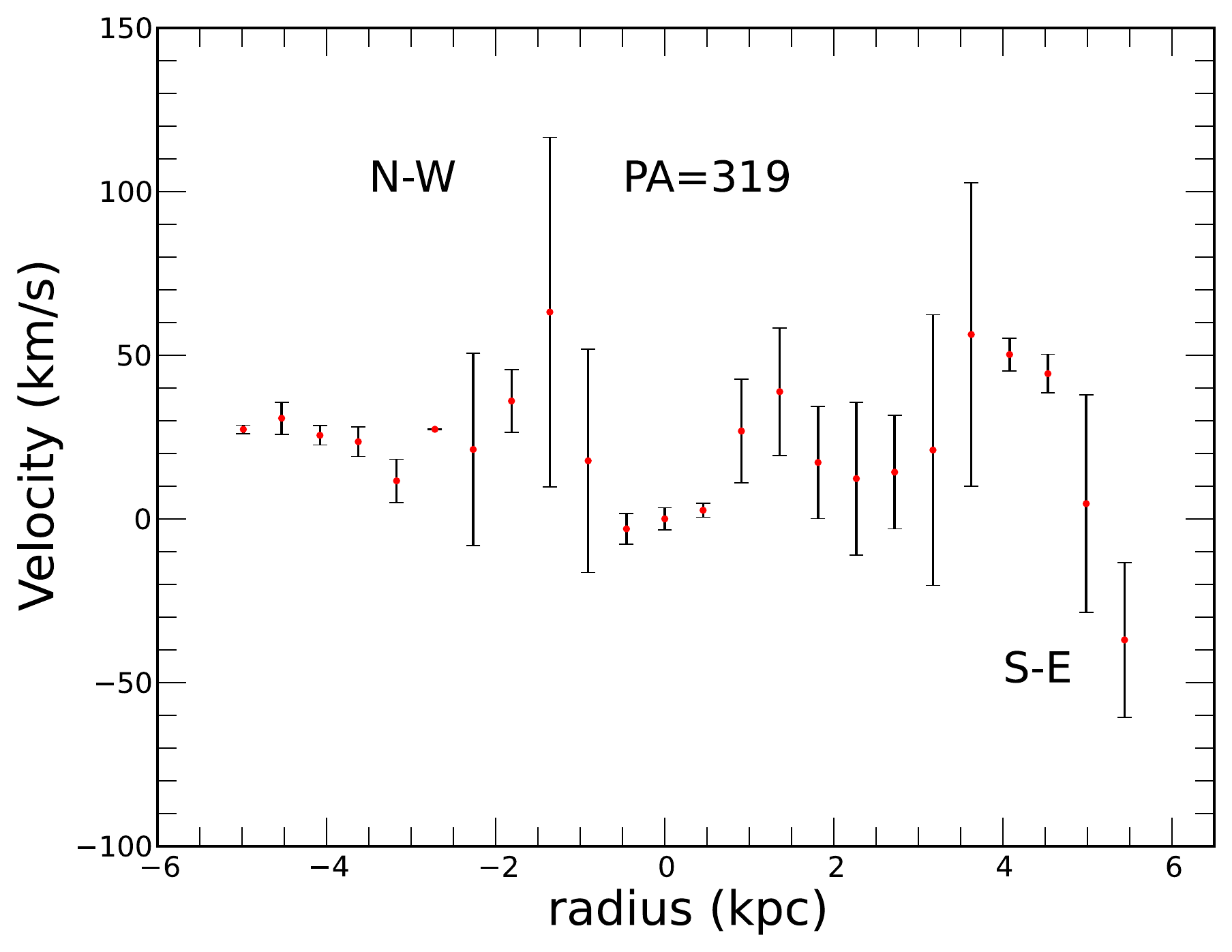}}
\subfigure{\includegraphics*[width=\columnwidth]{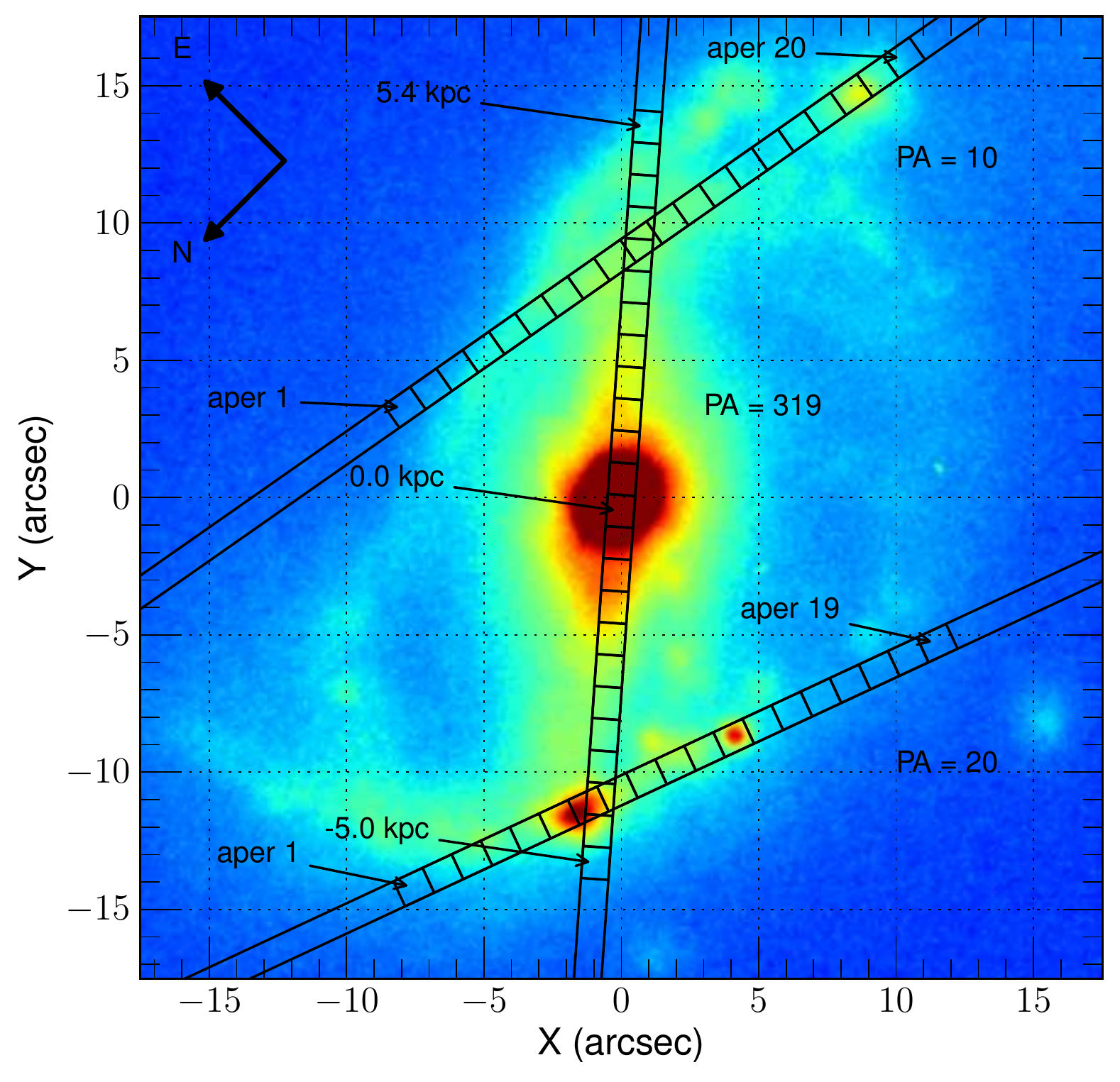}}
\subfigure{\includegraphics*[width=\columnwidth]{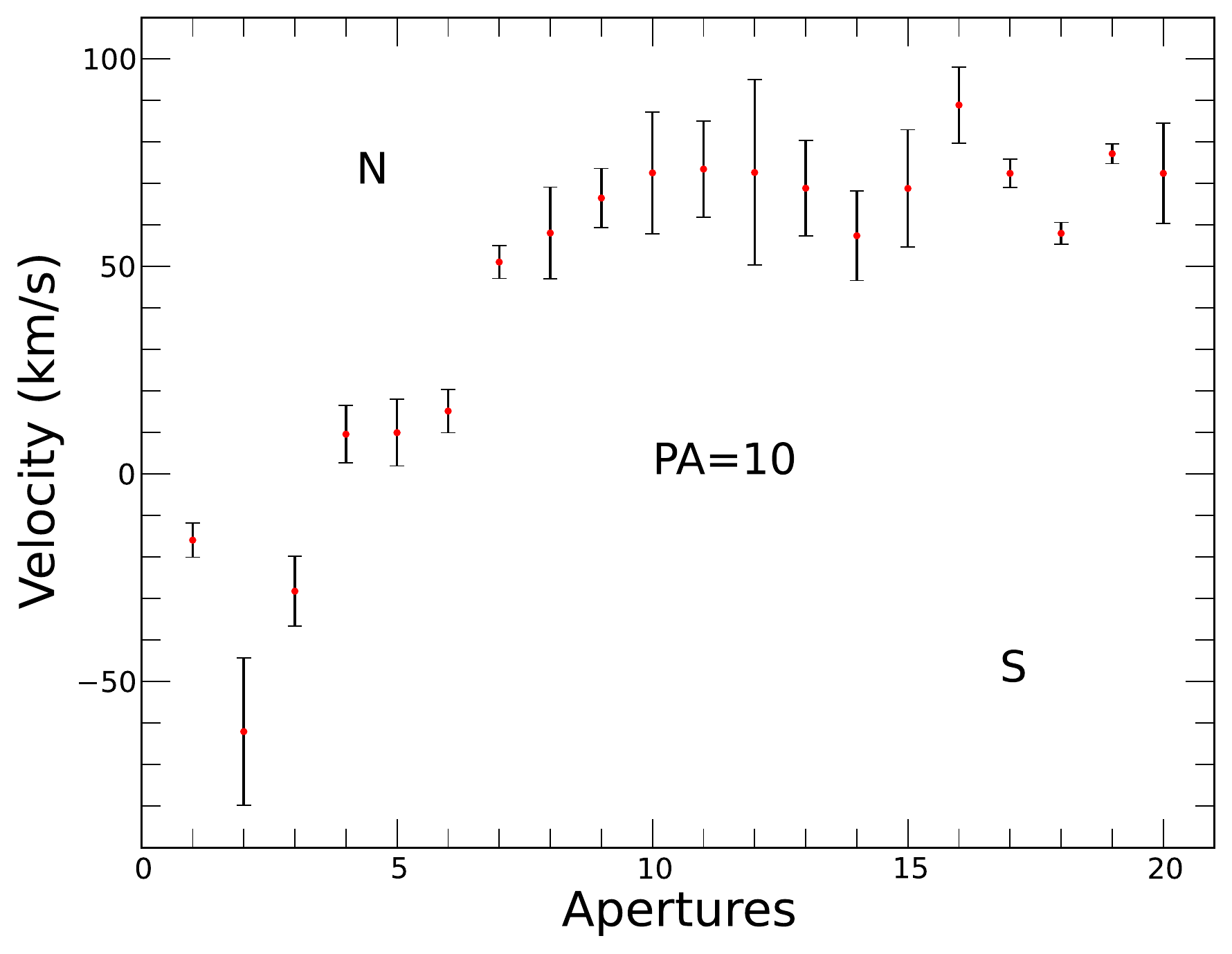}}			
\subfigure{\includegraphics*[width=\columnwidth]{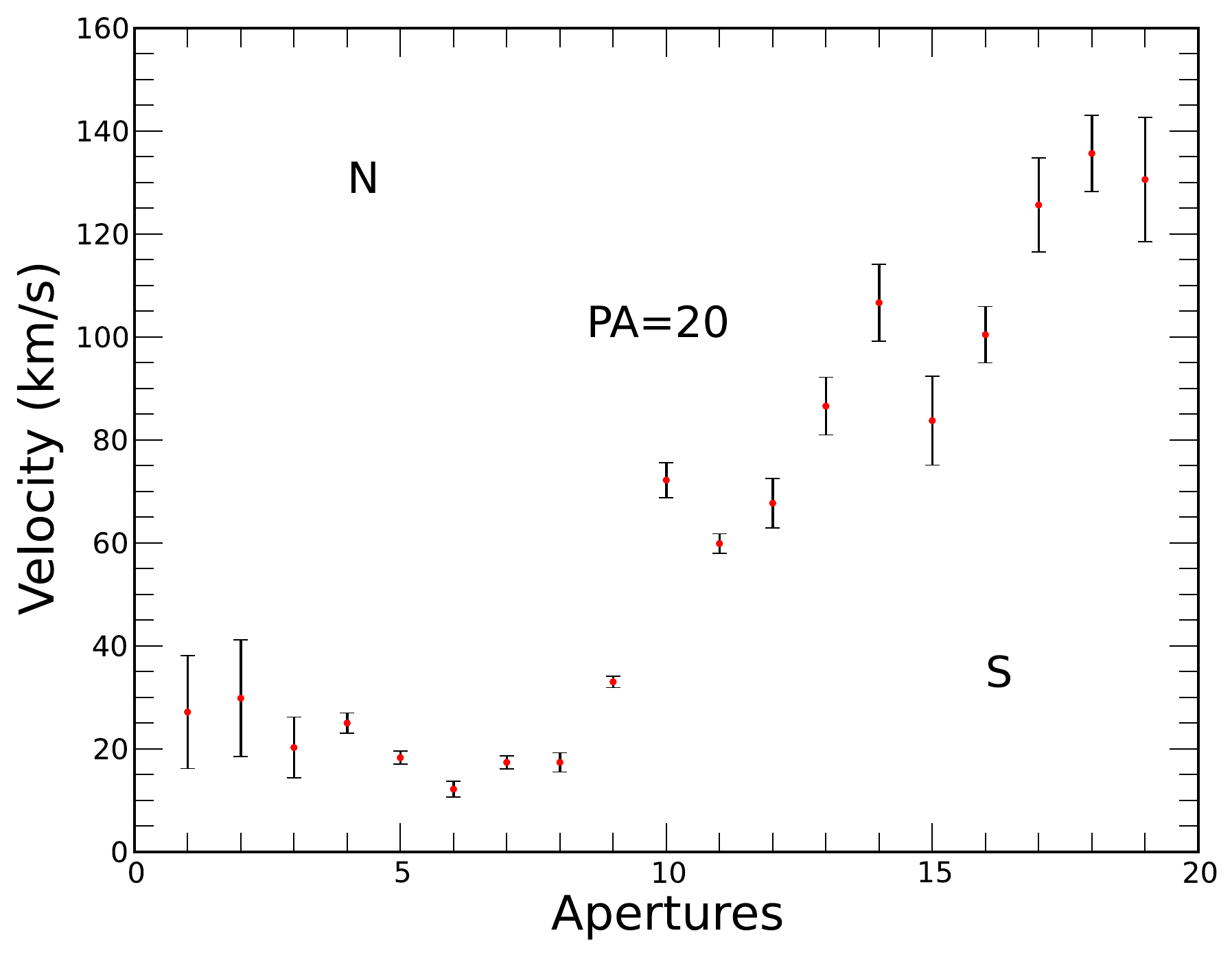}}						     
\caption{Same as Fig. \ref{curveam2058A},  for AM\,1228A (top-left panel) and slits 
with PA=319\degr (top-left), PA=10\degr (bottom-left) and PA=20\degr (bottom-right).}
\label{curveam1228A}
\end{figure*}

Figure \ref{curveam1228A} shows the RVPs for the slit positions at 
PA=319\degr, PA=10\degr and PA=20\degr\,, and location of the 
spectral extractions for AM\,1228A image. The RVP at PA=319\degr\, seems 
to be close to the zero-velocity line, with velocities between 0\,km\,s$^{-1}$ 
and 50\,km\,s$^{-1}$. In fact, as we discuss in Sect. \ref{massmodel}, 
there is a misalignment between the kinematic and photometric axes, like  
in AM\,2058B. On the other hand, the RVP at  PA=10\degr\, in the Northern part  
shows increasing velocity, from -60 up to 80\,km\,s$^{-1}$, 
while in the South, it becomes flat. Conversely,  the RVP at 
PA=20\degr\, is rather flat in the Northern part (with small oscillations 
smaller than 10 km\,s$^{-1}$)  at $\sim$ 20 km\,s$^{-1}$, rising linearly
up to 130 km\,s$^{-1}$ in the Southern part.   

\begin{figure*}
\subfigure{\includegraphics*[width=\columnwidth]{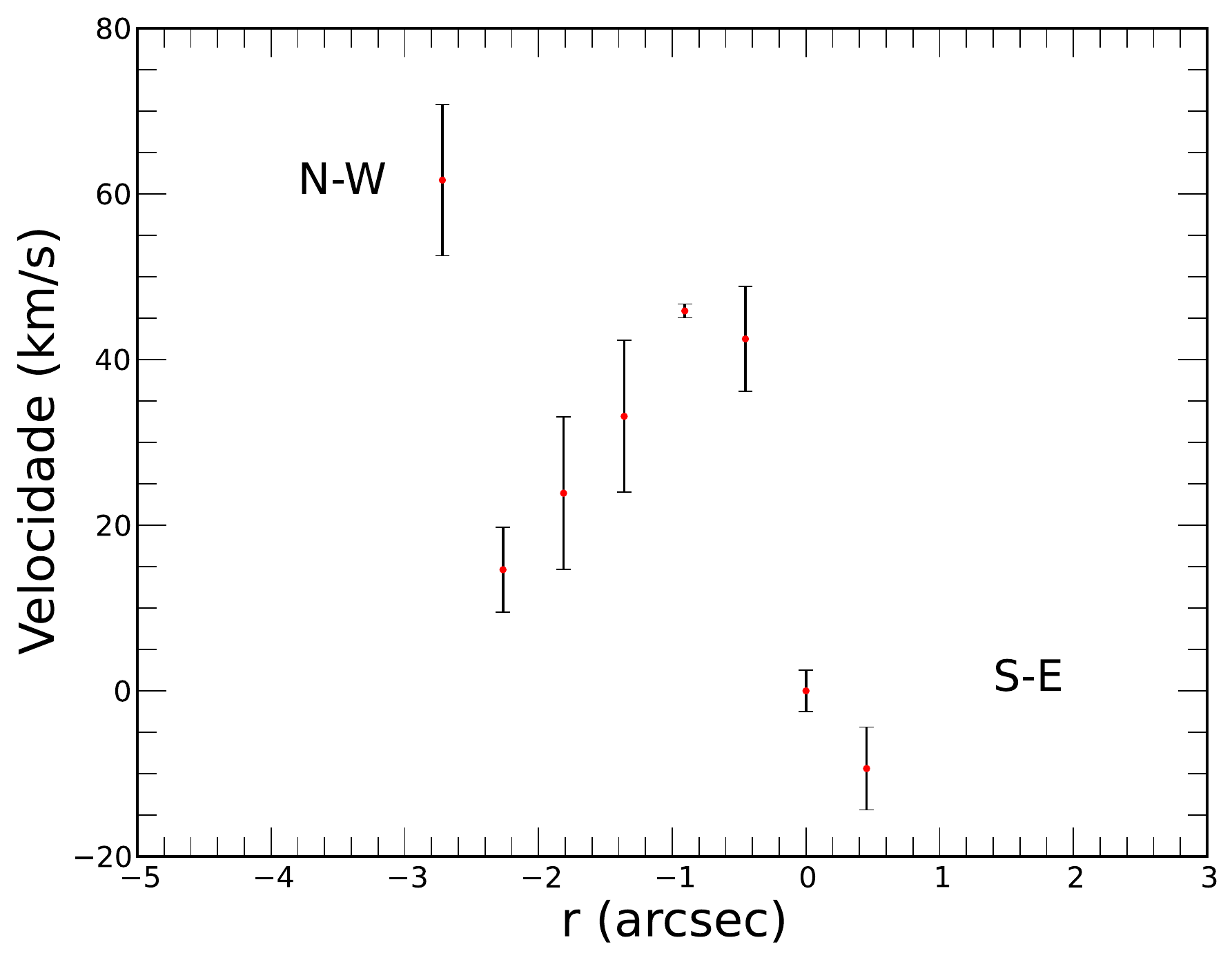}}
\subfigure{\includegraphics*[width=\columnwidth]{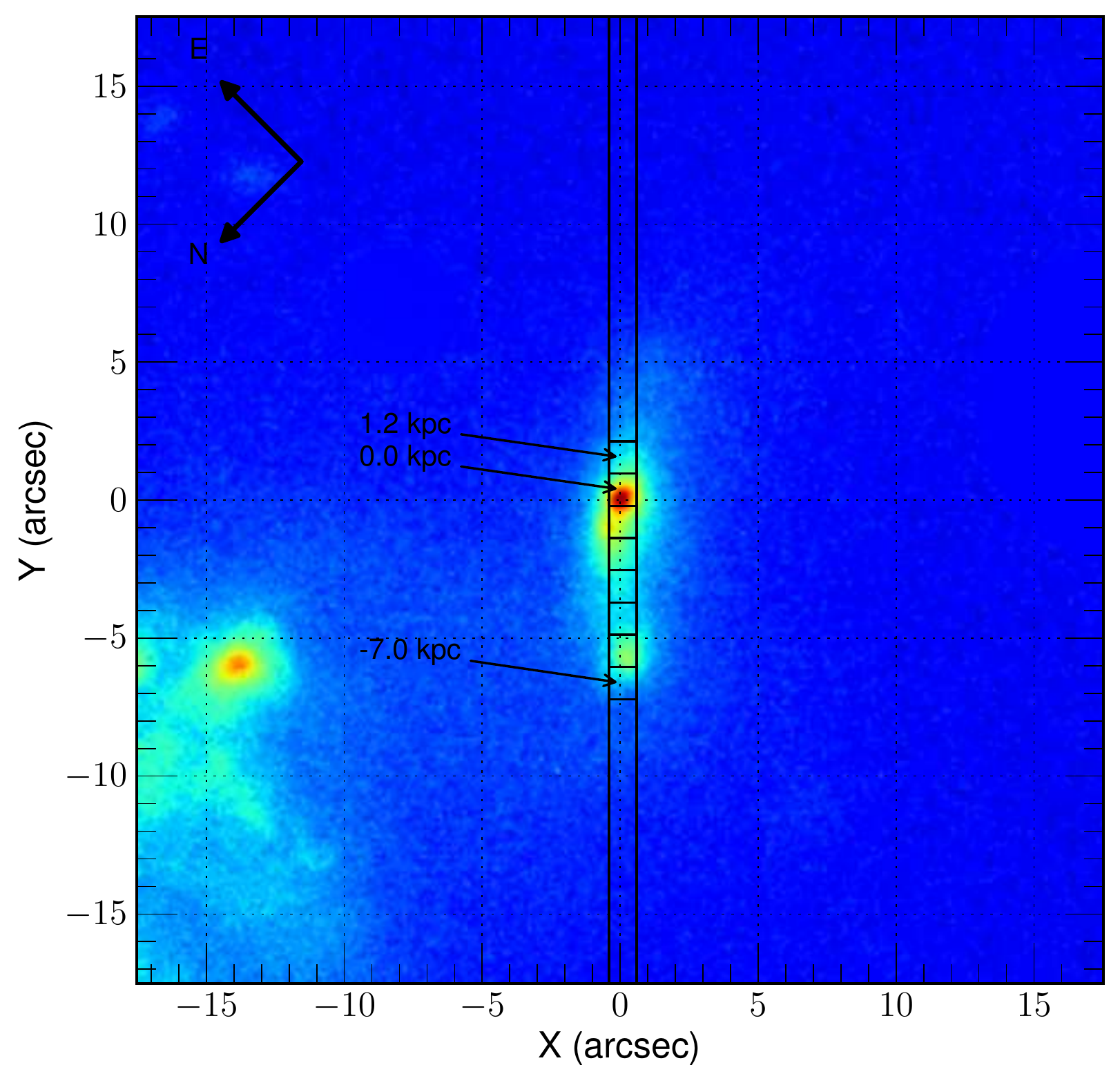}}				     
\caption{Same as Fig. \ref{curveam2058A} for AM\,1228B (right panel) and slit with PA=315\degr (left)}
\label{curveam1228B}
\end{figure*}

The RVP for AM\,1228B are  show in Fig. \ref{curveam1228B}. Similarly to 
AM\,2058B, the RVP for AM\,1228B  has few points due to  its small angular size. This RVP shows a very peculiar
form: it starts at North-West with a velocity of 60\,km\,s$^{-1}$, immediately drops to $\sim$ 15\,km\,s$^{-1}$,
then a linear increase up to  $\sim$ 15\,km\,s$^{-1}$ at $\sim$ 1\,kpc from the centre. 
Finally, at South-East direction, the measured velocities drop again, falling to $\sim$ $-10$\,km\,s$^{-1}$.

\begin{table}
\caption{Systemic Velocities}
\label{velsys}
\begin{tabular}{lcc}
\noalign{\smallskip}
\hline
\noalign{\smallskip}
Galaxy & Systemic Velocity (km\,s$^{-1}$) & PA Slit  (\degr)  \\
\noalign{\smallskip}
\hline
\noalign{\smallskip}
   AM\,2058A    & 12173$\pm$5    & 350 \\
   AM\,2058B    & 12309$\pm$4    & 94  \\
\hline
   AM\,1228A    & 5844$\pm$3    & 319 \\
   AM\,1228B    & 5937$\pm$3    & 4  \\

\noalign{\smallskip}
\hline
\noalign{\smallskip}
\end{tabular}
\end{table}

\section{Rotation curve models}
\label{massmodel}

The mass distributions of the main galaxies in the studied pairs  
are modelled as the sum of the bulge, disc and dark halo components. 
We assume that the mass distribution follows the deprojected 
luminosity distribution with constant M/L ratio for the bulge and disc.

For the bulge mass distribution, we use the rotation curve derived for a \citeauthor{sersic68} 
profile density. This profile is obtained  by an Abel integral equation \citep{binney87,simonneau04}, 
which relates bulge surface brightness (equation \ref{sersiclaw}) to density:

\begin{equation}
\label{density_sersic}
\rho (s) =  \frac{1}{\pi}\frac{k_{n}}{n}I_{b}\Upsilon_{b}\int_{s}^{\infty}\frac{exp[-k_{n}z^{\frac{1}{n}}]z^{\frac{1}{n}-1}}{\sqrt{z^{2}-s^{2}}}\, {\rm d}z,
\end{equation}
where $I_{b}$, $r_{e}$, $n$  and $k_{n}$ are those in equation \ref{sersiclaw}, 
and $s=(r/r_{e})$. $\Upsilon_{d}$ is the M/L for the bulge component.
The circular velocity ($V_{b}$) associated for the bulge is:

\begin{equation}
\label{curv_sersic}
V_{b}^{2}(r)=G\frac{M(r)}{r},
\end{equation}
where
\begin{equation}
\label{masssersic}
M(r)=4\pi\int_0^r r^2\rho(r)\, {\rm d}r. 
\end{equation}

For the disc,  the circular velocity ($V_{d}$) curve derived for an exponential disc 
is given by the following  equation \citep{freeman70,binney87} 

\begin{equation}
\label{curv_freeman}
V_{d}^{2}(r)=4\pi G \Upsilon_{d}I_{d}r_{d}y^{2}[I_{0}(y)K_{0}(y)-I_{1}(y)K_{1}(y)],
\end{equation}
where $I_{d}$ and $r_{d}$ are those in equation \ref{exponentialdisk} and $\Upsilon_{d}$ 
is the M/L for disc component. $y=r/2r_{d}$, $I_{n}$ and $K_{n}$ are modified Bessel 
functions of the first and second kinds, respectively.  

For the halo mass model, we use the density profile proposed by \citeauthor{navarro95}  
\citeyearpar[ hereafter NFW]{navarro95,navarro96,navarro97}. In this case the dark 
matter density is given by  

\begin{equation}
\label{profnfw}
\rho (r) =\frac{\rho_{0}\rho_{c}}{(\frac{r}{r_{s}})(1+\frac{r}{r_{s}})},
\end{equation}
where $r_s$ is a characteristic radius, $\rho_{c}$  is the present critical density and $\rho_{0}$ 
is the characteristic overdensity.
The latter is defined as $\rho_0 =\frac{200}{3}\frac{c^3}{[\ln(1+c-c/(1+c))]}$, 
where $c\equiv r_{200}/r_s$  is the halo concentration \citep{navarro96}. $r_{200}$ 
is the distance from the centre of the halo at which the mean density is 200 times
the $\rho_{c}$. The mass interior inside this radius is $M_{200}=\frac{4}{3}\pi200\rho_{c}r^3_{200}$.
The circular velocity ($V_h$) in the NFW profile  parametrized with $M_{200}$ and $c$ is:

\begin{equation}
\label{curv_nfw}
V^2_h(r)=\frac{GM_{200}}{g(c)r}\left[\ln(1+cr/r_{200})-\frac{cr/r_{200}}{1+cr/r_{200}}\right].
\end{equation}

The final rotation curve model is computed from the squared sum of the circular velocities 
of the bulge, disc and halo components:

\begin{equation}
\label{curv_final}
V_{c}^{2}(r)=V_{b}^{2}(r)+V_{d}^{2}(r)+V_{h}^{2}(r).
\end{equation}

This equation has 9 parameters, 5 photometric and 4 dynamic. The photometric parameters 
were already determined for the bulge  ($I_{b}$, $r_{e}$ and $n$) and disc ($I_{d}$ and $r_{d}$) 
in Sect. \ref{profphot}, and are fixed.  On the other hand, the dynamic 
parameters, the bulge and disc M/L ratios ($\Upsilon_{b}$ and $\Upsilon_{d}$, respectively) 
and the halo parameters ($M_{200}$ and $c$), are free. Since we have multiple observations 
with different  long-slit orientations on the main galaxies (see Figs. \ref{curveam2058A} 
and  \ref{curveam1228A} for AM\,2058A and AM\,1228A, respectively), we have fitted the  
projected $V_{c}$ in the plane of the sky for all positions simultaneously. Therefore, 
the observed radial velocity at position ($R,\phi$) on the sky plane  is related 
to the circular velocity [$V_{c}(r)$]  by the following equation 
\citep{elmegreen98,palunas00}.

\begin{equation}
\label{curv_dep}
V(R,\phi)=V_{sys} + V_c(r)\sin i \left[ \frac{\cos i\cos(\phi-\phi_0)}{\sqrt{1-\sin^2i\cos^2(\phi -\phi_0)}}\right],
\end{equation}
and
\begin{equation}
r=R\sqrt{1+\sin^2(\phi -\phi_0)\cos^2i},
\end{equation}
where  $i$ is the inclination of the galactic disc, $\phi_0$ is the PA of the
projected major axis, and $V_{sys}$ is the systemic velocity. The disc centre $(R_c,\phi_c)$ is an
implicit pair of parameters in the model. It is important to note that the term in
brackets is equal to one when  $V_c$ is  measured along the major axis, in which case,  $r=R$.
The latter equation introduces five additional parameters, namely:
$i$, $\phi_0$, $V_{sys}$, $R_c$ and $\phi_c$. The first two are determined by the fit
of the outer isophote of the disc (Sect. \ref{elme}), and thus, are fixed parameters, 
while the remaining three are free parameters in the rotation curve model.

\begin{table*}
\caption{Parameters derived from the phenomenological model}
\label{bertola}
\begin{tabular}{lccccccccc}
\noalign{\smallskip}
\hline
\noalign{\smallskip}
Galaxy &  A (km\,s$^{-1}$) &   c (kpc)  & p  &  $V_{sys}$ (km\,s$^{-1}$) & $\Delta\,x$ (kpc) & $\Delta\,y$ (kpc) & PA (kine)  &  PA (phot)  & $\Delta\theta$\\
\noalign{\smallskip}
\hline
\noalign{\smallskip}
AM\,2058A & 823.2 & 45.5 & 1.2  & 12164.3 & -0.02 & 0.07 & 194.5\degr & 198.9\degr & 4.4\degr  \\
\hline
AM\,1228A & 105.6  & 14.8 & 0.9 & 5887.2  & 0.42  & -0.15& 221.1\degr & 162.1\degr & 58.9\degr \\

\noalign{\smallskip}
\hline
\noalign{\smallskip}
\end{tabular}
\end{table*}

Note that the photometric major axis 
is not necessarily aligned  with the kinematic one. In fact, in a recent  paper, 
\citet{barrera14} studied the velocity maps for a sample of 80 
non-interacting spiral galaxies, and found that 10\% of those galaxies present 
kinematic misalignments larger than 22\degr. In order to indirectly determine 
the PA kinematics major axis, we fitted our data with a 
phenomenological potential given by \citet{bertola91},
with an on-the-sky projection

\begin{equation}
V(R,\phi)= V_{sys} + \frac{AR \cos(\phi-\phi_{0})\sin i\cos^pi }{  (R^2 \eta + c_0^2\cos^{2} i)^{p/2} },
\label{curv_bertola}
\end{equation}
with
\begin{equation}
\eta \equiv [\sin^{2}(\phi-\phi_{0}) + \cos^{2}(i)\cos^{2}(\phi-\phi_{0})],
\end{equation}
where $A$ and $c_0$ and $p$  are parameters that define the amplitude and shape of the curve.  
The remaining parameters, $V_{sys}$, $\phi_0$, $R_c$ and $\phi_c$, are the same as in
equation \ref{curv_dep}. The inclination remains constant due to the well known limitation to derive 
this parameter from kinematics. The parameter obtained by fitting the above equation to the  
AM\,2058A and AM\,1228A data are listed in Table \ref{bertola}. Instead of  
$\phi_0$ and  $R_c$, we give the difference
 between kinematic and photometric centres, in the sky plane, $\Delta\,x$ and $\Delta\,y$. 
In addition to these parameters, 
Table  \ref{bertola} also gives the angular difference found between the PA of the kinematic  
and photometric major axis. The $p$ parameter for both galaxies is close to 1, which is the expected value for 
flat rotation curves \citep{bertola91}. $V_{sys}$ values agree with the 
observation, while both galaxies show an offset between the photometric and kinematic centres of 
$\sim$ 0.2 and $\sim$ 0.4\,kpc, for AM\,2058A and AM\,1228A, respectively. However, these offsets 
are smaller than the seeing for each galaxy 
(0.94 and 0.45\,kpc, respectively).       
For AM\,2058A, there is a good agreement between the photometric and 
kinematic  axes orientation,  while for AM\,1228A, there is  a 
misalignment of 58\degr\, between the axes. One possible explanation  
is that the photometric PA,  derived
from the outermost isophotes of AM\,1228A's disc are twisted 
due to the common external tidal  structure present in this system.  
Another possibility would be  the well-known 
characteristic ``S''-shape in the zero-velocity curve, like that observed in the 
velocity field of the barred spirals \citep[e.g.,][]{peterson80,garcia01,emsellem06,barrera14}. 
However, this effect introduces asymmetries rather than misalignments between 
the photometric and kinematic axes orientation.

\section{Mass models}
\label{fitmodel}

In order to determine the mass distribution of the main galaxies of the studied 
pairs, we use the force method outlined in \citet{hernandez13}. This method
consists basically in  exploring the phase space generated by M/L ratios of the bulge ($\Upsilon_{b}$)
and disc ($\Upsilon_{d}$), the halo parameters ($M_{200}$, $c$) and geometrical  parameters ($V_{sys}$, $\phi_0$,    
$R_c$). Each  point in  this phase space represents a  model of the rotation curve given by 
equation \ref{curv_dep}, and associated with this model the $\chi^2$ resulting of the fit of
the data. The explored ranges for the $\Upsilon_{b}$, $\Upsilon_{d}$, $M_{200}$, $c$, $\phi_0$ and $R_c$
parameters are given in Table \ref{parexp}, again the kinematics centre is given in 
terms of the offset with respect to the geometrical centre,  $\Delta\,x$ and 
$\Delta\,y$. The choice of halo parameters is based  on the values found in cosmological simulations
with NFW's profile \citep{navarro96,bullock01}. With respect to the explored ranges of  
M/L for the bulge and disc, we chose values corresponding to the minimum and maximum disc
\citep[e.g.,][]{vanalbada85,carignan85,kent87}. On the other hand, the kinematic centres were chosen
to be inside the respective seeing boxes. Finally,  we explored 5 values of $V_{sys}$ for 
each galaxy: the radial velocity measured at the central and two adjacent  apertures, plus the mean values between them.

 \begin{table}
\caption{Explored ranges of the mass model parameters }
\label{parexp}
\begin{tabular}{lccc}
\noalign{\smallskip}
\hline
\noalign{\smallskip}
Parameter & Min. value & Max. value & $\Delta$ value   \\
\noalign{\smallskip}
\hline
\noalign{\smallskip}
$\Upsilon_{b}$ & 0.00 &  2.00  & 0.10    \\
$\Upsilon_{d}$ & 0.00 &  2.00  & 0.10  \\
$\log (M_{200}/10^{12}\,\mbox{M}_{\odot}$)  & -1.30      & 1.00  & 0.03  \\
$c$            &  5.0     & 60.0  & 1.00              \\
$\Delta\,x,y$ (kpc) for AM\,1228A  & -0.94  & 0.94   &  0.470  \\
$\Delta\,x,y$ (kpc) for AM\,2058A  & -0.45  & 0.45  &   0.225 \\
\noalign{\smallskip}
\hline
\noalign{\smallskip}
\end{tabular}
\end{table}

The RVPs used to fit the mass model for AM\,2058A are those observed at PA=350\degr and PA=42\degr. The RVP at PA=125\degr 
was excluded because it crosses along  the N-W arm and present kinematic irregularities (Sect. \ref{vel}).
On the other hand, all observed RVPs for AM\,1228A were used to fit the mass distribution model.  

\begin{figure*}
\subfigure{\includegraphics*[width=\columnwidth]{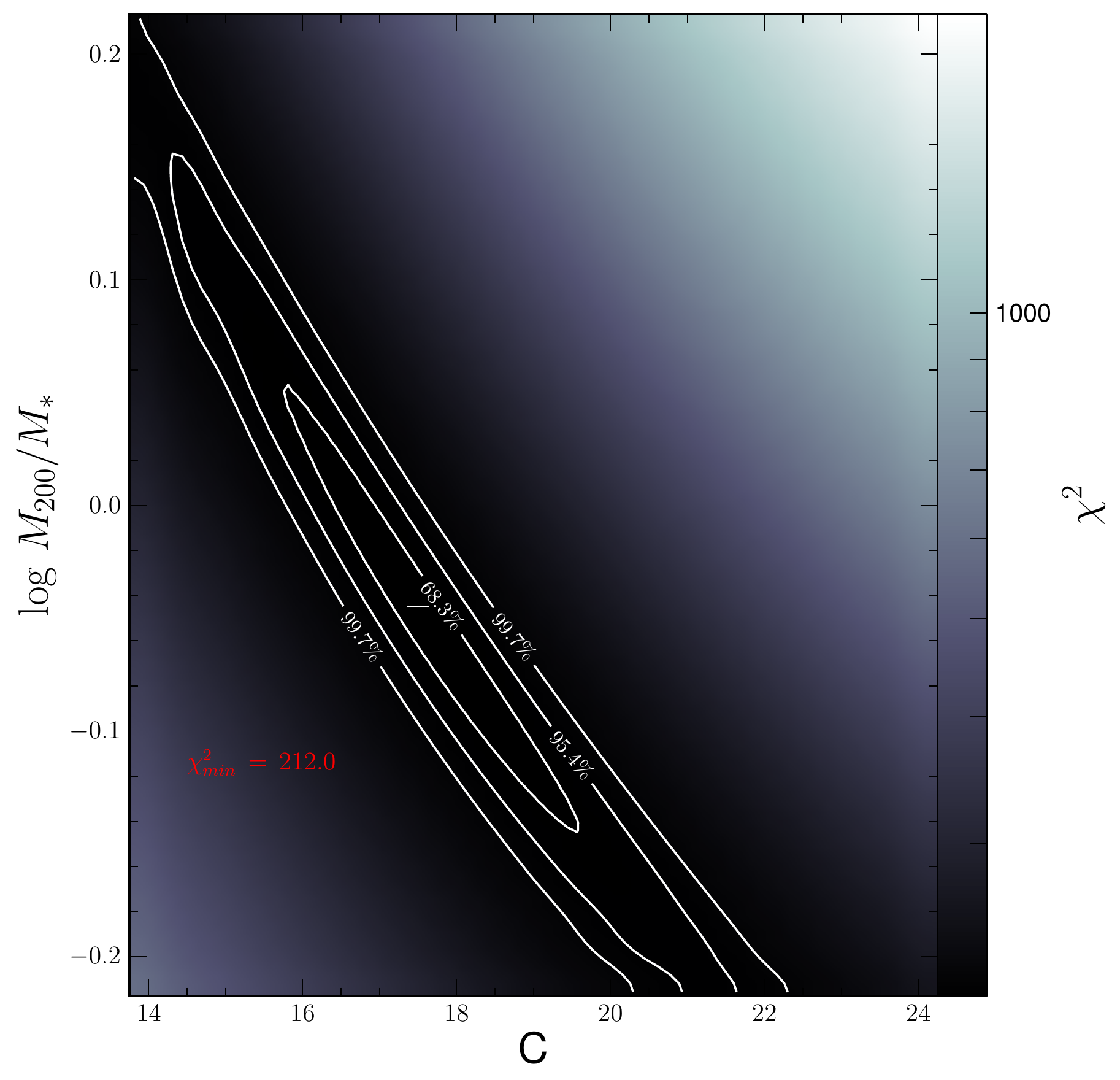}}					     
\subfigure{\includegraphics*[width=0.95\columnwidth]{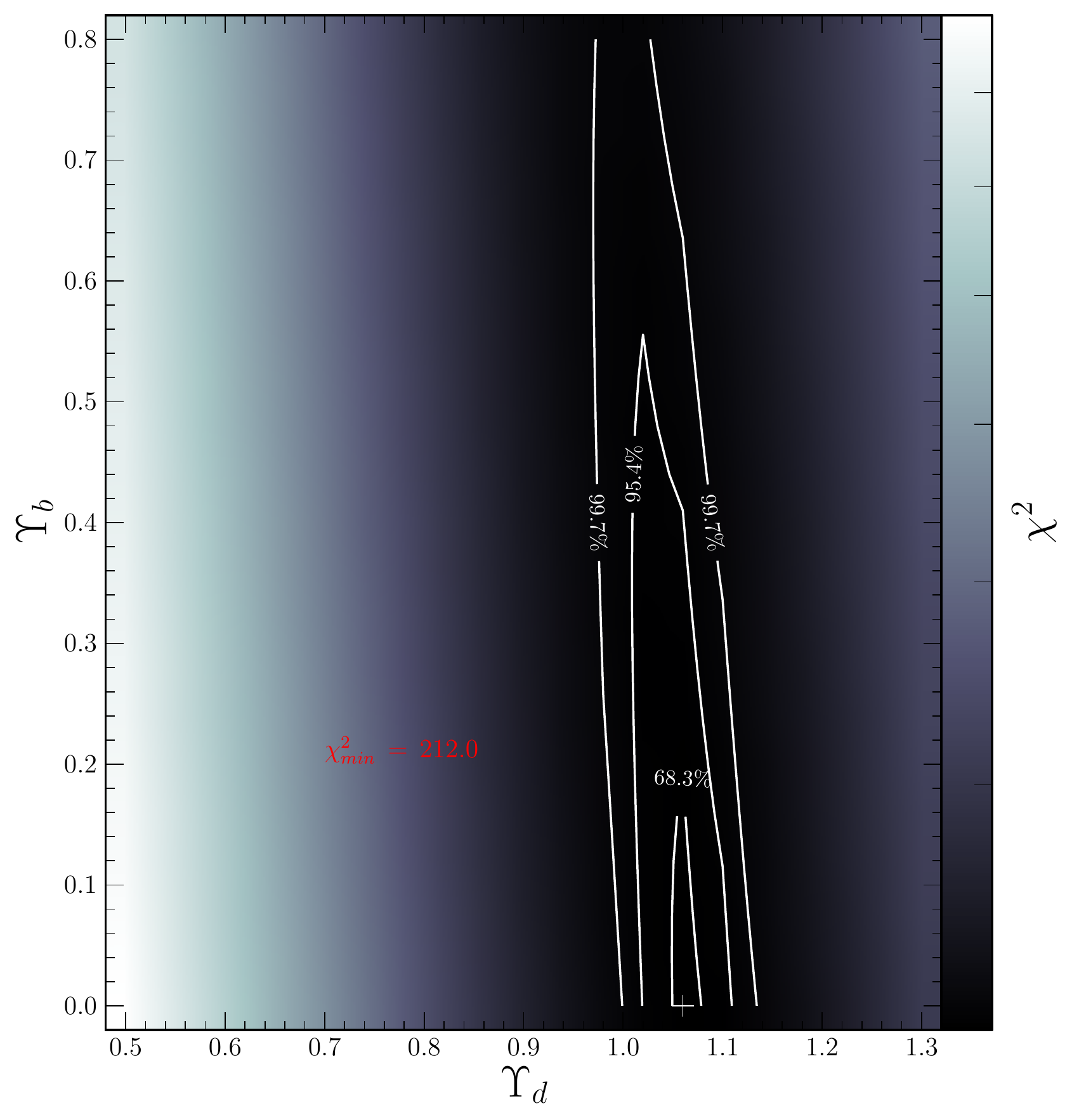}}
\subfigure{\includegraphics*[width=\columnwidth]{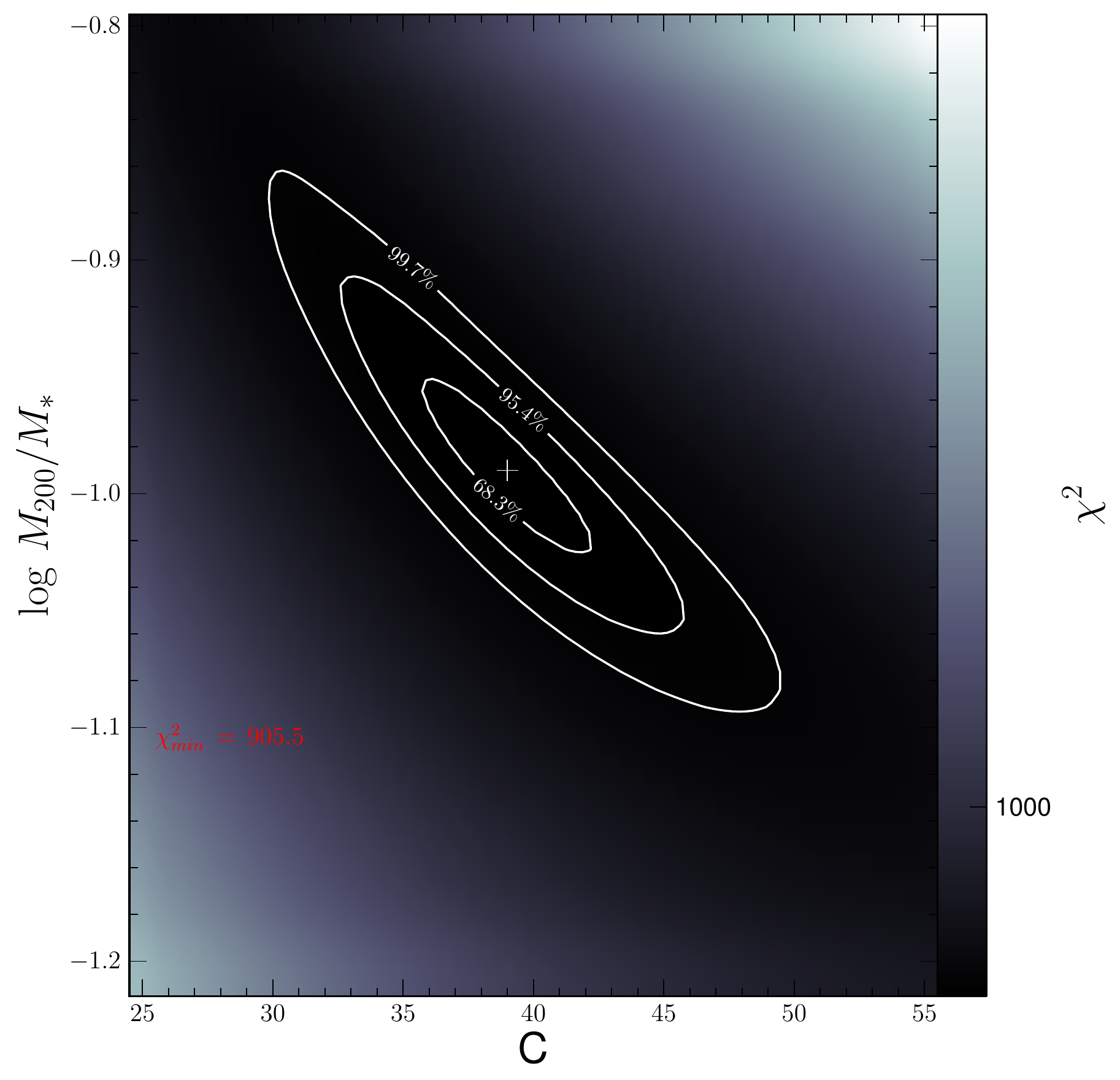}}					     
\subfigure{\includegraphics*[width=0.95\columnwidth]{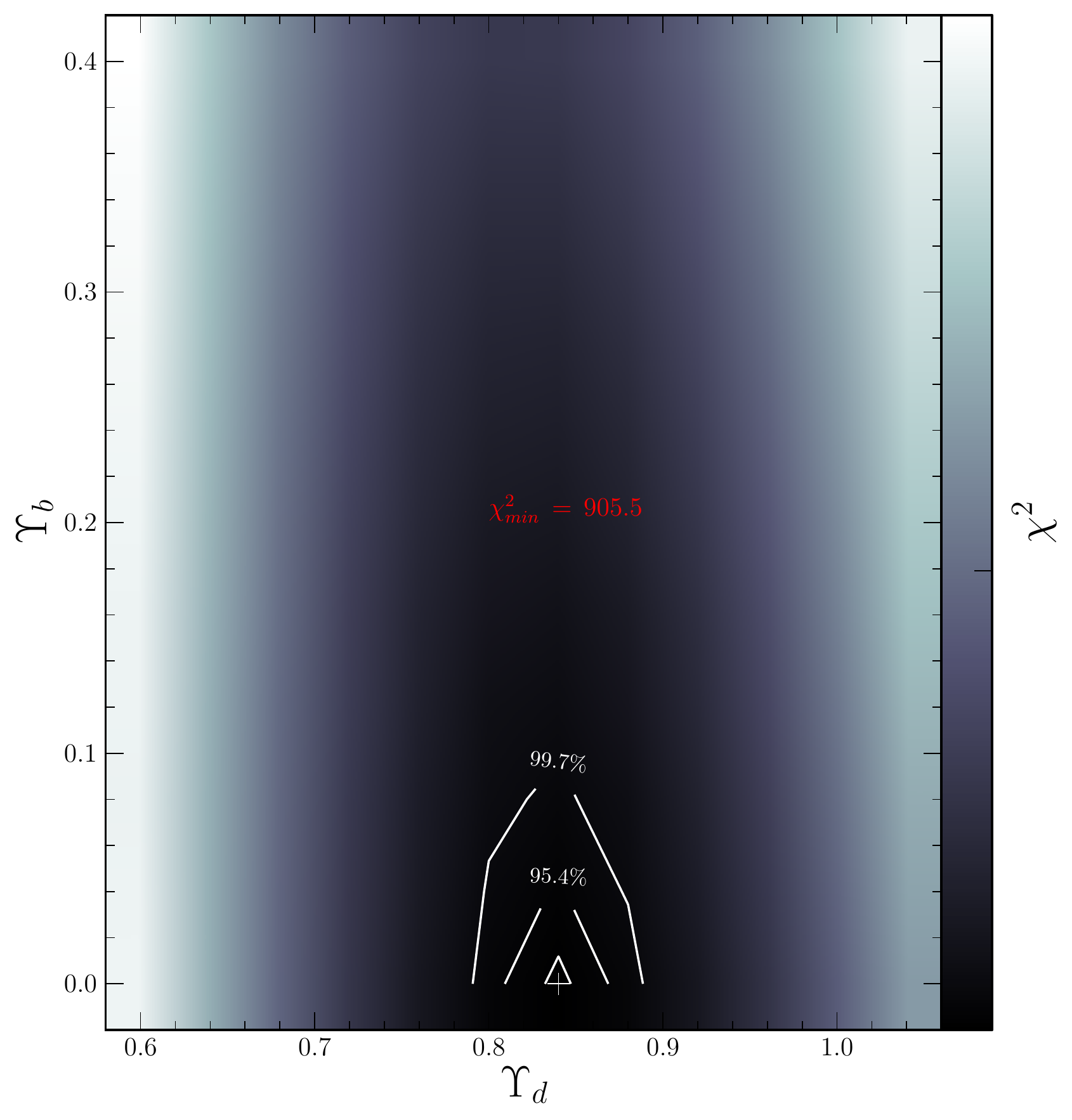}}							     
\caption{Left panels: $\chi^2$ space projections on the plane 
$\log(M_{200}/M_{\ast})$--$c$ for AM\,2058A (top) and AM\,1228A (bottom). 
Right panels: $\chi^2$ space projections 
on the plane $\Upsilon_{b}$--$\Upsilon_{d}$ for AM\,2058A (top) and AM\,1228A (bottom). 
Contours of $\Delta \chi^2$ corresponding to a probability of 68.3, 95.4 and  
99.7 per cent (1$\sigma$, 2$\sigma$, 3$\sigma$) for 1 degree of freedom. The 
plus symbol indicates the global minimum of $\chi^2$.}
\label{chimap}
\end{figure*}

The geometrical and  dynamic parameters for the best-fitting models 
for AM\,2058A and AM\,1228A, corresponding to the 
global minimum of $\chi^2$, are listed in Tables \ref{pargeo} and \ref{parmodels}, respectively. 
Uncertainties at 1$\sigma$ confidence (68\%) are also given.
Fig. \ref{chimap} shows the $\chi^2$ space projections of AM\,2058A and AM\,1228B on the planes 
$\log(M_{200}/M_{\ast})$--$c$ and $\Upsilon_{b}$--$\Upsilon_{d}$. These  
plots are useful to find the global minimum and its convergence pattern.
The convergence pattern in the plane $\log(M_{200}/M_{\ast})$--$c$  has a ``banana'' shape
due to the degeneracy between $M200$ and $c $; a decrease in  $c$ is balanced with an increase in 
$M_{200}$, and vice versa. The ``banana'' shape is more evident in the $\chi^2$ space projection of AM\,2058A
 (Fig. \ref{chimap}). Anyway, both convergence patterns are tight and deep, with a marked absolute minimum.
On the other hand, the shape of the converge pattern in the $\Upsilon_{b}$--$\Upsilon_{d}$ planes
is similar, in terms of the narrowness with  respect to $\Upsilon_{d}$ axis, in both galaxies. Regarding the 
$\Upsilon_{b}$ axis, the absolute minimum for both galaxies is 0.0, but the confidence curves of the AM\,1228A 
are tighter than in AM\,2058A. These results are not surprising, because both galaxies are late-type spirals
having B/T ratios rather low, $\sim$ 3\% (see Table \ref{contpar}). In general, the mass distribution for this type 
of galaxy is modelled without bulge \cite[e.g.,][]{vanalbada85,carignan85,begeman89,kuzio08}.

\begin{table}
\caption{Geometrical parameters for the best-fitting models for AM\,2058A and AM\,1228A}
\label{pargeo}
\begin{tabular}{lccccccccc}
\noalign{\smallskip}
\hline
\noalign{\smallskip}
Galaxy &  $V_{sys}$ (km\,s$^{-1}$) & $\Delta\,x$ (kpc) & $\Delta\,y$ (kpc) \\
\noalign{\smallskip}
\hline
\noalign{\smallskip}
AM\,2058A & 12157.3 & 0.47 & 0.94    \\
\hline
AM\,1228A & 5894.4  & 0.45  & -0.22  \\
\noalign{\smallskip}
\hline
\noalign{\smallskip}
\end{tabular}
\end{table}

\begin{table*}
\caption{Dynamic parameters for the best-fitting models for AM\,2058A and AM\,1228A}
\label{parmodels}
\begin{tabular}{ccccccccc}
\noalign{\smallskip}
\hline
\noalign{\smallskip}
Galaxy & $\Upsilon_{b}$ & $\Upsilon_{d}$ & $c$ &  $M_{200}/\mbox{M}_{\odot}$ &  $M_b/\mbox{M}_{\odot}$ & $M_d/\mbox{M}_{\odot}$ & $M_h/\mbox{M}_{\odot}$ & $M_t/\mbox{M}_{\odot}$\\
\hline
\noalign{\smallskip}
AM\,2058A & $0.00_{0.00}^{+0.28}$ & $1.06_{-0.32}^{+0.32}$ & $17.5_{-2.0}^{+2.0}$ &$0.902_{-0.275}^{+0.463}\times10^{12}$ & -& $8.47\times10^{10}$ & $9.03\times10^{10}$ & $1.75\times10^{11}$ \\   
\hline
AM\,1228A & $0.00_{0.00}^{+0.04}$ & $0.84_{-0.16}^{+0.08}$ & $39.0_{-3.0}^{+3.0}$ &$0.102_{-0.019}^{+0.043}\times10^{12}$ & - & $2.27\times10^{10}$ & $1.94\times10^{10}$ & $4.21\times10^{10}$ \\   
\noalign{\smallskip} 
\hline    
\noalign{\smallskip}
\end{tabular}
\end{table*}

The halo parameters found for AM\,2058A and AM\,1228A are 
compared with those reported for the MW, M\,31, and a late-type spiral galaxy model.  
Table \ref{compmodels} lists the parameters $c$, $R_{200}$ and $M_{200}$ for 
all those galaxies.  We see that halo parameters for AM\,2058A are similar to 
those of the MW and M\,31, while those for AM\,1228A  are quite different. 
The halo mass of AM\,2058A is roughly nine times larger than that of AM\,1228A. This difference 
may be related to galaxy size, since  the equivalent radius of 
the outermost isophote for AM\,2058A is 11.6\,kpc,  while for AM\,1228A is 5.7\,kpc. 

\begin{figure*}
\subfigure{\includegraphics*[width=\columnwidth]{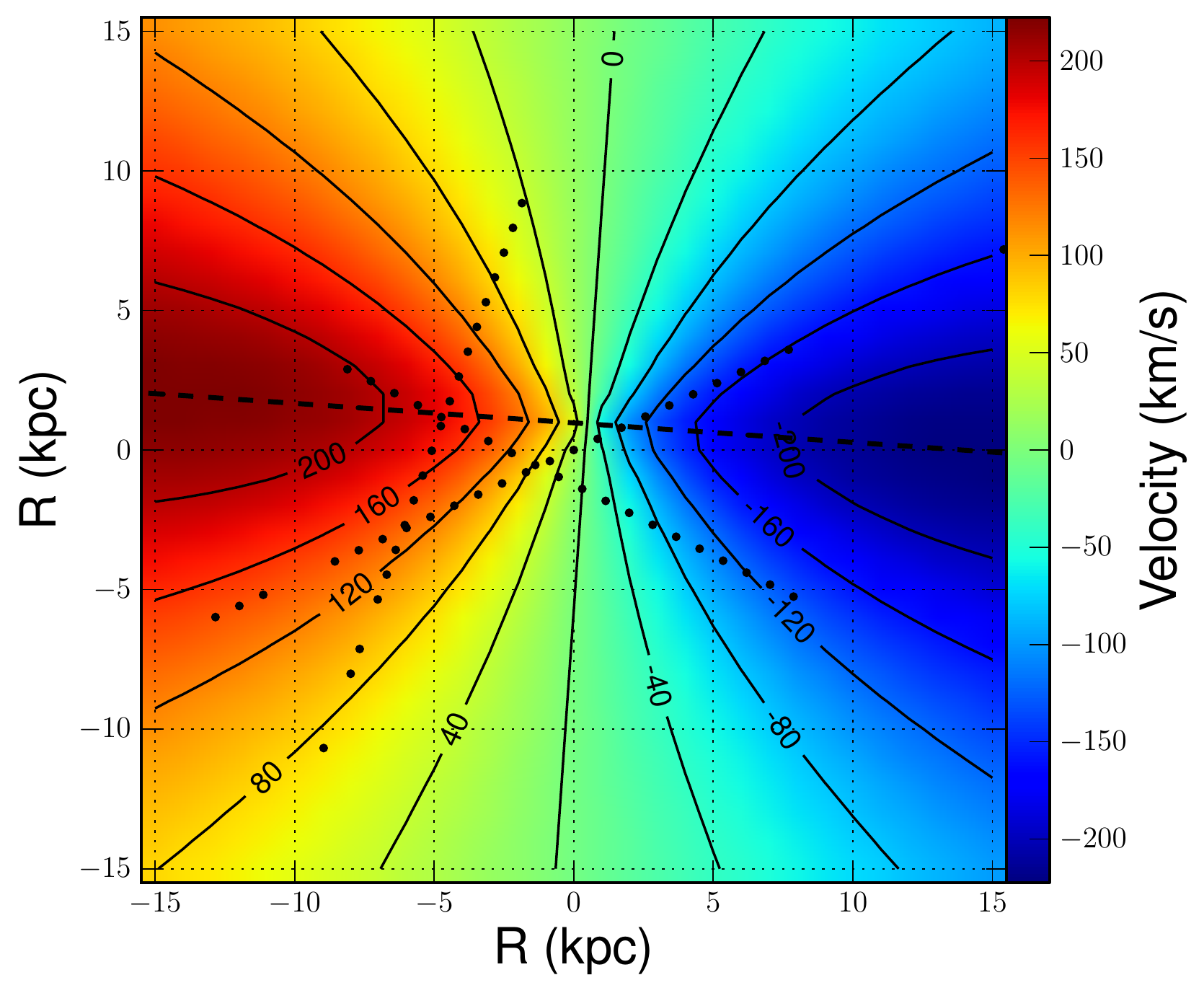}}
\subfigure{\includegraphics*[width=\columnwidth]{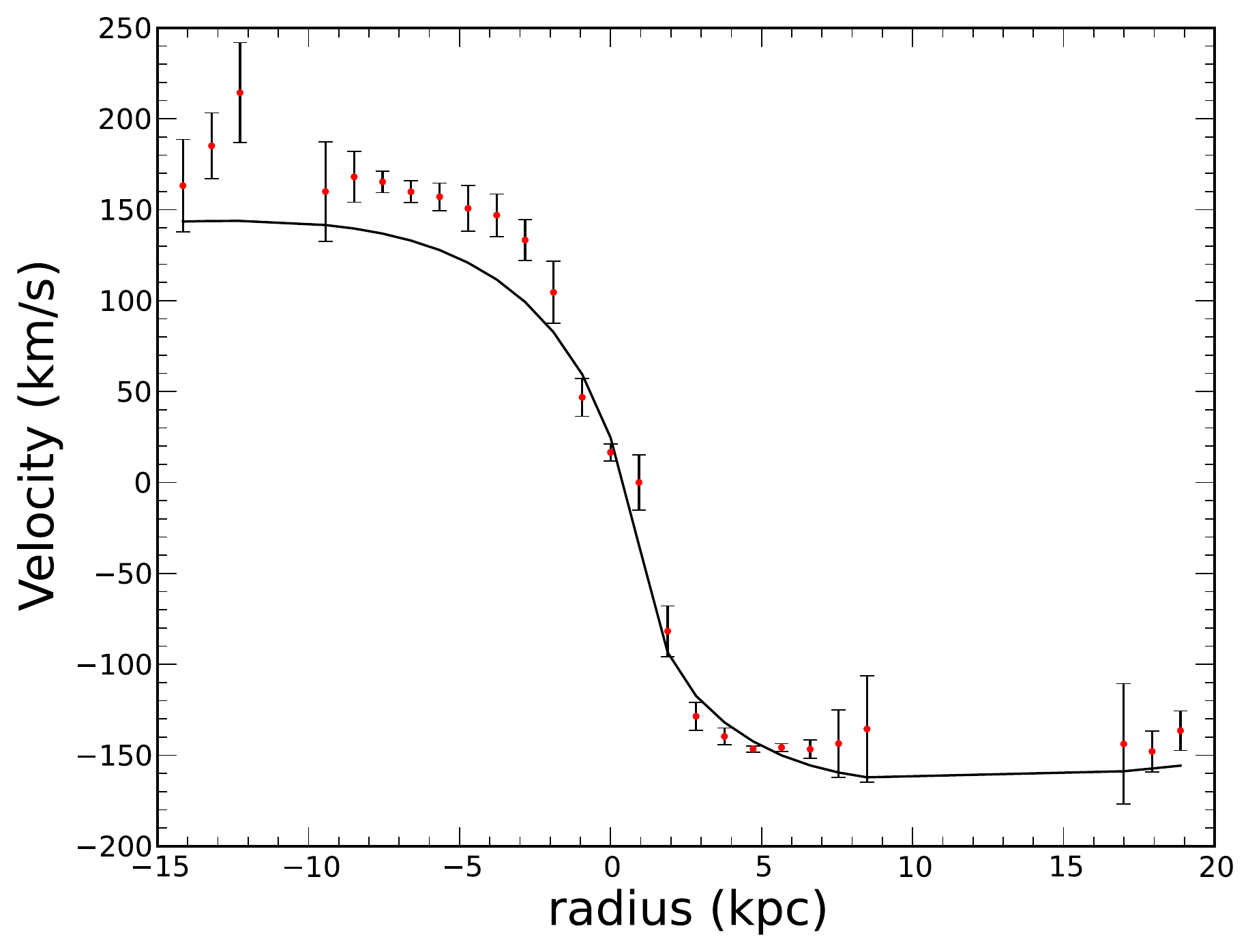}}
\subfigure{\includegraphics*[width=\columnwidth]{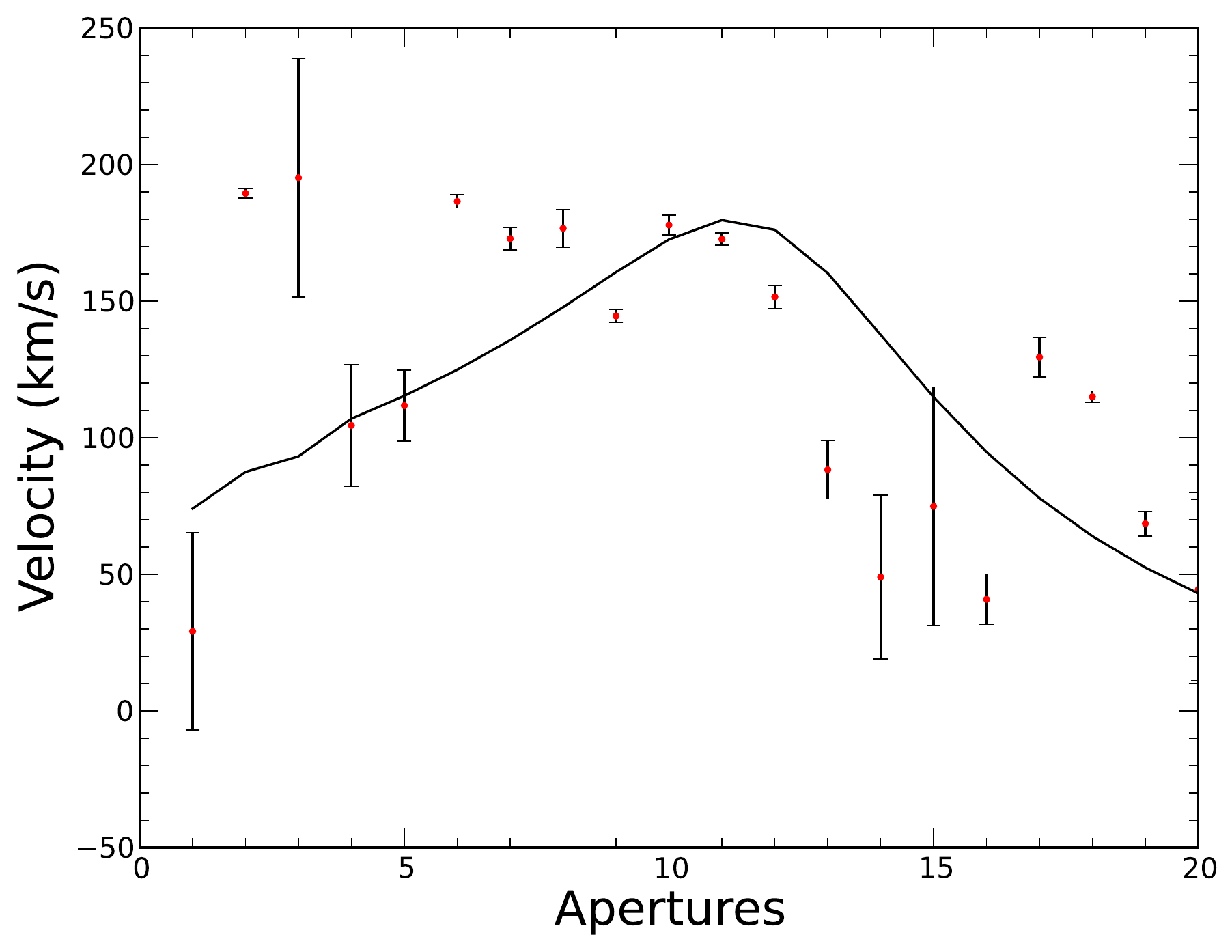}}
\subfigure{\includegraphics*[width=\columnwidth]{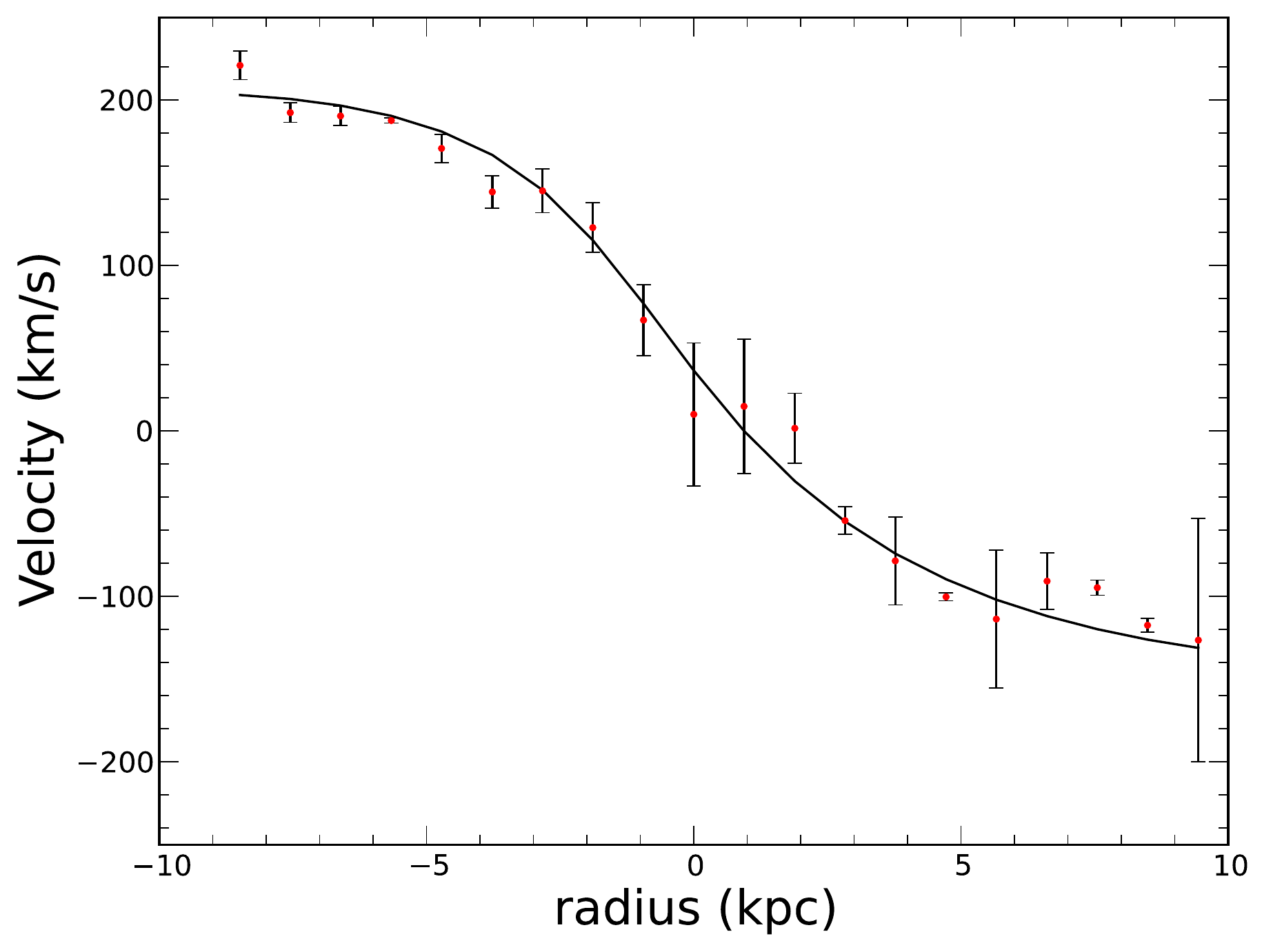}}	
\caption{The resulting velocity field (upper-left panel) from the  best-fitting model for AM\,2058A, and their projections 
overlaid on the observed radial velocity profiles along the slit positions  at PA=$350\degr$ (upper-right), PA=$125\degr$ (lower-left) and  PA=$42\degr$ (lower-right). The models of the observed radial velocity profiles are the continuous lines and observed data are red points with error bars.}
\label{fieldam2058}
\end{figure*}

Figure \ref{fieldam2058}  shows the velocity field modelled for AM\,2058A, together with its 
projections on  observed RVPs obtained at PA=350\degr, PA=42\degr\, and 
PA=125\degr. In general, there is a good match to the observations, in
particular, for the RVP along PA=42\degr. On the other hand, 
the model for RVP along PA=350\degr\, shows  a good agreement with the 
data in the approaching side, while in the receding side there is a departure 
between model and observations. This shift in velocity is of the order of 
$\Delta V \sim 20$\,km\,s$^{-1}$. We can interpret this departure in 
velocity as if this part of the galaxy is speeding up, and/or as if 
it is being deviated from the galactic plane due to interaction with AM\,2058B.
This type of irregularity has been reported in  two interacting 
systems, NGC\,5427 \citep{fuentes04} and AM\,1219-430 \citep{hernandez13}. 
It is also observed  in galaxies in high density environments, such 
as galaxy clusters \citep{dale01}.
Finally, the model for RVP along PA=42\degr\, follows the trend of 
the observed curve. However, some points have  $\Delta V >$  50\,km\,s$^{-1}$. 
Nevertheless, as commented in Sect. \ref{vel}, this behaviour is expected because 
the slit crosses the North-West arm (Fig. \ref{curveam1228A}).

\begin{figure*}
\subfigure{\includegraphics*[width=\columnwidth]{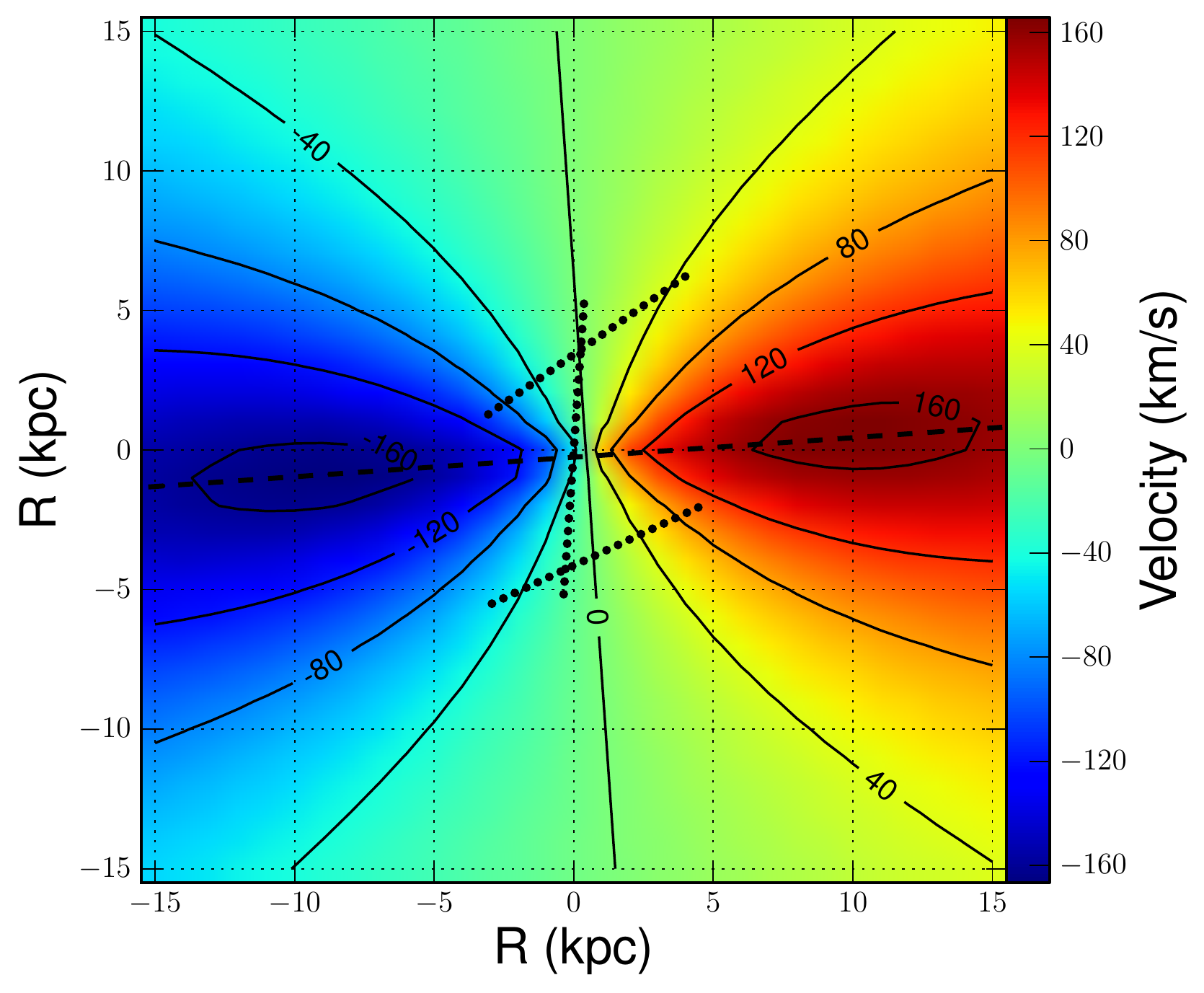}}
\subfigure{\includegraphics*[width=\columnwidth]{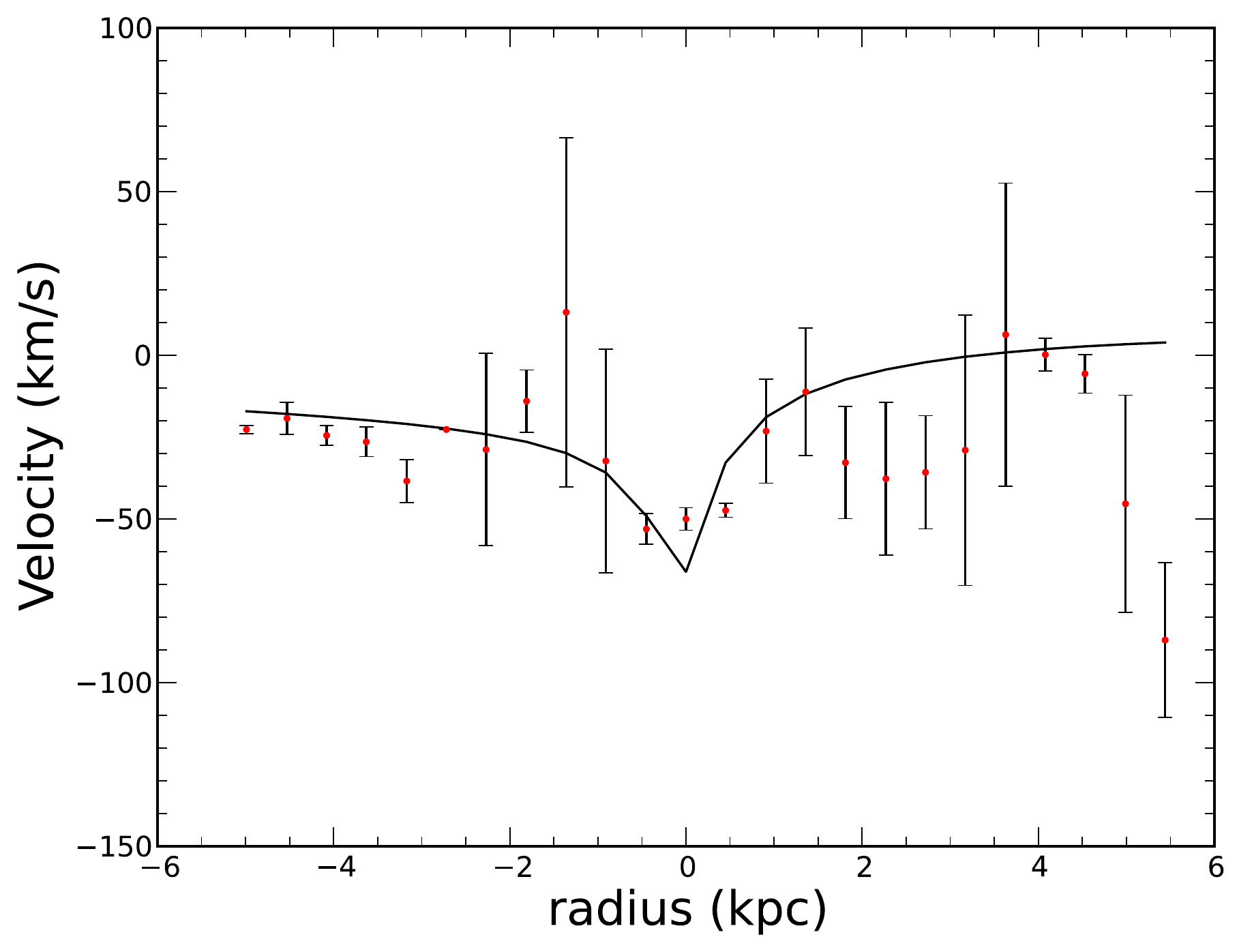}}
\subfigure{\includegraphics*[width=\columnwidth]{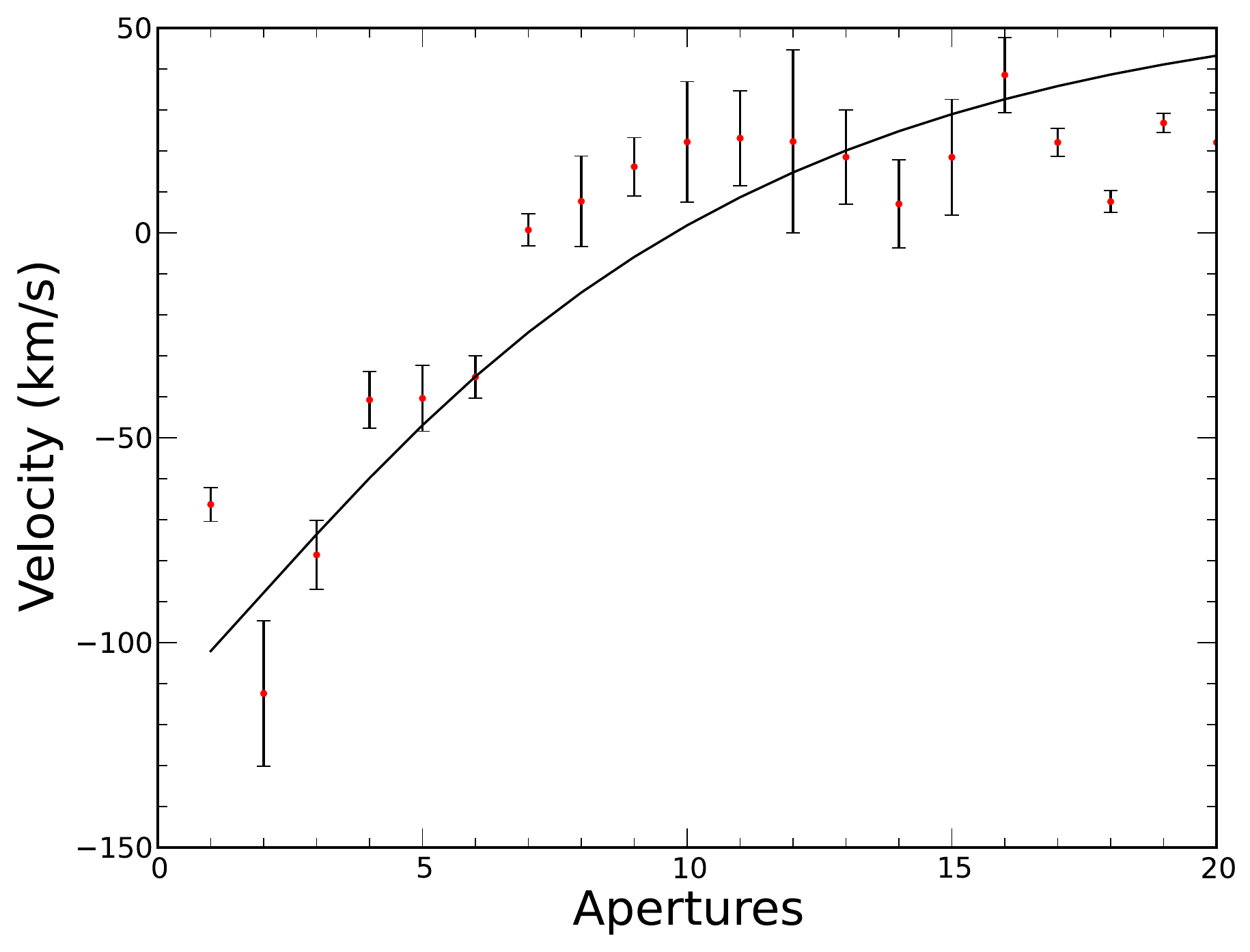}}			
\subfigure{\includegraphics*[width=\columnwidth]{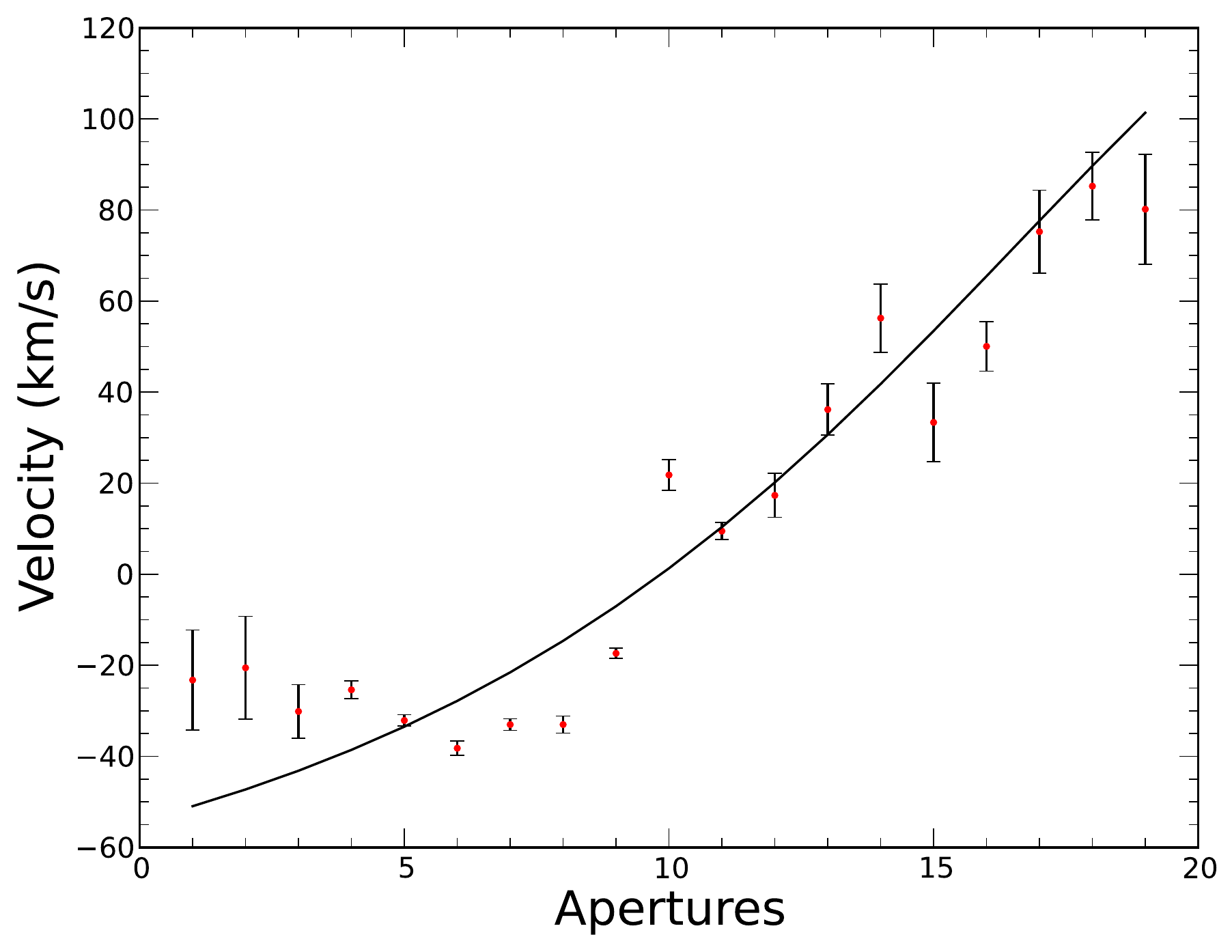}}						     
\caption{Same as  Fig. \ref{fieldam2058} for the best-fitting model of AM\,1228A, slit positions 
corresponding to PA=$319\degr$ (upper-right), PA=$10\degr$ (lower-left) and PA=$20\degr$ (lower-right).}
\label{fieldam1228}
\end{figure*}

\begin{figure*}
\centering
\subfigure{\includegraphics*[width=\columnwidth]{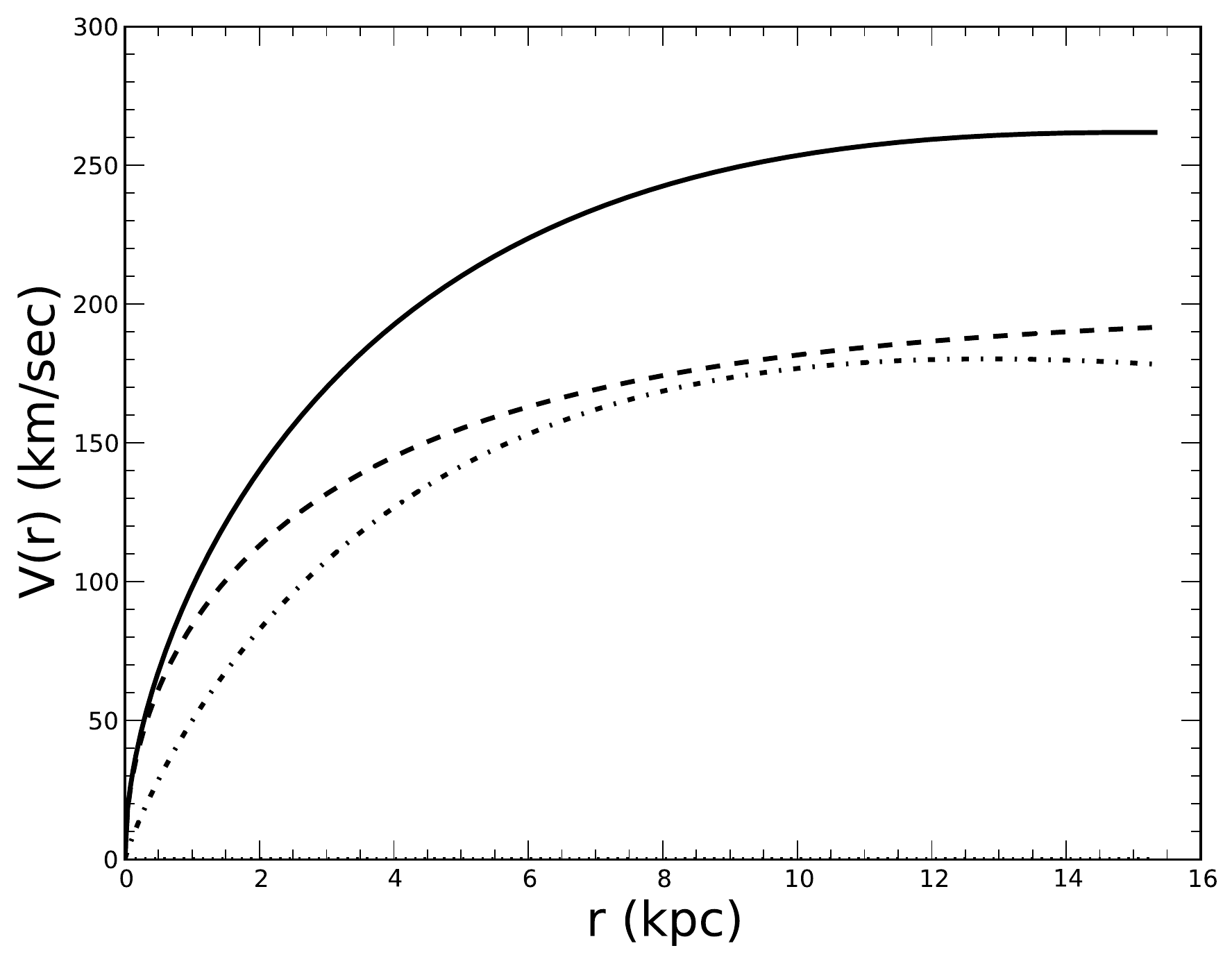}}
\subfigure{\includegraphics*[width=\columnwidth]{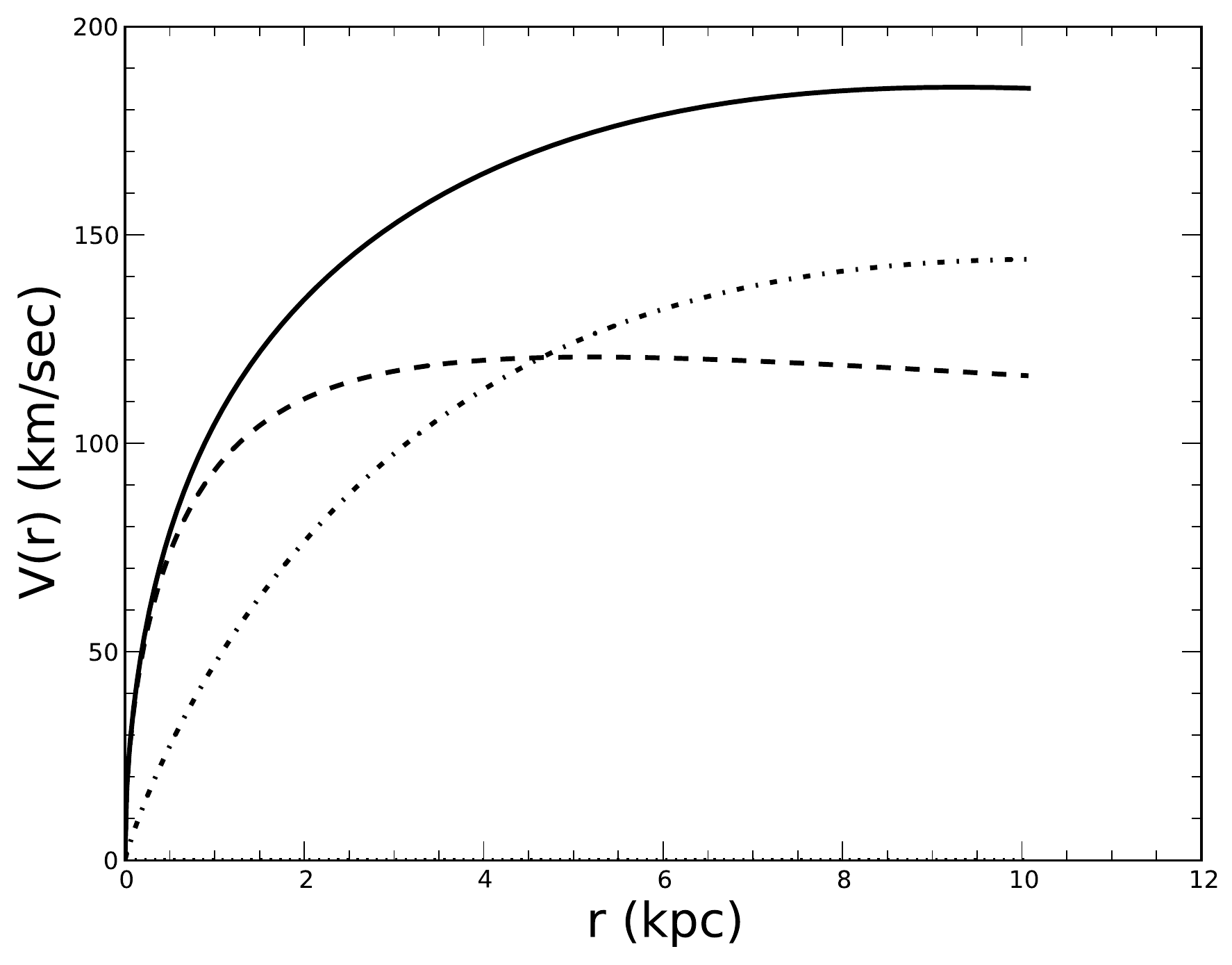}}				     
\caption{Final rotation curves (continuous lines) and components,  disc (dotted) 
and halo (dashed), from  the best-fitting models for AM\,2058A (left panel) and AM\,1228A (right).}
\label{modelbest}
\end{figure*}

Figure \ref{fieldam1228}  shows the resulting model for the velocity field of AM\,1228A, along with the 
projected  RVPs and data points for different slit positions. The observed data are  well 
represented by the model. However, the global minimum 
$\chi^2$  for AM\,1228A is much greater than that of AM\,2058A. 
This discrepancy may be due to two factors:
first,  as the model of AM\,1228A has more points to fit, it is expected that 
the $\chi^2$ be higher for this galaxy than that for  AM\,2058A. Secondly, the RVPs observed along 
AM\,1228A have more irregularities than those on AM\,2058A (Fig. \ref{curveam1228A}). Regarding the quality of the
modelled velocity field in specific RVPs, the RVP model 
along PA=319\degr\, follows the trend of the observed curve.  This RVP is close to the zero-velocity line 
of the modelled velocity field (Fig. \ref{fieldam1228}). On the other hand, the models for 
RVPs along  PA=10\degr\, and PA=20\degr\, also follow the trend of the observed curves, 
but do not reproduce completely the flat parts of these curves, the South and North parts, respectively. 
     
The final rotation curve models are shown in Fig. \ref{modelbest}, along with the disc and
halo components. For AM\,2058A, the disc and halo have similar weights along the overall radii
of the rotation curve, being the halo component somewhat more important than the disc component.
On the order hand, the middle part of the rotation curve of AM\,1228A ($0.0\lesssim r \lesssim 5.0$ kpc) is 
dominated by the halo component, while the disc becomes dominant 
in the outer parts  ($5.0 \gtrsim r$ kpc). It is worth mentioning that the 
disc component will dominate up to their peak at  
10.5\,kpc, after that, the curve will be dominated completely by the halo component.

The cumulative masses for the disc ($M_d$) and halo ($M_h$) components of the main galaxies, 
along with the total masses ($M_t$), are listed  in Table \ref{parmodels}.
These values are estimated  inside the equivalent radii of the outermost isophotes.  
The total masses of AM\,2058A and AM\,1228A are $1.75\times10^{11}$ and $4.21\times10^{10}M_{\odot}$,
respectively.
Thus, the ratio between the integrated masses of both galaxies is proportional to 
their physical sizes. We  found for AM\,2058A and AM\,1228A, the mass-to-light ratios, $M/L_r$, 
3.05 and 1.37, respectively. The $M/L_r$ value found for AM\,2058A is in agreement with the mean value, 
$M/L_r=4.5\pm1.8$,   derived  for a sample of 290 late-type spiral galaxies studied by 
\citet{broeils97}. The low $M/L_r$ value found for AM\,1228A may be accounted for by 
intense star formation.

\begin{table}
\caption{Comparison of the derived halo parameters for AM\,2058A and AM\,1228A  with those found for other galaxies}
\label{compmodels}
\begin{tabular}{lccccc}
\noalign{\smallskip}
\hline
\noalign{\smallskip}
Galaxy  & $c$ & $R_{200}$ (kpc) &  $M_{200}/\mbox{M}_{\odot}$ \\
\hline
\noalign{\smallskip}
AM\,2058A ($\chi^2_{\scriptsize{\mbox{min}}}$)    & 17 & 194  &  $0.902_{-0.275}^{+0.463}\times10^{12}$ \\
AM\,1228A ($\chi^2_{\scriptsize{\mbox{min}}}$)    & 39 & 94  &  $0.102_{-0.019}^{+0.043}\times10^{12}$ \\
MW $^{(a)}$            & 18 & 186  &  $0.8_{-0.2}^{+1.2}\times10^{12}$ \\
M\,31 $^{(b)}$                & 13 & 200  &  $1.04\times10^{12}$ \\
Simulation Sc $^{(c)}$        & 22 & 239  &  $0.79\times10^{12}$ \\
\noalign{\smallskip} 
\hline
\noalign{\smallskip}
\end{tabular}
{\bf Note:} values taken from, $^{(a)}$ \citet{battaglia05}, $^{(b)}$ \citet{tamm12} and 
$^{(c)}$ ERIS simulation for the formation of late-type spiral galaxies \citep{guedes11}.
\end{table}

\section{Conclusions}
\label{final}

A detailed study of the morphology, kinematics and dynamics of the minor mergers AM\,2058-381 and AM\,1228-260 was performed. 
The work is based in $r'$ images  and long-slit spectra in the  wavelength range from 4\,280 to 7\,130\AA\ , obtained with  
the  GMOS at Gemini South. The main results are the following:

\begin{enumerate}

\item AM\,2058A is  $\sim$ 5 times more luminous than
AM\,2058B, while AM1228A is  $\sim$ 20 times more luminous than AM\,1228B. In addition, AM\,2058-381 is a very luminous
minor merger when  compared to the MW system. In contrast, the main and secondary galaxies of the pair AM\,1228-260 have
similar luminosities similar to MW and LMC, respectively.

\item For AM\,1228-260  we detected a common isophote enclosing   the   members, which 
contributes  with 20\% of the total luminosity of the pair.
 For the main galaxy of AM\,2058-381, we detected two symmetric, long tidal tails, having
 only 5\%  of the system total luminosity.

\item The main galaxies, AM\,2058A and AM\,1228A, were decomposed in bulge, bar, ring and disc, while
the secondary galaxies, AM\,2058B and AM\,1228B, in bulge and disc.
The disc parameters derived for these galaxies agree with the average values found for galaxies 
with no sign of ongoing interaction or disturbed morphology \citep{fathi10a,fathi10b}. 
This indicates that the  symmetrization method is adequate to recover the unperturbed disc of the interacting galaxies.  

\item The studied galaxies have pseudo-bulges, with  a \citeauthor{sersic68} index $n<2$.
On the other hand, the B/T for AM\,2058A, AM\,1228A and 
AM\,1228B are very small (B/T $<0.1$), which is typical of 
late-type spirals. For AM\,2058B, B/T  is 0.34, which is similar to the early-type galaxies.   

\item The receding side of the RVP  along PA=350\degr\ of AM\,2058A 
departs  from the velocity field model. This departure can
be interpreted as if this part of the galaxy is speeding up, and/or as if 
it is being deviated from the galactic plane due to interaction with AM\,2058B.
There is a strong evidence that AM\,2058B be a 
tumbling body, rotating along its major axis.

\item The observed RVPs of AM\,1228A indicate that there is a 
misalignment between kinematic and photometric major axes.
Only a small fraction of non-interactions galaxies present this feature 
\citep{barrera14}. The observed  RVP at PA=319\degr\, for 
AM\,1228B is quite  perturbed, very likely due to the interaction 
with AM\,1228A.      

\item The NFW halo parameters ($M_{200}$ and $c)$ found for AM\,2058A are 
similar to those reported for the MW and M\,31, while the halo mass of  AM\,1228A 
is nine times smaller than that of AM\,2058A. It was found a M/L$_{r'}$ of 3.05 and 1.37 
for AM\,2058A and AM\,1228A, respectively. The M/L$_{r'}$ of  AM\,2058A is in agreement 
with the  mean value derived  for late-type spiral galaxies  \citep{broeils97}, while the low 
M/L$_{r'}$ obtained for AM\,1228A may be due to the intense star formation ongoing in this galaxy.

\end{enumerate}

The parameters obtained in this paper will serve  
as a starting point in future numerical simulations to reproduce the dynamical
histories and predict the evolution of the encounter of these pairs.

\section*{Acknowledgements}

We thank anonymous referee for important
comments and suggestions that helped to
improve the contents of this manuscript.  This work is based on observations obtained at the Gemini Observatory,
which is operated by the Association of Universities for Research
in Astronomy, Inc. (AURA), under a cooperative agreement with the NSF
on behalf of the Gemini partnership: the National Science Foundation (United States), 
the National Research Council (Canada), CONICYT (Chile), the Australian Research Council 
(Australia), Minist\'erio da Ciencia e Tecnologia (Brazil) and SECYT (Argentina). 
This work has been partially supported by the Brazilian institutions Conselho 
Nacional de Desenvolvimento Cient\'ifico e Tecnol\'ogico (CNPq) and 
Coordena\c c\~ao de Aperfei\c coamento de Pessoal de N\'ivel Superior (CAPES).
A.C.K.  thanks to support  FAPESP, process 2010/1490-3. I.R. thanks to
support FAPESP, process 2013/17247-9.


\begin{thebibliography}{}

\bibitem[\protect\citeauthoryear{Alfaro et al. }{2001}]{alfaro01} Alfaro, E.~J., P\'erez, E., Gonz\'alez-Delgado, R.~M., Martos, M.~A., Franco, J., 2001, ApJ, 550, 253
\bibitem[\protect\citeauthoryear{Arp \& Madore}{1987}]{arp87} Arp, H. \& Madore, B. 1987, .A Catalogue of Southern Peculiar Galaxies and Associations. Cambridge University Press, Cambridge
\bibitem[\protect\citeauthoryear{Barber\`a, Athanassoula \& Garc\'ia-G\'omez}{2004}]{barbera04} 	Barber\`a,~C., Athanassoula, E., Garc\'ia-G\'omez, C.,  \ 2004, A\&A, 415, 849
\bibitem[\protect\citeauthoryear{Barnes \& Hibbard}{2009}]{barnes09} Barnes, J.~E. \& Hibbard, J.~E. \ 2009, AJ, 137, 3071
\bibitem[\protect\citeauthoryear{Barton et al.}{2000}]{barton00}	Barton, E. ~J., Geller, M.~J., \& Kenyon, S.~J.,  \ 2000, ApJ, 530, 660	
\bibitem[\protect\citeauthoryear{Barrera-Ballesteros et al. }{2014}]{barrera14}	Barrera-Ballesteros, J.~K. et al., \ 2014, A\&A, 568, 70
\bibitem[\protect\citeauthoryear{Battaglia et al.}{2005}]{battaglia05} Battaglia, G., et al., \ 2005, MNRAS, 364, 433 
\bibitem[\protect\citeauthoryear{B\'edorf \& Portegies Zwart}{2012}]{bedorf12}	B\'edorf, J. \& Portegies Zwart,~S. \	2013, MNRAS, 431, 767
\bibitem[\protect\citeauthoryear{Begeman}{1989}]{begeman89} Begeman, K.~G., \ 1989, A\&A, 223, 47
\bibitem[\protect\citeauthoryear{Berentzen et al.}{2003}]{berentzen03} Berentzen, I., Athanassoula, E., Heller, C.~H., Fricke, K.~J., \ 2003, MNRAS, 341, 343
\bibitem[\protect\citeauthoryear{Bertola et al.}{1991}]{bertola91} Bertola, F., Bettoni, D., Danziger, J., Sadler, E., Sparke, L., \& de Zeeuw, T.\ 1991, ApJ, 373, 369
\bibitem[\protect\citeauthoryear{Binney \& Tremaine }{1987}]{binney87} Binney, J., \& Tremaine, S. \ 1987, Galactic Dynamics. Princeton University Press, Princeton, NJ
\bibitem[\protect\citeauthoryear{Blais-Ouellette et al.}{2001}]{blais01} Blais-Ouellette, S., Amram, P., \& Carignan, C. \ 2001, AJ, 121, 1952
\bibitem[\protect\citeauthoryear{Blanton et al.}{2003}]{blanton03} Blanton, M.~R et al. 2003, ApJ, 592, 819 
\bibitem[\protect\citeauthoryear{Broeils \& Courteau}{1997}]{broeils97} Broeils, A.~H., Courteau, S., \ 1997, in  Persic M. \&  Salucci P., eds, ASP Conf. Ser. Vol. 117, Dark and Visible Matter in Galaxies and Cosmological Implications. Astron. Soc. Pac., San Fransisco, p. 74
\bibitem[\protect\citeauthoryear{Bullock et al.}{2001}]{bullock01} Bullock, J. S., Kolatt, T. S., Sigad, Y., Somerville, R. S., Kravtsov, A. V.,
Klypin, A. A., Primack, J. R., \& Dekel, A. \ 2001, MNRAS, 321, 559
\bibitem[\protect\citeauthoryear{Buta}{1996}]{buta96}  Buta, R. 1996, in Buta R., Crocker, D. A., Elmegreen, B. G., eds, ASP Conf. Ser. Vol. 91, IAU Colloq. 157: Barred Galaxies, Astron.Soc.Pac., San Francisco, p. 11
\bibitem[\protect\citeauthoryear{Cabrera-Lavers \& Garz\'on}{2004}]{cabrera04} Cabrera-Lavers, A.; Garz\'on, F. 2004, AJ, 127, 1386
\bibitem[\protect\citeauthoryear{Carignan}{1985}]{carignan85} Carignan, C. \ 1985, AJ, 299, 59
\bibitem[\protect\citeauthoryear{Cen}{2014}]{cen14} 	Cen, R. \ 2014, ApJ, 785L, 15
\bibitem[\protect\citeauthoryear{Cole et al.}{2000}]{cole00}	Cole, S., Lacey, C.~G., Baugh, C.~M., Frenk, C.~S., 	2000, MNRAS, 319, 168
\bibitem[\protect\citeauthoryear{Cox et al.}{2008}]{cox08}	Cox, T.~J., Jonsson, P., Somerville, R.~S., Primack, J.~R.; Dekel, A., \ 2008, MNRAS, 384, 386
\bibitem[\protect\citeauthoryear{Dale et al.}{2001}]{dale01} Dale, D.~A., Giovanelli, R., Haynes, M.~P.,  Hardy E., Campusano L. E., 2001, AJ, 121, 1886
\bibitem[\protect\citeauthoryear{Dalcanton}{2007}]{dalcanton07} Dalcanton, J.~J. 2007, ApJ, 658, 941.
\bibitem[\protect\citeauthoryear{D{\'{\i}}az et al.}{2000}]{irapa2000} D{\'{\i}}az, R., Rodrigues, I., Dottori, H., \& Carranza, G.\ 2000, AJ, 119, 111 
\bibitem[\protect\citeauthoryear{Donzelli \& Pastoriza}{1997}]{donzelli97} Donzelli, C.~J., \& Pastoriza, M.~G.\ 1997, ApJS, 111, 181 
\bibitem[\protect\citeauthoryear{Elmegreen \& Elmegreen}{1985}]{elmegreen85} Elmegreen, B.~G., Elmegreen, D.~M., \ 1985, ApJ, 288
\bibitem[\protect\citeauthoryear{Elmegreen, Elmegreen \& Montenegro}{1992}]{eem92} Elmegreen, B.~G., Elmegreen, D.~M., \& Montenegro, L. 1992, ApJS, 79, 37
\bibitem[\protect\citeauthoryear{Elmegreen }{1998}]{elmegreen98}	Elmegreen, D.~M. \ 1998, Galaxies and Galactic Structure. Prentice Hall, Englewood Cliffs, NJ 
\bibitem[\protect\citeauthoryear{Eliche-moral et al.}{2011}]{eliche11} Eliche-Moral, M.~C., Gonz\'alez-Garc\'ia, A.~C., Balcells, M., Aguerri, J.~A.~L., Gallego, J., Zamorano, J., Prieto, M., 	\ 2011, A\&A, 533, 104
\bibitem[\protect\citeauthoryear{Emsellem et al. }{2006}]{emsellem06} Emsellem, E., Fathi, K., Wozniak, H., Ferruit P., Mundell C. G., Schinnerer E., \ 2006, MNRAS, 365, 367
\bibitem[\protect\citeauthoryear{Fathi et al.}{2010}]{fathi10a} Fathi, K., Allen M., Boch T., Hatziminaoglou E., Peletier R. F., \ 2010, MNRAS, 406, 1595
\bibitem[\protect\citeauthoryear{Fathi}{2010}]{fathi10b} Fathi, K., 2010, ApJ, 722, L120
\bibitem[\protect\citeauthoryear{Ferreiro \& Pastoriza}{2004}]{ferreiro04} Ferreiro, D.~L., \& Pastoriza, M.~G.\ 2004, A \& A, 428, 837 
\bibitem[\protect\citeauthoryear{Ferreiro, Pastoriza \& Rickes}{2008}]{ferreiro08} Ferreiro, D.~L., Pastoriza, M.~G.\ 2008, \& Rieks, M., A \& A, 481, 645 
\bibitem[\protect\citeauthoryear{Fisher \& Drory}{2008}]{fisher08} Fisher, D.~B., \& Drory, N. \ 2008, AJ, 136, 773
\bibitem[\protect\citeauthoryear{Freeman}{1970}]{freeman70} Freeman, K.~C. \ 1970, ApJ, 160, 811
\bibitem[\protect\citeauthoryear{Freeman}{1966}]{freeman66} Freeman, K.~C. \ 1966, MNRAS, 133, 47
\bibitem[\protect\citeauthoryear{Fuentes-Carrera et al.}{2004}]{fuentes04} Fuentes-Carrera, I., et al.\ 2004, A\&A, 415, 451 
\bibitem[\protect\citeauthoryear{Gadotti}{2008}]{gadotti08} Gadotti, D.~A. \ 2008, MNRAS, 384, 420
\bibitem[\protect\citeauthoryear{Gadotti}{2009}]{gadotti09} Gadotti, D.~A. \ 2009, MNRAS, 393, 1531
\bibitem[\protect\citeauthoryear{Garc\'ia-Barreto \& Rosado}{2001}]{garcia01} Garc\'ia-Barreto, J.~A., Rosado, M., \ 2001, AJ, 121, 2540
\bibitem[\protect\citeauthoryear{Garc\'ia-Barreto, Carrilo \& Vera-Villamizar}{2003}]{garcia03} Garc\'ia-Barreto, J~ A., Carrillo, R., Vera-Villamizar, N. \ 2003, AJ, 126, 1707
\bibitem[\protect\citeauthoryear{Guedes et al.}{2011}]{guedes11}	Guedes, J., Callegari, S., Madau, P., \& Mayer, L. \	2011, ApJ, 742, 76	
\bibitem[\protect\citeauthoryear{Grosbol}{1985}]{grosbol85}	Grosbol, P.~J., 1985, A\&AS, 60, 261
\bibitem[\protect\citeauthoryear{Graham \& Worley}{2008}]{graham08} Graham, A.~W., \& Worley, C.~C. \ 2008, MNRAS, 388, 1708
\bibitem[\protect\citeauthoryear{Hernquist \& Mihos}{1995}]{hernquist95} Hernquist, L., \& Mihos, J. C. \ 1995, ApJ, 448, 41
\bibitem[\protect\citeauthoryear{Hernandez-Jimenez et al.}{2013}]{hernandez13} Hernandez-Jimenez, J. A., Pastoriza, M. G., Rodrigues, I. Krabbe, A. C., Winge Cl\'audia., Bonatto, C. 2013, MNRAS, 435, 3342
\bibitem[\protect\citeauthoryear{Hibbard et al.}{2001}]{hibbard01} Hibbard, J.~E., van der Hulst, J.~M., Barnes, J.~E., Rich, R. ~M. \ 2001, AJ, 122, 2969
\bibitem[\protect\citeauthoryear{Hopkins et al.}{2010}]{hopkins10} Hopkins P.~F. et al., 2010, ApJ, 715, 202
\bibitem[\protect\citeauthoryear{Jedrzejewski}{1987}]{jedrzejewski87}	Jedrzejewski, R. ~I. 1987, MNRAS, 226, 747
\bibitem[\protect\citeauthoryear{Kennicutt et al.}{1987}]{kennicutt87} Kennicutt, R.~C., Jr., Roettiger, K.~A., Keel, W.~C., van der Hulst, J.~M., \& Hummel, E.\ 1987, AJ, 93, 1011 
\bibitem[\protect\citeauthoryear{Kent}{1987}]{kent87} Kent, S.~M. \ 1987,  AJ, 93, 816
\bibitem[\protect\citeauthoryear{Kewley et al.}{2010}]{kewley10}  Kewley L. J., Rupke D., Jabran Hahid H., Geller M. J., Barton E. J., 2010, ApJ, 721, L48.
\bibitem[\protect\citeauthoryear{Kormendy \& Kennicutt}{2004}]{kormendy04} Kormendy, John; Kennicutt, Robert C., Jr. \ 2004, ARA\&A, 42, 603
\bibitem[\protect\citeauthoryear{Krabbe et al.}{2008}]{krabbe08} Krabbe, A.~C., Pastoriza, M.~G., Winge, C., Rodrigues, I., \& Ferreiro, D. ~L. 2008, MNRAS, 389, 1593
\bibitem[\protect\citeauthoryear{Krabbe et al.}{2011}]{krabbe11} Krabbe, A.~C., Pastoriza, M.~G., Winge, C., Rodrigues, I., Dors, O. ~L.,\& Ferreiro, D. ~L. 2011, MNRAS, 416, 38
\bibitem[\protect\citeauthoryear{Krabbe et al.}{2014}]{krabbe14} Krabbe, A.~C., Rosa, D.~A., Dors, O.~L., Pastoriza M. G., Winge C., H\"agele G. F., Cardaci M. V., Rodrigues I., 2014, MNRAS 437, 1155
\bibitem[\protect\citeauthoryear{Kronberger et al.}{2006}]{kronberger06} Kronberger, T., Kapferer, W., Schindler, S., B{\"o}hm, A., Kutdemir, E., \& Ziegler, B.~L.\ 2006, A \& A, 458, 69 
\bibitem[\protect\citeauthoryear{Kuzio de Naray, McGaugh \& de Blok}{2008}]{kuzio08}	Kuzio de Naray, R., McGaugh, S.~S., de Blok, W.~J.~G., 2008, ApJ, 676, 920
\bibitem[\protect\citeauthoryear{Lambas et al.}{2003}]{lambas03}	Lambas, D.~G., Tissera, P.~B., Alonso, M.~S., Coldwell, G., \ 2003, MNRAS, 346, 1189
\bibitem[\protect\citeauthoryear{Lambas et al.}{2012}]{lambas12} Lambas D.~G., Alonso S., Mesa V., O’Mill A.~L., \ 2012, A\&A, 539, A45 
\bibitem[\protect\citeauthoryear{Larson \& Tinsley}{1978}]{larson78}	Larson, R.~B., Tinsley, B.~M., \ 1978, ApJ, 219, 46
\bibitem[\protect\citeauthoryear{Lucy}{1974}]{lucy74} Lucy, L.~B. \ 1974, AJ, 79, 745
\bibitem[\protect\citeauthoryear{Mendes de Oliveira et al.}{2003}]{mendes03} Mendes de Oliveira, C., Amram, P., Plana, H., \& Balkowski, C. 2003, AJ, 126,2635
\bibitem[\protect\citeauthoryear{Mihos \& Bothun}{1997}]{mihos97}  Mihos, J. C., \& Bothun, G. D. \ 1997, ApJ, 481, 741
\bibitem[\protect\citeauthoryear{Mihos \& Herquist}{1994}]{mihos94} 	Mihos, J.~C., Hernquist, L. \ 1994, ApJ, 425, L13
\bibitem[\protect\citeauthoryear{Monet et al.}{2003}]{monet03}	Monet, D. ~G. et al. \ 2003, AJ, 125, 984
\bibitem[\protect\citeauthoryear{Naab \& Burkert}{2003}]{naab03} Naab, T., Burkert, A., \ 2003, ApJ, 597, 893
\bibitem[\protect\citeauthoryear{Navarro, Frenk \& White}{1995}]{navarro95} Navarro, J. F., Frenk, C. S., \& White, S. D. M. \ 1995, MNRAS, 275, 56
\bibitem[\protect\citeauthoryear{Navarro, Frenk \& White}{1996}]{navarro96} Navarro, J. F., Frenk, C. S., \& White, S. D. M. \ 1996, ApJ, 462, 563
\bibitem[\protect\citeauthoryear{Navarro, Frenk \& White}{1997}]{navarro97} Navarro, J. F., Frenk, C. S., \& White, S. D. M. \ 1997, ApJ, 490, 493
\bibitem[\protect\citeauthoryear{Pastoriza, Donzelli \& Bonatto}{1999}]{pastoriza99} Pastoriza, M.~G., Donzelli, C.~J., \& Bonatto, C., 1999, A\&A, 347, 55
\bibitem[\protect\citeauthoryear{Palunas \& Williams}{2000}]{palunas00} Palunas, P., Williams, T.~B., 2000, AJ, 120, 2884
\bibitem[\protect\citeauthoryear{Perez, Michel-Dansac \& Tissera}{2011}]{perez11} Perez, J., Michel-Dansac, L., Tissera, P.~B., \ 2011, MNRAS, 417, 580
\bibitem[\protect\citeauthoryear{Peterson \& Huntley}{1980}]{peterson80} Peterson, C.~J. \&  Huntley, J.~M. \ 1980, ApJ, 242, 913
\bibitem[\protect\citeauthoryear{Pogge \& Martini}{2002}]{pogge02} Pogge, R.~W. \& Martini, P. \ 2002, ApJ, 569, 624	
\bibitem[\protect\citeauthoryear{Qu et al.}{2011}]{qu11} Qu Y., Di Matteo P., Lehnert M. D., van Driel W., Jog C. J., 2011, A\&A, 535, A5
\bibitem[\protect\citeauthoryear{Quinn, Herquist, \& Fullagar}{1993}]{quinn93}	Quinn, P.~J., Hernquist, L., Fullagar, D.~P., \ 1993 ApJ, 403, 74
\bibitem[\protect\citeauthoryear{Richardson}{1972}]{richardson72} Richardson, W.~H. \ 1972, J. Opt. Soc. Am., 62, 55
\bibitem[\protect\citeauthoryear{Robotham et al.}{2012}]{robotham12} 	Robotham, A.~S.~G. et al. 2012, 	MNRAS, 424, 1448
\bibitem[\protect\citeauthoryear{Rodrigues et al.}{1999}]{irapa1999} Rodrigues, I., Dottori, H., Brinks, E., \& Mirabel, I.~F.\ 1999, AJ, 117, 2695
\bibitem[\protect\citeauthoryear{Rosa et al.}{2014}]{rosa14} Rosa, D.~A., Dors, O.~L., Krabbe, A.~C., H\"agele, G.~F., Cardaci, M.~V., Pastoriza, M.~G., Rodrigues, I., Winge, C., 2014, MNRAS, 444, 2005
\bibitem[\protect\citeauthoryear{Rubin et al.}{1991}]{rubin91} Rubin, V.~C., Hunter, D.~A., \& Ford, W.~K. 1991, ApJS, 76, 153
\bibitem[\protect\citeauthoryear{Rubin et al.}{1999}]{rubin99} Rubin, V.~C., Waterman, A.~H., \& Kenney, J.~D.~P. 1999, AJ, 118, 236
\bibitem[\protect\citeauthoryear{Salo \& Laurikainen}{1993}]{salo93} Salo, H., \& Laurikainen, E. 1993, ApJ, 410, 586
\bibitem[\protect\citeauthoryear{Schlafly \& Finkbeiner}{2011}]{schlafly11} Schlafly, E.~F. \& Finkbeiner, D.~P. \ 2011, ApJ, 737, 103 
\bibitem[\protect\citeauthoryear{Schwarzkopf \& Dettmar}{2000}]{schwarzkopf00} Schwarzkopf, U., \& Dettmar, R.~J. \ 2000, A\&A, 361, 451
\bibitem[\protect\citeauthoryear{S\'ersic}{1968}]{sersic68} S\'ersic J.~L. \ 1968, Atlas de Galaxias Australes. Observatorio Astron\'omico, C\'ordoba
\bibitem[\protect\citeauthoryear{Simonneau \& Prada}{2004}]{simonneau04} Simonneau, E., Prada, F. \ 2004, Rev. Mex. Astron. Astrofis., 40, 69
\bibitem[\protect\citeauthoryear{Somerville, Primack \& Faber}{2001}]{somerville01}Somerville, R.~S., Primack, J.~R., Faber, S.~M.\ 	2001, MNRAS, 320, 504
\bibitem[\protect\citeauthoryear{Spergel et al.}{2007}]{spergel07} Spergel, D., et al. 2007, ApJS, 170, 377
\bibitem[\protect\citeauthoryear{Tamm et al.}{2012}]{tamm12}	Tamm, A., Tempel, E., Tenjes, P., Tihhonova, O. \& Tuvikene, T. \ 2012, A\&A, 546, 4
\bibitem[\protect\citeauthoryear{Thies \& Kohle}{2001}]{thies01} Thies, C., \& Kohle, S. 2001, A\&A, 370, 365 
\bibitem[\protect\citeauthoryear{van Albada et al.}{1985}]{vanalbada85} van Albada, T.~S., Bahcall, J.~N., Begeman, K., Sancisi, R. \ 1985, AJ, 295, 305
\bibitem[\protect\citeauthoryear{van den Broek et al.}{1991}]{vandenbroek91} van den Broek, A.~C., van Driel, W., de Jong, T., Goudfrooij, P., Lub, J., de Grijp, M.~H.~K. 1991, A\&AS, 91, 61
\bibitem[\protect\citeauthoryear{Walker, Mihos \& Herquist}{1996}]{walker96}	Walker, I.~R., Mihos, J.~C., Hernquist, L., 	\ 1996, ApJ, 460, 121
\bibitem[\protect\citeauthoryear{Wechler et al.}{2002}]{wechsler02} Wechsler, R. H., Bullock, J. S., Primack, J. R., Kravtsov, A. V., \& Dekel, A. \ 2002, ApJ, 568, 52
\bibitem[\protect\citeauthoryear{Weinzirl et al.}{2009}]{weinzirl09} Weinzirl, T., Jogee, S., Khochfar, S., Burkert, A., Kormendy, J. \ 2009, ApJ, 696, 411
\bibitem[\protect\citeauthoryear{Winge et al.}{in preparation}]{winge15} Winge, Cl\'adia, et al. in preparation 
\bibitem[\protect\citeauthoryear{Woods \& Gueller}{2007}]{woods07} Woods D.~F., Geller M. ~J. 2007, ApJ, 134, 527
	
\end{thebibliography}
\end{document}